**MINISTÈRE
DE L'ÉDUCATION
NATIONALE
ET DE LA JEUNESSE**

*Liberté
Égalité
Fraternité*

**Direction du numérique
pour l'éducation**

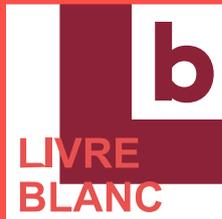

LIVRE
BLANC

# Enseigner et apprendre à l'ère de l'intelligence artificielle

**GTnum LINE #Scol_IA**
2020-2022

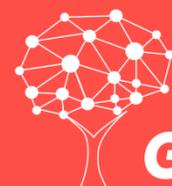

**GTnum**

# ENSEIGNER ET APPRENDRE A L'ERE DE L'IA
## Acculturation, intégration et usages créatifs de l'IA en éducation
## LIVRE BLANC

**Edition**
Margarida Romero, Université Côte d'Azur
Laurent Heiser, Université Côte d'Azur
Alexandre Lepage, Université de Montréal

**Auteurs des chapitres**
Alexandre Lepage, Université de Montréal
Anne Gagnebien, Université de Toulon
Audrey Bonjour, Université de Toulon
Aurélie Lagarrigue, INRIA Mnemosyne
Axel Palaude, INRIA Mnemosyne
Caroline Boulord, La Scientothèque
Charles-Antoine Gagneur, Conservatoire national des arts et métiers de Paris
Chloé Mercier, INRIA Mnemosyne
Christelle Caucheteux, Life Bloom Academy
Dominique Guidoni-Stoltz, Institut Agrosup Dijon
Florence Tressols, Maison de l'intelligence artificielle
Frédéric Alexandre, INRIA Mnemosyne
Jean-François Céci, Université de Poitiers
Jean-François Metral, Institut Agrosup Dijon
Jérémy Camponovo, DRANE de l'Académie de Nice
Julie Henry, Université de Namur
Laurent Fouché, Life Bloom Academy
Laurent Heiser, Université de Toulon et Université Côte d'Azur
Lianne-Blue Hodgkins, Life Bloom Academy
Ludovic Dibiaggio, Skema Business School
Margarida Romero, Université Côte d'Azur et INRIA Mnemosyne
Marie-Hélène Comte, INRIA Mnemosyne
Michel Durampart, Université de Toulon
Patricia Corieri, La Scientothèque
Paul Olry, Institut Agrosup Dijon et Conservatoire national des arts et métiers de Paris
Pauline Reboul, Université de Toulon
Philippe Bonfils, Université de Toulon
Sami Ben Amor, Université de Toulon
Simon Collin, Université du Québec à Montréal
Solange Ciavaldini-Cartaut, Université Côte d'Azur
Thierry Viéville, INRIA Mnemosyne et Université Côte d'Azur
Victoire Batifol, Life Bloom Academy
Yann-Aël Le Borgne, La Scientothèque et Université Libre de Bruxelles

**Pilotage du GTnum**
Margarida Romero, Laurent Heiser et Maryna Rafalska, Université Côte d'Azur

# Sommaire

# AVANT-PROPOS

Au cours des deux dernières années, le Groupe de Travail numérique (GTnum) « Renouvellement des pratiques numériques et usages créatifs du numérique et IA » (Scol_IA) a rassemblé des dizaines de personnes de tout horizon pour étudier le potentiel pédagogique et les limites de l'intelligence artificielle (IA) appliquée au domaine de l'éducation. Ce contexte de recherche est singulier à deux égards : d'abord, le déploiement de l'IA en éducation est déjà entamé et la réflexion pédagogique qui l'accompagne est animée par une certaine urgence d'agir. En s'impliquant dans la réflexion sur l'IA à l'école, l'équipe cherche à donner des repères pour que l'intégration de l'IA à l'école soit réussie et qu'elle profite aux élèves comme aux enseignants. Ensuite, ce contexte est marqué par une interdisciplinarité qui apporte une perspective multiple à cet ouvrage. En guise d'exemple, nous aborderons dans l'ouvrage la modélisation de l'apprenant, un défis des neurosciences computationnelles pouvant bénéficier des perspectives d'autres disciplines, dont les sciences de l'éducation et de la formation. Pendant deux ans, nous nous sommes affairés à faire dialoguer des disciplines, et incidemment des personnes, qui s'ancrent dans des univers épistémologiques différents, au premier plan desquels se retrouvent les sciences informatiques et de celles de l'éducation. À leurs côtés, la psychologie, l'économie et la sociologie. Nous avons également engagé des acteurs éducatifs de l'écosystème azuréen comme la Maison de l'intelligence artificielle ou Terra Numerica, en collaboration avec la Direction régionale académique du numérique éducatif (région Provence-Alpes-Côte-d'Azur). Les praticiens professeurs de terrain ont permis de renforcer cette démarche et facilité la collaboration avec des élèves du primaire et du secondaire au cours de ce GTnum. Nous avons mis en œuvre des mécanismes de collaboration permettant d'éclairer l'IA en éducation depuis plusieurs perspectives.

Fruit de cette collaboration entre les différents acteurs éducatifs et de recherche, ce livre blanc tient compte de l'expérience partagée au cours de ces deux ans. Cet ouvrage est à la fois destiné aux praticiens de l'éducation et aux chercheurs ; il présente la diversité des formes que peut prendre l'IA en éducation parmi lesquelles se trouvent notamment les tuteurs intelligents, les tableaux de bord pour la réussite éducative et les aides à la décision pour les enseignants. Il aborde aussi la question de l'éducation à l'IA pour tous. Des réflexions sont proposées sur le caractère disruptif de l'IA par rapport à la forme scolaire actuelle. Nous avons tenté de donner un sens à l'IA en éducation malgré les contradictions qu'elle porte, entre la recherche du modèle d'apprenant universel et la personnalisation des parcours d'apprentissage, entre l'agentivité des enseignants et l'automatisation de leurs actions, entre l'innovation technologique et la préservation du patrimoine scolaire.

# INTRODUCTION

Margarida Romero[1]
Ludovic Dibiaggio[2]
Laurent Heiser[1]
Alexandre Lepage[3]

[1] Laboratoire d'innovation pour le numérique en éducation, Université Côte d'Azur
[2] Skema Business School
[3] Université de Montréal

Les défis sociaux auxquels nous faisons face appellent à la l'actualisation des modèles pédagogiques. A contrario de certains domaines techniques, où l'utilisation de systèmes d'IA peut produire des résultats aux avantages consensuels, l'éducation est un contexte particulièrement marqué par la richesse des relations humaines. De ce fait, il faut apprendre à distinguer deux types de situations : d'une part, celles qui nécessitent une intervention humaine pour compléter les résultats proposés par un algorithme (par exemple pour des décisions liées à l'orientation, le décrochage, le mentorat ou encore l'obtention d'un certificat) et, d'autre part, les situations qui ne nécessitent plus aucune intervention humaine (résultats fournis par un algorithme adaptatif).

D'emblée, il convient de présenter quelques définitions pour le lecteur non spécialiste. Le terme « intelligence artificielle » (IA) recouvre un ensemble de théories et de techniques qui traite de problèmes dont la résolution fait normalement appel à l'intelligence humaine. Aujourd'hui, l'IA s'appuie principalement – mais pas seulement – sur l'apprentissage automatique, que certains connaissent sous le terme anglais de *machine learning*. L'apprentissage automatique exploite des méthodes mathématiques pour analyser de grands ensembles de données (dites données massives ou *big data*) et identifier des tendances (corrélations, similarités, *patterns*). De cette analyse découlent des modèles permettant de prédire, parfois avec des taux d'exactitude impressionnants, certaines variables comme les résultats scolaires ou le risque de décrochage. Les modèles d'apprentissage automatique peuvent aussi être employés pour des problèmes de classification où l'on cherchera à identifier l'appartenance à un profil d'apprenant en vue d'adapter des actions.

## L'IA, une jeune science

Depuis les années 1950, l'IA est un domaine de recherche foisonnant mélangeant mathématiques appliquées et psychologie. Les techniques ont évolué, plusieurs définitions ont été proposées, mais l'idée de repousser les limites des actions humaines pouvant être automatisées demeure centrale. On parlera tantôt de prise de décisions, tantôt d'aide à la décision, mais il demeure que l'IA affiche l'ambition de modéliser certaines facultés mentales de l'être humain. De nos jours, avec la multiplication des ensembles de données massives, le raffinement des méthodes d'analyse et la visibilité de certains projets qui stimulent l'imaginaire collectif comme les voitures autonomes, les applications potentielles de l'IA dans différents domaines de connaissances connaissent un essor important.

Ce livre blanc a l'ambition de livrer un éclairage scientifique sur les outils et modèles d'apprentissage automatique, à la fois supports pédagogiques et outils expérimentaux permettant d'approfondir notre compréhension du fonctionnement de l'apprentissage humain. Il présente une science en train de se faire, ouverte à la discussion, et non un discours établi et prescriptif. Les débats théoriques et les rapports d'expériences réalisés dans des environnements scolaires ou dans l'enseignement professionnel nous racontent un métier en évolution, celui d'enseignant.

Les différentes expériences menées dans le cadre du GTnum Scol_IA mettent aussi en évidence les difficultés rencontrées par les acteurs éducatifs. Nous devons tenir compte des compétences de ces derniers et de l'acculturation sous-jacente à l'intégration de l'IA, tant dans les systèmes formels d'éducation que dans les systèmes informels.

L'ouvrage est structuré en deux parties. La première concerne les pratiques émergentes d'enseignement de l'IA auprès de différents publics principalement scolaires. Ainsi, les chapitres 1 à 4 présentent des réalisations issues de l'éducation ou de la médiation scientifique qui peuvent inspirer l'actualisation des pratiques enseignantes dans les écoles. La seconde partie de l'ouvrage propose un recul réflexif pour aborder une question fondamentale : quelle place devrait occuper l'IA en éducation ? Les chapitres 5 à 9 présentent des situations portant tant sur des contextes d'éducation formelle et informelle, mais également des activités pour l'acculturation et la formation à l'IA.

Les usages de l'IA en éducation sont abordés à partir du modèle #5c21 (Romero et al., 2017). Selon ce modèle, tout usage du numérique se situe sur un spectre en cinq niveaux allant de la consommation passive à la co-création participative de connaissances. Nous cherchons en particulier à mettre en valeur les usages de l'IA comme moyen de conduire des activités renforçant la créativité en tenant compte des contextes et niveaux éducatifs différents (Heiser et al., 2022). C'est pourquoi nous proposons une échelle des usages créatifs de l'IA en éducation. Aux premier et deuxième niveaux se situent les usages de l'IA qui n'engagent ni l'apprenant ni l'enseignant de manière créative. Aux troisième et quatrième niveaux sont représentés les usages de l'IA qui soutiennent la créativité individuelle ou collective. Finalement, au cinquième niveau, les usages de l'IA en éducation soutiennent la résolution créative de problèmes à visée participative ou sociétale. Nous espérons que le livre aidera les enseignants à imaginer des activités poursuivant cette ambition.

# PARTIE 1 ///
# RETOURS
# D'EXPERIENCE

# 1 /// UN MOOC POUR INITIER A L'IA : « INTELLIGENCE ARTIFICIELLE AVEC INTELLIGENCE »

Frédéric Alexandre[1]
Marie-Hélène Comte[1]
Aurélie Lagarrigue[1]
Chloé Mercier[1]
Axel Palaude[1]
Margarida Romero[1 2]
Thierry Viéville[1 2]

[1] Institut national de recherche en sciences et technologies du numérique, Équipe Mnémosyne
[2] Laboratoire d'innovation pour le numérique en éducation, Université Côte d'Azur

Pour appréhender l'IA au quotidien, il faut former les citoyens dès la fin du primaire et tout au long de la vie à la compréhension de ses fondamentaux. Le MOOC « Intelligence artificielle avec intelligence » (IAI) est une formation hybride et participative permettant à des citoyens de s'initier à l'IA de manière à la fois théorique et expérimentale, par l'essai de différentes technologies comme la reconnaissance d'images. La formation permet de mieux comprendre, pour mieux appréhender cette IA désormais présente au quotidien.

## Introduction

Tout le monde est concerné par les technologies numériques. Les enjeux de l'acculturation au numérique doivent désormais tenir compte de ce qu'on appelle l'intelligence artificielle (IA), car les nouvelles technologies s'en réclamant sont de plus en plus nombreuses et accessibles. Il est important de permettre à chacun de comprendre le fonctionnement des mécanismes de l'IA afin de développer un regard critique et créatif par rapport à ses usages actuels et futurs. Dans ce chapitre, nous présentons un cours en ligne ouvert massivement (*Massive open online course – MOOC*) hybride et participatif intitulé *Intelligence artificielle avec intelligence* (IAI) dont l'objectif principal est de permettre à tout le monde, au-delà des publics scolaires, de développer une compréhension de la manière dont l'IA est intégrée dans notre vie. À l'origine, ce projet s'inspire de l'ambition de la Finlande de former 1 % de sa population à l'IA (Roos et Storchan, 2020), mais aussi du succès de projets antérieurs visant, par exemple, l'initiation des enseignants à la pensée informatique (Mariais et al. 2019). Début 2022, plus de 32 000 personnes avaient suivi la formation et ont témoigné d'un taux de satisfaction de 94 % (Alexandre et al., 2020). Par cette initiative, nous participons au vaste projet d'université citoyenne ubiquitaire en culture numérique et en sciences numériques (Atlan et al. 2019).

En premier lieu, ce chapitre rapporte la réflexion qui a conduit à la réalisation d'un MOOC. Il sera question de l'importance de développer une culture de l'IA pour prendre part aux réflexions éthiques qui l'entourent. Il sera aussi question des partenaires et du financement qui ont permis de faire de ces idées une réalité. En second lieu, la structure du cours et les activités qui le composent seront présentées. Finalement, nous partagerons les retombées du MOOC sur les participants et celles sur les milieux scolaires.

# Pourquoi une formation citoyenne sur l'IA

L'IA concerne tous les citoyens d'abord parce qu'elle est de plus en plus présente dans tous les secteurs de la société et ensuite parce qu'il est nécessaire de la comprendre pour participer aux débats éthiques qu'elle engendre. Il est normal de se questionner sur la pertinence de confier à des algorithmes des tâches qui mènent à des décisions cruciales, par exemple en matière de justice, d'embauche, ou d'autres situations à fortes conséquences humaines. L'utilisation de l'IA pour ces cas sensibles doit être soutenue par une réflexion éthique, et c'est notamment sur cette base que s'est construit le MOOC. Réfléchir à l'acceptabilité ou non de l'IA dans certaines situations implique de maîtriser des notions fines comme l'interprétabilité[1] et l'explicabilité[2], ou encore les causes des biais dans les mécanismes d'IA venant des données ou des algorithmes. Aucun sujet technique ne peut être abordé sans que ces aspects éthiques ou sociétaux le soient aussi, comme c'est le cas de certaines formations en robotique pédagogique (p. ex. le projet *Robotination* où des enfants doivent construire un robot à partir de leurs représentations et de matériel électronique). D'un point de vue éthique, la responsabilité est toujours *humaine*, par exemple si on laisse l'algorithme décider, c'est notre décision de le faire : déléguer la décision à un algorithme au lieu de la prendre soi-même, c'est un choix et c'est un humain qui doit faire ce choix. Si une personne choisit de *faire confiance* à une machine avec un algorithme d'IA, elle fait surtout confiance à son propre jugement quant aux performances de ce mécanisme (voir Alexandre et al., 2022).

INTELLIGENCE HUMAINE ET INTELLIGENCE ARTIFICIELLE

Ensuite, toutes les personnes sont concernées par l'IA étant donné que des tâches cognitives de plus en plus complexes sont réalisées par des programmes. Cela nous amène aussi à questionner ce que nous considérons comme étant les caractéristiques que l'on attribue à l'intelligence humaine (Houdé, 2019 ; Romero, 2018). On se pose souvent la question « symétrique » de savoir si une machine peut être ou devenir intelligente : le débat est interminable, car il suffit de changer la définition de ce que l'on appelle intelligence pour répondre « oui, pourquoi-pas » ou au contraire « non, jamais ». De façon simplifiée, selon une acception communément véhiculée, le but de l'IA est de faire faire à une machine ce qui aurait été jugé intelligent si réalisé par un humain.

En revanche, avec la mécanisation de processus cognitifs, ce qui paraissait intelligent il y a des années ne l'est plus nécessairement. Par exemple, le calcul mental est moins associé à une faculté humaine extraordinaire depuis l'apparition de calculettes, même si leur usage n'est pas toujours un gage d'économie temporelle (Virgo et al., 2017). De même, l'IA a le potentiel de soulager les humains de travaux intellectuels que l'on peut désormais automatiser. Cela oblige à réfléchir à l'intelligence humaine en fonction et au-delà de ce que nous appelons la pensée informatique, cette compétence permettant de résoudre des problèmes complexes en mobilisant des solutions informatiques souvent algorithmiques. Par exemple, nous savons que plus le problème à résoudre est spécifique, plus une méthode algorithmique sera efficace, possiblement plus que la cognition humaine, tandis qu'à l'inverse plus le problème à résoudre est général, moins un algorithme pourra intrinsèquement être performant, quel que soit le sujet d'application. Il se trouve que les systèmes biologiques eux aussi ont cette restriction, l'intelligence humaine n'est donc peut-être pas aussi générale qu'on ne le pense et se développe dans des domaines d'applications spécifiques (Alexandre, Viéville et Comte, 2022).

---

[1] L'interprétabilité « consiste à fournir une information représentant à la fois le raisonnement de l'algorithme et la représentation interne des données dans un format interprétable par un expert en apprentissage automatique ou en sciences des données » (Chraibi-Kaadoud, 2020).
[2] L'applicabilité vise à justifier de la façon la plus précise possible un résultat donné par un modèle, auprès des personnes concernées par le résultat, donc au-delà des experts. Par exemple, dans le cas d'un algorithme de reconnaissance de chat, le système doit être capable de prédire si effectivement l'animal observé est un chat et doit être en mesure de dire quels ont été les critères déterminants dans cette décision » (Talbi, 2022, p.1).

Un autre aspect de l'IA qui rend pertinente une formation citoyenne et éthique à son sujet est l'omniprésence des usages dans tous les domaines de la vie et les transformations qui en découlent. Que des robots assistent des personnes âgées pourrait être considéré comme un progrès, permettant de les maintenir chez eux, à leur domicile et en toute dignité. Mais si cela est vu uniquement comme un levier de réduction des coûts de prise en charge, ou un moyen de nous désengager d'une tâche parmi les plus humaines qui soit, à savoir s'occuper des autres, alors la machine nous déshumanisera. Cet exemple montre surtout, comme la crise sanitaire l'a fait au cours ces dernières années, que des circonstances exceptionnelles nous obligent à revoir en profondeur les équilibres que nous pensions acquis pour notre société. Quand – et cela est en train d'advenir – la plupart des tâches professionnelles d'aujourd'hui auront été mécanisées et automatisées, la société devra être organisée autrement.

Nous vivons au temps des algorithmes (Abiteboul et Dowek, 2017). Quelle place voulons-nous accorder aux algorithmes dans la vie courante ? Est-ce que cela nous conduit à repenser cette cité ? Comment mieux nous préparer aux usages de l'IA ? Ce sont les questions pour lesquelles le MOOC a voulu préparer les participants. En formant les citoyens, ils auront « les moyens de construire un outil qui rend possible la construction d'un monde meilleur, d'un monde plus libre, d'un monde plus juste » écrivent Gilles Dowek et Serge Abiteboul en conclusion du *Temps des algorithmes*.

## LE PUBLIC CIBLE

À l'origine, le public visé par le MOOC est constitué de l'ensemble des personnes œuvrant à l'éducation des enfants et des adolescents : enseignants, animateurs et parents. Ces personnes doivent comprendre pour, à leur tour, en initier d'autres à ce qu'est l'IA. Par exemple, les enseignants de sciences de 1ère et de terminale doivent être en mesure de piloter des activités abordant la résolution de problèmes par un ordinateur ou l'apprentissage automatique. Les enseignants d'informatique ont la possibilité d'aborder l'IA de façon transversale à plusieurs disciplines, en raison de ses nombreuses applications, ou à organiser des ateliers extrascolaires sur le sujet.

Le MOOC s'adresse également à toutes les personnes qui veulent découvrir ce qu'est l'IA et se faire une vision claire des défis et enjeux posés, ceci en comprenant comment ça marche. Pour atteindre cet idéal d'accès universel par tous, la formation est gratuite et attestée.

## FINANCEMENT ET PARTENAIRES

Le projet a été soutenu par des fonds publics et des partenaires industriels pour un total de 80K€. Plus spécifiquement, il est le fruit d'une collaboration entre La Ligue de l'Enseignement,

Magic Maker, EducAzur, la Direction du numérique pour l'éducation (DNE), l'Institut national de recherche en sciences et technologies du numérique et des laboratoires de recherche comme le Laboratoire d'innovation avec le numérique en éducation de l'Université Côte d'Azur. Cette coopération était essentielle étant donné la nature interdisciplinaire de l'IA. En associant des compétences académiques en sciences du numérique, neurosciences cognitives et sciences de l'éducation, l'équipe a rassemblé des connaissances pour contribuer à illustrer les liens entre l'IA et l'intelligence humaine.

DIFFUSION

Faire connaître l'existence de la formation a représenté un défi, car l'appel à prendre du temps pour se former est moins sensationnel que d'autres lorsqu'il est question d'IA. La notoriété du MOOC a été construite principalement par les retours des personnes qui ont pu en bénéficier. L'enjeu d'une telle formation est d'attirer un nouveau public : difficile de faire prendre conscience de l'intérêt à des gens peu ou pas intéressés, alors qu'une fois lancés, les participants sont facilement convaincus. Le MOOC a donc été principalement connu grâce à la collaboration avec la Direction du numérique pour l'éducation et l'Université Numérique d'Ingénierie et de Technologie.

# Approche pédagogique et activités du MOOC

Le MOOC adopte une approche pédagogique ludique et expérientielle. Des capsules vidéo ont été produites, parfois avec des comédiens professionnels et parfois avec des experts de production de contenus. Une variété de ressources complémentaires y sont proposées pour donner le choix à l'apprenant des formats qui lui conviennent. Certaines sont plus théoriques, d'autres plus ludiques, certaines plus approfondies et d'autres plus vulgarisées. Dès les premiers modules, l'IA est expliquée et des mythes à son sujet sont déconstruits (Lagarrigue et Viéville, 2021). Les deux principaux paradigmes en IA sont présentés, soit l'IA symbolique et l'IA connexionniste, de même que quelques repères historiques pour comprendre son évolution au cours du XXe siècle. Les participants sont mis en action dans des activités concrètes : ils manipulent des réseaux de neurones, essaient de faire reconnaître leurs dessins par une IA, sont invités à entraîner des modèles d'apprentissage automatique eux-mêmes. Ils sont invités à réfléchir collectivement, via les forums de discussion, à des questions soulevées par le développement de l'IA. Le MOOC IAI a aussi proposé des webinaires, des rencontres en ligne ou en présentiel. Ces possibilités d'échanges entre participants ont été un point fort de la formation. La Figure 1 montre une des capsules vidéo ludiques et la Figure 2 illustre un exemple d'activité où les participants sont invités à choisir une image dans une banque et à observer si un programme d'IA peut déterminer ce qu'elle représente, ainsi que le niveau de confiance envers la prédiction.

> Le MOOC IAI a une approche ludique et expérientielle qui vise à engager le participant dans des activités concrètes comme l'expérimentation d'un réseau de neurones pour démystifier son fonctionnement.

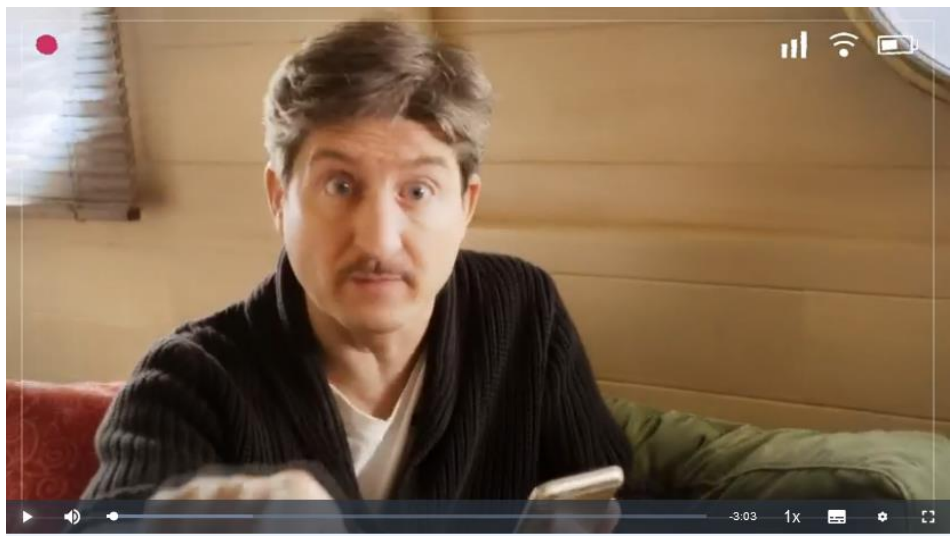

**Figure 1. Une capsule vidéo ludique du MOOC IAI**

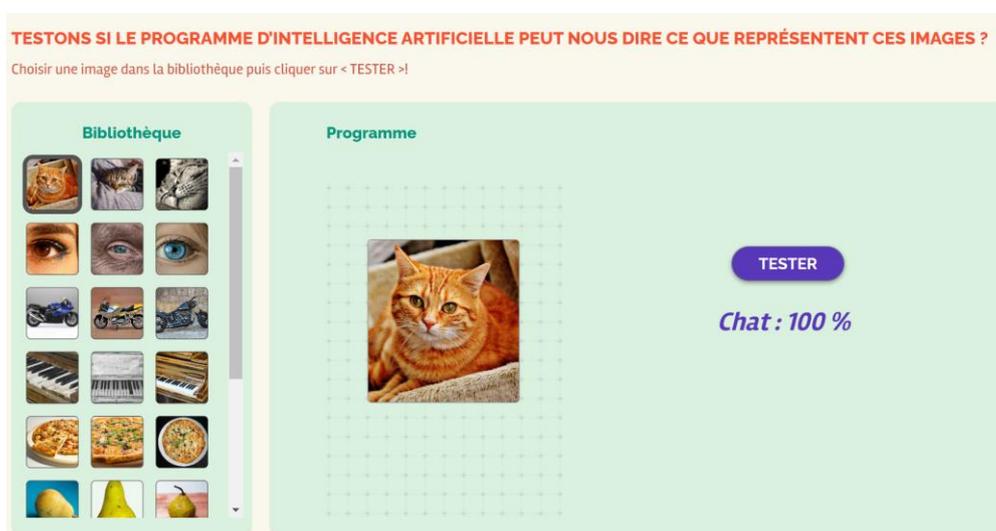

**Figure 2. Exemple d'activité développée par Class'Code où les participants expérimentent un programme de reconnaissance d'images**

# Retombées sur les participants et le milieu scolaire

Depuis son lancement, plus de 2 500 attestations de suivi ont été délivrées parmi les 32 000 participants. Il y a plus de 10 000 personnes sur le forum et près de 14 000 messages échangés, soit entre les participants ou avec l'équipe pédagogique. Le taux de satisfaction des participants l'égard de la formation atteint 94 %. Les webinaires ont attiré entre 50 et 100 personnes et ont été revisionnés par plusieurs centaines d'autres.

Au terme du MOOC, les participants sont invités à suivre la formation *Elements Of AI* (disponible en français), ou encore à se diriger vers *lumni.fr* afin de renforcer certains concepts. Ces ressources visent à accroître la confiance[3] dans le développement des innovations liées à l'IA et aider au développement d'un esprit critique[4] sur ces sujets.

---

3 Lors d'une déclaration commune en Août 2018, la France et la Finlande ont affirmé leur volonté partagée de « jouer un rôle actif pour promouvoir une vision de l'IA juste, solidaire et centrée sur l'humain, à la fois fondée sur la confiance et facteur de confiance ».
4 Motivé par la déclaration commune franco-finlandaise de « promouvoir une vision de l'IA juste, solidaire et centrée sur l'humain » nous pensons que la première étape est d'instruire et donner les moyens de s'éduquer.

Les ressources produites dans le cadre du MOOC ont été réinvesties directement en appui du programme scolaire en Terminale, notamment dans le cadre du cours NSI. Ces ressources ciblent les premiers fondements de l'IA : apprentissage automatique (avec l'exemple des réseaux de neurones artificiels, ou des approches bayésiennes[5]) (Viéville et Salaun 2020). On peut considérer ceci comme une base de culture scientifique pour toutes et tous dans le domaine. Les vidéos, placées sous licence libre CC-BY 4.0, peuvent être réutilisées par les enseignants.

## Ressources complémentaires

Le MOOC IAI est ouvert à tous pour se former à l'intelligence artificielle avec intelligence de manière ludique et pratique. www.fun-mooc.fr/fr/cours/lintelligence-artificielle-avec-intelligence

ClassCode Pixees. https://pixees.fr/classcode-v2/

ChatGPT et les nouveaux enjeux de l'IA. https://pixees.fr/chatgpt-les-nouveaux-enjeux-de-lia/

Elements of AI (formation complémentaire suggérée). https://course.elementsofai.com/fr-be

*Ce texte est partiellement repris de Roos et Storchan (2020) pour la partie écrite par un des auteurs de ce chapitre avec l'autorisation du journal Le Monde (blogue Binaire).*

# Références

---

[5] L'inférence bayésienne est une méthode d'inférence statistique « qui a pour objectif de calculer le degré de confiance à accorder à une cause hypothétique. Cette technique algorithmique prend comme point de départ le théorème de Bayes, qui présente les principes permettant de calculer une probabilité conditionnelle. Le théorème détermine la probabilité qu'un événement se produise en considérant la probabilité d'un autre événement qui s'est déjà produit » (Rojas-Vazquez, 2020, p.1).

# 2 /// COMMENT LA SCIENTOTHEQUE A L'UNIVERSITE LIBRE DE BRUXELLES INITIE ENSEIGNANTS ET ELEVES A L'IA DEPUIS 2020

Caroline Boulord[1]
Yann-Aël Le Borgne[1] [2]
Patricia Corieri[1]

[1] La Scientothèque, Université Libre de Bruxelles, Belgique
[2] Machine Learning Group, Université Libre de Bruxelles, Belgique

Un défi majeur du développement de l'intelligence artificielle (IA) est de tendre vers une plus grande inclusion des populations vulnérables et une distribution équitable de ses bénéfices potentiels. Ce chapitre présente les pistes explorées en ce sens par la Scientothèque, association ancrée dans la région bruxelloise et œuvrant depuis 20 ans pour l'accessibilité des nouvelles technologies aux populations marginalisées. Nous y décrivons, d'une part, les dispositifs pédagogiques mis en place depuis 2020 visant l'enseignement de l'IA et la formation des enseignants et, d'autre part, les retours d'expérience d'ateliers réalisés sur le thème de l'IA avec des jeunes issus de milieux précarisés. Au-delà de ces activités de formation et d'animation, le chapitre vise à illustrer l'importance du réseau associatif dans la création de ressources éducatives innovantes et dans la réduction de la fracture numérique.

## Présentation de la Scientothèque : l'égalité des chances par les sciences

La Scientothèque est une association sans but lucratif installée au sein de l'Expérimentarium, musée de Physique de l'Université Libre de Bruxelles. Depuis sa fondation en 2001, sa mission principale est de contribuer à la diminution des inégalités sociales (voir la Figure 3). Deux réalités ont mené à la mise en place de l'association : d'une part, le risque accru de décrochage scolaire et la difficulté d'accès aux études supérieures pour les jeunes issus de milieux précarisés (Coslin, 2012) et, d'autre part, la désaffection pour les études scientifiques de la part du public féminin. Il a d'ailleurs été démontré qu'il était moins évident d'attirer les filles vers les projets à caractère scientifique ou technologique (Blanchard, 2021). La stratégie consistant à organiser des activités sur le thème de l'IA dans un cadre scolaire a l'avantage de permettre d'atteindre autant les filles que les garçons et représente un moyen de compenser ce biais lié au genre.

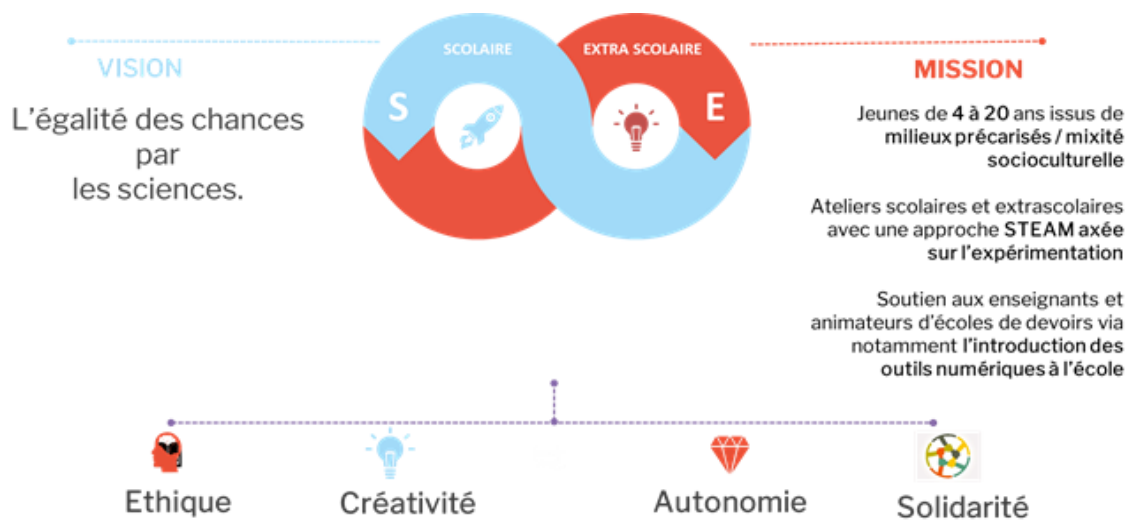

**Figure 3. Vision, mission et valeurs de La Scientothèque**

L'objectif premier de l'association est d'accompagner les jeunes de 4 à 20 ans lors d'ateliers scolaires et extrascolaires d'expérimentation où se côtoient sciences, technologies, ingénierie, arts et mathématiques (approche pluridisciplinaire STEAM). Dans cette optique, la Scientothèque apporte également un soutien aux enseignants et au personnel d'animation en charge des devoirs, notamment concernant l'introduction des outils numériques à l'école. Plus récemment, la crise sanitaire a eu « un effet loupe » sur les inégalités sociales, et de ce fait sur le phénomène de « fracture numérique » (Lucas, 2020 ; Fenoglio, 2021), raison pour laquelle les activités autour du numérique ont pris ces dernières années une place prépondérante au sein des ateliers et des actions menées par la Scientothèque.

La devise de la Scientothèque est « l'égalité des chances par les sciences ». La spécificité de la mission de l'association est en effet de mener et promouvoir des activités selon la démarche STEAM pour lutter contre l'échec scolaire et les discriminations sociales, culturelles, mais aussi de genre.

> La démarche de la Scientothèque consiste à lutter contre le décrochage scolaire et les inégalités sociales, les deux étant liés, en proposant des activités aux jeunes selon la pédagogie par projet STEAM.

Au fil du temps, avec le développement de son expertise et la visibilité grandissante donnée à ses projets, la Scientothèque a été de plus en plus sollicitée par les professionnels de l'éducation, dont les enseignants et les acteurs associatifs bruxellois, pour animer des ateliers STEAM[6] à destination des jeunes dans leurs classes, pour partager les ressources mises au point ou encore pour développer des formations. Le positionnement de l'association est d'accompagner pour autonomiser, tout en développant un écosystème d'apprentissage collaboratif.

Les principaux partenaires scolaires en Belgique sont l'école Marguerite Yourcenar à Laeken, l'Institut des Ursulines et le centre sportif Victoria à Koekelberg, l'Institut Saint-Charles à Molenbeek, l'école Escale à Woluwe Saint-Lambert ainsi que des écoles de devoirs. La

---

[6] Acronyme anglais signifiant Science, Technologie, Ingénierie, Art et Mathématiques.

Scientothèque a également été un des partenaires du projet *Fablab Mobile Brussels*[7] rassemblant les principaux Fablabs de Bruxelles, ainsi que d'autres initiatives de formation aux technologies et à la création. Un des aboutissements de ce projet a été la création du *FabULaB'Kid* dont l'aménagement a été achevé en mars 2020. L'association s'inscrit dans le développement de dispositifs innovants d'accompagnement des enseignants grâce à des projets européens multipartenaires : le projet CAI[8] visant à rompre la solitude de l'enseignant en créant plusieurs possibilités d'échanges (plateformes de co-construction de ressources, discussion d'échange de pratique, webinaire), le projet ESERO Belgium[9] financé par l'agence spatiale européenne et Belspo, dont la Scientothèque est coordinatrice pour la communauté francophone et germanophone, au côté de la KU Leuven qui assure la coordination globale et pour les communautés flamandes, œuvrant dans la diffusion et la promotion des STEM en lien avec les thématiques spatiales en milieu scolaire, et enfin le projet européen Dexterlab[10] en collaboration avec différentes universités (France, Grèce, Espagne et Belgique) consistant à développer des ressources et un catalogue d'activités expérimentales éducatives en sciences basées sur la culture du *fait maison* et du *do it yourself*.

# L'approche pédagogique de la Scientothèque en matière d'IA

Forte de 20 années d'expérience dans la pédagogie STEAM et, plus spécifiquement de 8 ans dans le développement d'activités liées à la programmation et aux *Fablabs*, la Scientothèque a récemment choisi de développer des projets pédagogiques autour du thème de l'IA pour les jeunes de 8 à 18 ans. Les deux axes principaux guidant les actions de l'association en matière d'éducation à l'IA sont les suivants : fournir aux jeunes des outils pour développer leur pensée critique et une réflexion éthique sur le sujet, et soutenir les enseignants dans la découverte et la transmission de la culture numérique et son appropriation.

La méthodologie de la Scientothèque pour aborder le thème de l'IA est basée sur celle de la pédagogie par projet. Les activités sont conçues de manière à mettre l'expérimentation au centre et favoriser une participation active des jeunes au processus de création et de compréhension tout en les amenant à collaborer avec les autres. Les compétences et les connaissances développées grâce à cette approche s'inscrivent à plus long terme dans la mémoire et concourent à les remettre au cœur des apprentissages (Papert et Harel, 1991). Il s'agit d'une méthode de remédiation scolaire innovante pour prévenir le décrochage scolaire.

## UN CATALOGUE DE RESSOURCES PEDAGOGIQUES SUR L'IA

En 2020, un premier travail de recension des ressources pédagogiques sur l'IA a été réalisé, tant celles portant sur des apprentissages éthiques que techniques. Ce catalogue a permis de rassembler et structurer, au sein d'une base de données, les ressources déjà mises au point par différentes organisations afin de dresser un état de l'art dans ce domaine. Le catalogue, recensant plus de 200 sources pédagogiques et sites d'intérêts, a été rendu accessible sous licence Creative Commons BY-SA et sous un format d'édition collaborative[11].

---

7 Les Fablabs sont des laboratoires de fabrication numérique et servent à l'expérimentation de matériel informatique, numérique ou électronique, au prototypage. Ce sont des lieux qui soutiennent l'apprentissage créatif par des situations authentiques.
8 CAI – Communauté d'apprentissage de l'informatique, voir https://cai.community/
9 Le projet ESERO Belgique qui propose ressources et formation sur l'aérospatial, voir https://eserobelgium.be
10 Le projet DexterLab, voir www.thedexterlab.eu
11 Voir le catalogue de plus de 200 ressources : https://lascientotheque.github.io/ressources-ia

À partir de ce catalogue, un ensemble complet de scénarios pédagogiques a été développé à destination des enseignants travaillant auprès d'élèves âgés de 8 à 14 ans. Par exemple, une activité vise à découvrir les inventions qui ont marqué l'histoire de l'IA grâce à un jeu de cartes, une à identifier les liens avec l'intelligence biologique à travers des expérimentations, une autre à comprendre la notion d'algorithme et les mécanismes d'apprentissage par renforcement par des activités débranchées. Certaines impliquent la programmation de robots Thymio[12] ou bien à l'aide du logiciel Scratch, alors que d'autres ciblent des aspects éthiques et amènent à débattre des conséquences de l'IA sur la société. Les fiches pédagogiques ont été conçues par La Scientothèque ou sont issues et adaptées de ressources libres identifiées préalablement, et sont mises à disposition en ligne[13]. Un des avantages de cet ensemble d'activités tient à son caractère modulaire : il est possible de suivre l'ordre suggéré par le programme ou de composer un parcours adapté à ses besoins en sélectionnant tout ou partie des fiches pédagogiques.

## ACCOMPAGNEMENT DES ENSEIGNANTS

Suivant les théories de la pédagogie constructiviste s'inscrivant dans la tradition piagétienne (Piaget, 1998) et de l'andragogie (Knowles et al., 2015), l'apprentissage des adultes est de plus en plus considéré comme autodirigé, voire autodéterminé. De nos jours, les développements théoriques dans le domaine de la pédagogie insistent de plus en plus sur le rôle proactif de l'adulte apprenant, et ceci est particulièrement illustré par les technologies numériques en constante évolution qui nécessitent un mouvement constant de va-et-vient entre l'apprentissage des compétences et leur application. Heureusement, ces technologies numériques nous donnent un accès individuel à une panoplie d'outils d'apprentissage, multipliant les possibilités d'exercer des choix dans son développement professionnel.

Il a été constaté que l'impact des formations était moins grand lorsqu'elles ne sont pas suivies d'un accompagnement sur le terrain. En effet, une fois en classe, la personne enseignante se retrouve seule, face à une activité qu'elle n'a pas l'impression de maîtriser. Quant aux animations réalisées par des intervenants externes, qui peuvent présenter un intérêt ponctuel, elles génèrent rarement une évolution des pratiques de l'enseignant dans sa classe. À l'instar de ce que nous mettons en place pour les jeunes, il faut un contexte facilitant la mise en œuvre d'un nouveau savoir-faire. De plus, en rejoignant des enseignants et des associations sans but lucratif extérieures et en les mettant en relation, nous participons à créer un réseau à travers lequel les membres peuvent interagir, s'entraider et continuer à se former.

Les formations destinées aux enseignants n'ont d'impact significatif et durable que si elles sont suivies d'un accompagnement et de possibilités d'échanges au sein de réseaux collaboratifs. La Scientothèque développe de tels dispositifs de suivi et de mise en relation : permanences, cocréation de ressources, conférences et mise en contact via des groupes de partage.

Afin d'enrichir le processus traditionnel de formation en cours de carrière et en s'inspirant des développements pédagogiques mentionnés ci-dessus, la Scientothèque développe un ensemble de dispositifs pour la formation à l'IA des personnes enseignantes. Elle a notamment créé des permanences pour accompagner la mise en œuvre des ressources STEAM dans les classes et organisé des conférences scientifiques sur des thématiques de l'IA. Des enseignants ont aussi été mis en contact avec des doctorants ou des scientifiques du domaine de l'IA. Finalement, la Scientothèque pilote des réseaux de collaboration inter enseignants (p. ex. tous les mercredis en visioconférence ou via les réseaux sociaux) en plus de donner accès à des plateformes de partage de pratiques comme la plateforme CAI et des groupes Facebook.

---

12 Le robot Thymio est un petit appareil pourvu de deux roues indépendantes, de capteurs et de lumières. Il peut être programmé pour la détection d'objets et la détection de lignes au sol.
13 L'ensemble de l'offre de services de la Scientothèque en lien avec l'IA : www.lascientotheque.be/pour-les-pros/nos-ressources-steam/intelligence-artificielle/

# Retours d'expérience

Une sélection d'activités a été proposée à deux groupes de jeunes de 10-12 ans durant l'année scolaire 2020-2021 lors d'ateliers extrascolaires hebdomadaires à l'Institut des Ursulines à Koekelberg. Ce projet, financé par la région de Bruxelles-Capitale, visait à diminuer le décrochage scolaire pour des élèves vivant dans des quartiers précarisés. Tout au long des séances, les jeunes ont pu, grâce à une approche ludique et collaborative, découvrir l'histoire des machines, établir des liens avec l'intelligence biologique, comprendre ce qui se cache derrière le mot « algorithme », observer et programmer des robots Thymio, s'initier à la programmation sur la plateforme Scratch, et débattre des conséquences de l'IA sur la société. Au fil des semaines, ils ont eu l'occasion d'aborder différentes disciplines scientifiques, renforçant ainsi certaines compétences déjà mises en place par l'école et se sont approprié la démarche scientifique : formulation de questionnements, confrontation d'hypothèses en collaboration avec les autres, mises à l'essai et retour sur la démarche. Ils ont dû mobiliser des compétences scientifiques et mathématiques (logique algorithmique, biologie du cerveau et du système nerveux, logiciels de programmation, robotique, principe d'apprentissage par renforcement, éthique).

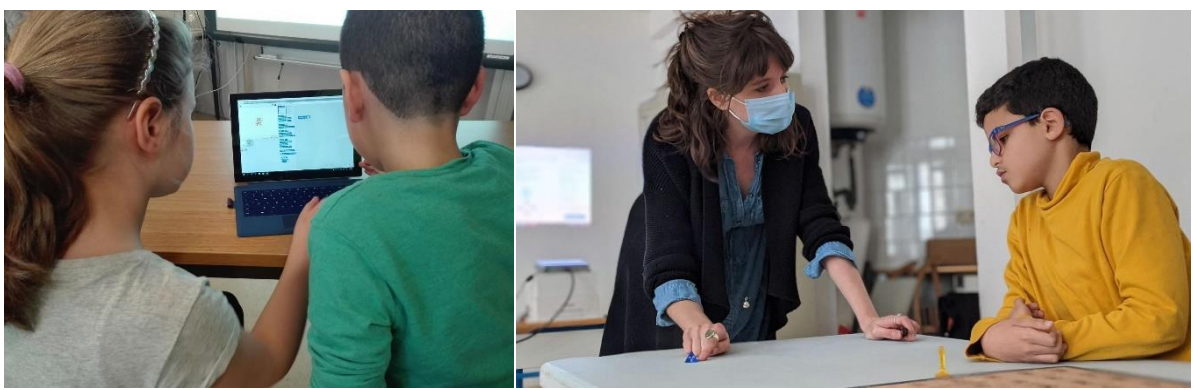

**Figure 4. Activités à la Scientothèque pour la démystification de la programmation**

Aborder de manière concrète et ludique les concepts de base de l'IA, pourtant abstraits et complexes, permet aux jeunes d'acquérir des compétences techniques et des outils de réflexion critique sur le sujet dès l'âge de 8 ans.

La dimension expérimentale des activités a permis aux jeunes de comprendre les concepts, pourtant abstraits et complexes à aborder pour leur âge. En démystifiant par la programmation ou le jeu ce qu'est l'IA, les jeunes ont acquis des outils qui leur permettront d'entamer une réflexion critique sur les technologies. Ils savent désormais s'informer pour mieux comprendre et s'exprimer sur leurs propres usages, et sont sensibilisés à la présence, parfois invisible, mais pourtant réelle, de l'IA dans leur quotidien numérique. Les activités avaient aussi un caractère collaboratif. La nécessité de collaborer au sein des sous-groupes a amené le jeunes à prendre en compte l'avis de l'autre, que ce soit pour résoudre collectivement un problème donné ou via des jeux de groupes ou débats (p. ex. le jeu du labyrinthe pour appréhender le *Q-learning*, un modèle d'apprentissage par renforcement). Notons également l'apport d'un atelier sur le renforcement de la maîtrise de la langue française, particulièrement important pour les jeunes primo-arrivants. Certains élèves ont même émis le souhait de s'orienter vers des carrières en sciences ou en informatique à l'avenir.

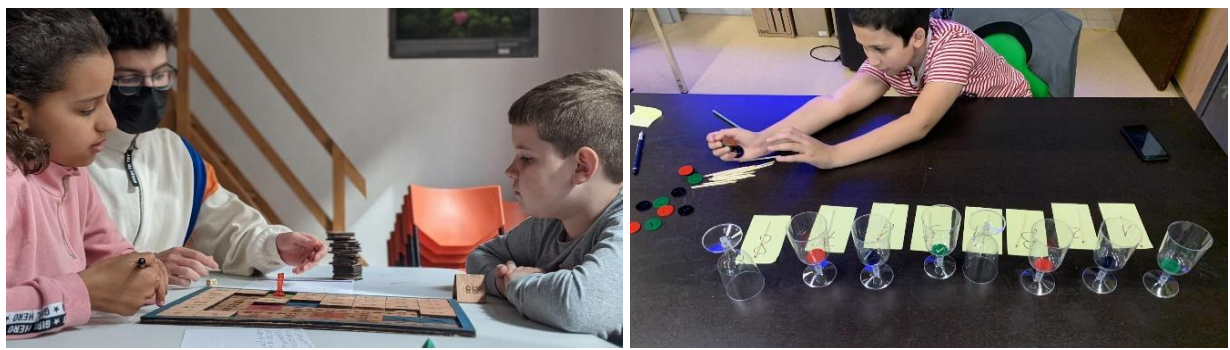

Figure 5. Activités débranchées et ludiques à la Scientothèque

Enfin, les jeunes ont été amenés à réaliser en petits groupes des courtes vidéos, suivant un scénario établi par leurs soins, pour rendre compte de ce qu'ils ont retenu des expériences réalisées. Plusieurs formes ont été explorées : interview, témoignages, conférences scientifiques. Cela a permis aux jeunes de développer leur capacité à mettre des mots sur ce qu'ils ont appris et à prendre la parole en public. Les enfants ont apprécié le fait de pouvoir conserver une trace des ateliers et étaient enthousiastes et fiers à l'idée de pouvoir montrer la vidéo à leurs camarades, enseignants et leurs proches. Les capsules ont été utilisées pour la réalisation du montage de la vidéo finale disponible sur la chaîne Youtube de la Scientothèque[14].

Une évaluation avec les jeunes à la fin de l'année a confirmé qu'ils avaient pris plaisir à participer aux ateliers. Ils ont exprimé à l'unanimité avoir apprécié les activités et appris beaucoup. Les élèves ont montré un véritable enthousiasme pour les activités proposées, qu'elles soient débranchées ou sur ordinateur. Les témoignages recueillis sont un bon indicateur que l'objectif pédagogique visé, à savoir donner du sens et de l'intérêt aux matières scientifiques et mathématiques en s'amusant, est atteint : *« J'ai beaucoup aimé quand on a fait le jeu de domino sur le langage binaire »*, *« J'ai appris qu'on pouvait faire des matières qu'on n'aime pas forcément en jouant, c'était cool !»*, *« Je sais maintenant ce qu'est un algorithme »* ou *« L'IA, en fait c'est des maths »* nous disent différents élèves.

# Perspectives

Les initiatives en matière de création de ressources pédagogiques, d'ateliers et de formations sur l'IA présentées dans ce chapitre sont nées de deux principaux besoins. Le premier est la nécessité croissante d'une éducation à l'IA, en particulier pour les jeunes publics, comme en témoignent les récents rapports sur le sujet de l'UNESCO (2021) ou de la Commission Européenne (Tuomi, 2018). Cette éducation vise d'une part à permettre aux jeunes d'acquérir et de développer une solide compréhension de l'IA – ce qu'elle est, comment elle fonctionne et comment elle est susceptible d'influencer leurs vies, et d'autre part à assurer que l'IA ne creuse pas les inégalités existantes.

---

14 Extraits vidéo des ateliers 2020-2021 de la Scientothèque sur l'IA auprès des jeunes, voir www.youtube.com/watch?v=4PgT3yHWbsE

Le second est la nécessité de créer ces ressources et de les intégrer dans les programmes éducatifs. En effet, la technologie de l'IA représentant un nouveau sujet pour les écoles, il n'existe encore que peu d'initiatives visant à définir les cadres de compétences et programmes d'enseignement associés pour les jeunes publics. Un aperçu de ces initiatives à l'international a été rendu disponible par l'UNESCO (2022), mettant en particulier en évidence le retard des pays francophones ou néerlandophones, comparativement à d'autres régions du monde telles que les Etats-Unis, l'Asie, ou le Moyen-Orient, où de tels programmes d'enseignements commencent à être rendus disponibles.

Dans ce contexte, les dispositifs mis en œuvre par La Scientothèque se révèlent être parmi les premières initiatives concrètes sur le sujet en Belgique. L'intérêt de notre démarche a ainsi été reconnu au niveau fédéral par le ministère Stratégie et Appui, avec le récent soutien d'un projet visant une ambition plus large : AI4InclusiveEducation[15]. Le projet, coordonné par la Scientothèque et impliquant un consortium d'acteurs associatifs et universitaires, visera à développer des contenus éducatifs pilotes, en français et en néerlandais, pour initier à l'IA, la programmation, les données et la robotique. Ce contenu éducatif sera présenté, optimisé et enfin validé dans les associations partenaires et dans les réseaux éducatifs, et sera ensuite diffusé plus largement en libre accès dans l'optique de servir de référence aux personnels enseignants francophones et néerlandophones.

## Ressources complémentaires

Site Web de la Scientothèque www.lascientotheque.be

Le catalogue recensant plus de 200 sources pédagogiques et sites d'intérêts sur l'IA, accessible sous licence Creative Commons BY-SA. https://lascientotheque.github.io/ressources-ia

Des fiches pédagogiques d'activités autour de l'IA à destination des enseignants conçues par La Scientothèque ou issues et adaptées de ressources libres, sont mises à disposition en ligne. www.lascientotheque.be/pour-les-pros/nos-ressources-steam/intelligence-artificielle

# Références

---

15 Le projet AI4InclusiveEducation visant le développement de ressources en français et en néerlandais auquel participe la Scientothèque, voir www.digit-all.be

# 3 /// SENSIBILISER 75 % DES ELEVES DE COLLEGE DES ALPES-MARITIMES A L'IA : LE PROJET ARC-EN-CIEL PILOTÉ PAR LA MAISON DE L'INTELLIGENCE ARTIFICIELLE

Florence Tressols[1]
Jérémy Camponovo[2]

[1] Maison de l'intelligence artificielle, France
[2] Délégation régionale académique du numérique pour l'éducation, Académie de Nice, France

Le projet Arc-en-ciel piloté par la Maison de l'intelligence artificielle (MIA) à Sophia Antipolis poursuit l'objectif de sensibiliser à l'intelligence artificielle (IA) 75 % des collégiens des Alpes-Maritimes d'ici 2023. Il est structuré en quatre chantiers : les activités pédagogiques à la MIA, les stages scolaires ou périscolaires, la sensibilisation aux biais et stéréotypes de genre en IA, et les activités hors les murs sur le territoire départemental. Autant les réalisations que l'étendue des partenariats qui les ont rendues possibles sont abordées dans le chapitre.

## Qu'est-ce que le projet Arc-en-ciel?

Le projet Arc-en-ciel est une initiative du *SMART Deal* des Alpes-Maritimes dont le pilotage a été confié à la Maison de l'intelligence artificielle (MIA) située dans la technopole Sophia Antipolis dans le département des Alpes-Maritimes en France. La cible principale de ce projet est de sensibiliser, entre 2020 et 2023, 75 % des collégiens du département à l'IA. Plus spécifiquement, il poursuit trois objectifs : (1) contribuer à l'éducation des collégiens des Alpes-Maritimes via une approche pratique et thématique, (2) partager des usages concrets de l'IA par les industries et entreprises locales, et (3) mettre en lumière les compétences des personnes œuvrant dans le domaine de l'IA pour susciter des vocations.

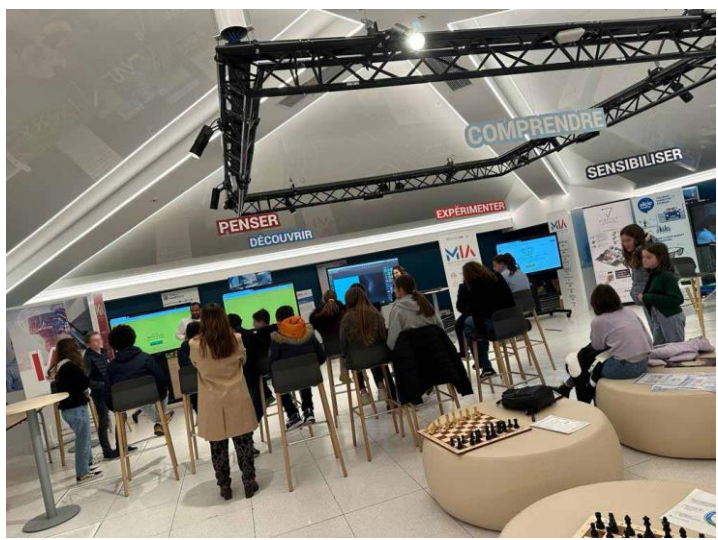

**Figure 6. Un des espaces de la MIA à Sophia Antipolis, la salle de démonstration où les personnes sont invitées à découvrir et expérimenter des cas d'usages de l'IA**

Ce chapitre présente d'abord les réalisations du projet Arc-en-ciel depuis son démarrage en 2020, notamment l'organisation de stages scolaires et périscolaires, des visites de la MIA par des groupes scolaires et des activités de médiation scientifique réalisées directement dans les collèges. Dans un deuxième temps, le contexte plus général dans lequel le projet a été réalisé sera présenté. Il sera question des différents partenariats et ententes qui ont mené à l'ouverture de la MIA en 2020, de sa raison d'être et des collaborations qu'elle entretient avec le monde de la recherche, principalement sur les questions d'enseignement de l'IA et d'équité des genres.

# Les réalisations du projet Arc-en-ciel

La banque d'activités de médiation scientifique, représentant une cinquantaine d'activités considérées comme stables sur différents thèmes de l'IA, est incontestablement l'une des réalisations phares du projet. Les activités scientifiques et culturelles sont choisies par l'équipe de médiation en fonction de publics cibles et des discussions préalables avec les enseignants responsables des groupes scolaires. Le choix des activités repose sur des paramètres variés : le message principal à véhiculer, le message secondaire, le public cible, les techniques d'IA, le nombre de personnes requis pour l'animation, allant même jusqu'à inclure de l'information sur le poids des équipements qui doivent être transportés comme l'éclairage. Actuellement, des activités sont en cours dans une phase de mise à l'essai et de perfectionnement, ce qui devrait accroître le nombre de médiations.

À l'issue de la première année (2020-2021), 2 982 élèves du collège (soit 33 % des collèges du département des Alpes-Maritimes) ont fait l'expérience d'au moins une activité : visite sur place à la MIA, avec démonstration sur des cas d'usages de l'IA, visite dans l'établissement scolaire, stage scolaires et/ou périscolaires, ou encore formation dispensée par la Délégation régionale académique du numérique éducatif (DRANE) en collaboration avec la MIA. Des formations ont aussi été offertes sur place à 30 enseignants (par la DRANE et l'organisme Terra Numerica situé à Sophia Antipolis). Dans le contexte de ses Mercredis enseignants, la Délégation académique à l'éducation artistique et culturelle a organisé en 2021-2022 trois visites à la MIA. Précisons que la MIA a aussi participé à des réunions de professeurs documentalistes: c'est ainsi que la conception de l'exposition « IA by MIA Arc-en-ciel », à destination d'usages pédagogique dans les écoles, a été facilitée.

Formellement, le projet Arc-en-ciel se décompose en quatre chantiers qui sont présentées succinctement ci-après.

## CHANTIER #1 : ATELIERS PÉDAGOGIQUES EN LIGNE ET PARCOURS *IN SITU*

Ce chantier vise le développement de la banque d'activités pédagogiques et des scénarios pédagogiques de visite des espaces de la MIA. En contexte de médiation scientifique, les contraintes de temps sont importantes et les activités doivent être conçues de façon à atteindre les objectifs d'apprentissage sans possibilité de prolongation. Pour cette raison, plusieurs activités sont destinées à être réalisées en 20 minutes auprès de demi-classes, permettant une alternance entre quelques types d'activités auprès d'un groupe-classe à l'intérieur d'une période d'environ deux heures, incluant les transitions.

Ces activités se définissent comme de nouvelles ressources pour les enseignants qui disent ne pas se sentir encore suffisamment outillés pour le faire en classe (Lin et Van Brummelen, 2021). Ainsi la MIA se charge-t-elle de sensibiliser notamment aux enjeux éthiques de l'IA, en permettant aux élèves de découvrir, comprendre et expérimenter plusieurs cas d'usage de l'IA, comme l'indiquent la liste des activités présentée ci-dessous :

| ACTIVITES | ACTIVITES (SUITE) |
|---|---|
| Une histoire de l'IA de l'Antiquité à nos jours | Automatisme ou autonomie avec le robot AlphAI |
| Usage de drones | Le phénomène des hypertrucages (*Deep Fakes*) |
| Apprentissage automatique pour classifier des images | Je programme mon neurone artificiel en Python |
| Jeu d'échecs et algorithme | Je programme mon neurone artificiel en Scratch |
| Voyage en train pour découvrir l'aide à la décision | Dilemmes de la *Moral Machine* du MIT |
| Expérimentation de Machine Learning for Kids | Robot d'assistance thérapeutique PARO |
| Reconnaissance de chiffres via réseau de neurones | Quiz « Intelligent » vs « Pas intelligent » |

**Tableau 1. Exemples d'activités développées et offertes à la MIA**

## CHANTIER #2 : ENTREPRENDRE À L'ÈRE DE L'IA

L'objectif de ce chantier est d'offrir aux élèves des occasions de développer leurs capacités entrepreneuriales. Plus spécifiquement, ce chantier cherche à développer l'agentivité des personnes dans un environnement de plus en plus marqué par l'IA afin qu'elles puissent contribuer elles-mêmes, dans le futur, à transformer cet environnement (Engeström et Sannino, 2013). Elles sont ainsi placées dans une posture de création face au numérique et non plus seulement dans une posture de consommation d'outils d'IA. Elles doivent « se prendre en main » et prendre conscience qu'il est possible de participer activement au développement de ce domaine. Ce chantier les incite à se positionner par rapport à l'IA, ne serait-ce qu'en prenant conscience de leur niveau d'intérêt, de leurs craintes ou de leurs croyances, de leurs appréhensions ou de leur confiance à en apprendre les rudiments. Cette approche permet aussi de faire évoluer les représentations des élèves (déconstruire et construire) (Ghotbi et Ho, 2021).

Le chantier « Entreprendre à l'ère de l'IA » a pris la forme de stages scolaires et périscolaires. Les stages scolaires, offerts à des élèves de 3e année, étaient des stages d'observation. Les participants étaient jumelés à une entreprise pour y observer comment l'IA est exploitée et explorer une profession dans le domaine. Les stages périscolaires, d'une durée d'une semaine, ont été réalisés principalement pendant les périodes de vacances (p. ex. vacances de la Toussaint ou du printemps). Pendant une semaine intensive, les participants sont mis à l'œuvre pour la réalisation d'un projet d'envergure impliquant l'IA. Les activités des cinq jours sont partagées entre des activités d'expérimentations pratiques de l'IA, des activités débranchées pour comprendre la logique algorithmique, et des activités d'idéation qui mettent en œuvre des méthodes de *Design Thinking*. Au début de la semaine, une problématique est proposée. Par exemple, une des problématiques était « Comment l'IA peut être utile au sport ? », une autre « Comment l'IA peut-elle contribuer à la maîtrise des consommations énergétiques ? ». L'équipe de médiation propose alors aux participants de réaliser des cartes d'empathie, c'est-à-dire de tenter de se placer dans la peau des acteurs concernés par la problématique. Dans le cas de l'utilisation de l'IA dans le sport, les stagiaires étaient invités à se placer dans la peau des entraîneurs, des personnes pratiquant des sports ou bien des spectateurs). Deux prototypes ont été réalisés, un avec des blocs LEGO© (voir Figure 7) et un autre en format numérique. La semaine culmine, le vendredi après-midi, avec la présence d'un jury bienveillant qui évalue les résultats et qui remet des prix aux équipes. Ce partage permet aux élèves de découvrir, comprendre et expérimenter l'IA, en plus de développer des compétences interpersonnelles. Celles qui sont les plus souvent citées par les stagiaires sont la persévérance, la coopération et la créativité.

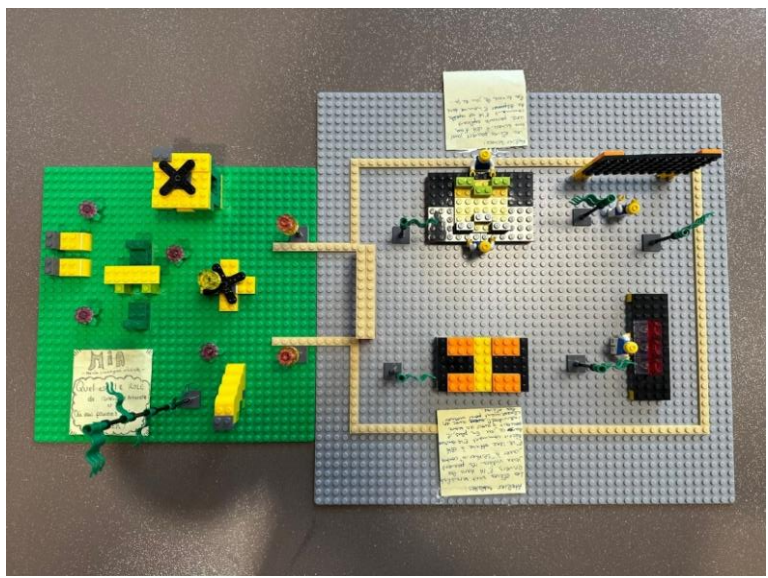

**Figure 7. Exemple de maquette réalisée dans le cadre d'un stage périscolaire**

## CHANTIER #3 : SE CONSTRUIRE SANS PREJUGE

Le domaine des technologies est encore l'objet de stéréotypes portant sur les hommes et les femmes. Toutes les activités du projet Arc-en-ciel sont très sensibles à cet enjeu et se veulent des vecteurs de changement des mentalités et des pratiques. L'objectif de ce chantier est donc de sensibiliser les équipes enseignantes, les équipes de médiation pédagogique et les élèves aux biais de genre dans le domaine de l'IA. Par exemple, Norouzi et al. (2020), dans le cadre d'un projet d'initiation à l'IA, ont été confrontés au fait que les filles avaient beaucoup moins confiance que les garçons en leurs compétences en programmation informatique. À la MIA, nous observons aussi des manifestations de stéréotypes persistants.

Ce chantier, piloté par une membre du Club égalité des Alpes-Maritimes travaillant chez l'organisme Alter Egaux, a permis de valoriser trois réalisations : d'abord, le livret « Toustes numériques » à destination des élèves et des équipes enseignantes[16]. Cette mallette prend la forme d'un guide d'une trentaine de pages dans lequel les personnes sont invitées à identifier des stéréotypes de genre et à se questionner sur leurs propres préjugés. Elle propose des activités pédagogiques courtes destinées à sensibiliser une classe à ces stéréotypes. La seconde réalisation est composée des études réalisées en partenariat avec le Laboratoire d'innovation pour le numérique en éducation de l'Université Côte d'Azur et, à nouveau, Alter Egaux. Les élèves de passage à la MIA ont été invités à compléter un court questionnaire avant et après leur visite, incluant des questions relatives à leur perception des stéréotypes de genre. Des membres d'Alter Egaux ont aussi créé des outils d'observation des activités de médiation scientifique et partagé leurs analyses avec l'équipe. Finalement, un jeu interactif intitulé « La mixité dans les métiers de l'IA » a été réalisé pendant les stages périscolaires.

---

16 La mallette « Toustes numériques » est disponible gratuitement en ligne sur le site Web d'Alter Egaux : www.alteregaux.org/mallette-toustes-numeriques

La phase préliminaire des observations d'Alter Egaux a permis d'établir un certain nombre de constats. D'abord, au sein des groupes mixtes, en raison de la socialisation sexiste différenciée des garçons et des filles, les garçons ont tendance à monopoliser la parole et l'attention et les filles à s'effacer. On retrouve ce même phénomène lors des interventions de la MIA, accentué par le prisme des exemples donnés (majoritairement des domaines à forte prédominance masculine : football, secteur militaire, etc.), l'absence de langage épicène et de rôle modèle féminin. De plus, il a été observé que les garçons se placent à l'avant de la salle et sont donc davantage dans le champ de vision des animateurs comme une preuve de leur intérêt plus important que celui des filles pour ces sujets avant même le début des interventions. Les animateurs ont observé que la plupart des garçons souhaitent expérimenter, manipuler, tester et se placent donc en premier lors des phases de manipulation. Les garçons passent souvent en premier tant dans la prise de parole que dans la manipulation des objets. Les animateurs observent un impact plus important sur les garçons que sur les filles, indépendamment du sexe de l'animateur.

Soulignons finalement que, dans le contexte de la journée internationale des droits des femmes célébrée le 8 mars 2022, la MIA est intervenue dans des classes avec WHAT06 (*Women Hackers Action Tank – département 06*) et Alter Egaux pour échanger avec les élèves, filles et garçons, sur la mixité des métiers scientifiques, numériques et intégrant l'IA. La rencontre de femmes inspirantes dans le domaine des technologies a contribué à faire tomber certaines barrières. Une des animations pédagogiques portait sur les compétences interpersonnelles et prenait la forme d'un débat interactif suivant la rencontre de professionnels de l'IA. Cette activité montre que les valeurs de coopération, flexibilité, persévérance, analyse, organisation, créativité, leadership et précision n'ont pas de genre. En 2022, dix classes de 3e année de collège ont ainsi été sensibilisées.

## CHANTIER #4 : IA HORS LES MURS

Initié en juin 2021, ce dernier chantier du projet Arc-en-ciel vise la population du département des Alpes-Maritimes éloignée de la MIA (littoral, moyen et haut pays). Concrètement, des équipes de la MIA et de ses partenaires sont appelées à se déplacer pour aller à la rencontre de différents publics et encourager le dialogue sur des questions relatives à l'IA. Le chantier est copiloté par une membre des Petits Débrouillards.

Les activités réalisées dans le cadre de la première année sont nombreuses et témoignent du succès de l'approche collaborative et partenariale de la MIA : d'abord, un projet pilote d'intervention en classe a été mené dans les vallées sinistrées par la tempête Alex sous le leadership d'*Ingénieurs pour l'école* de la Délégation régionale académique de la formation professionnelle initiale et continue. De même, à partir d'août 2021, des animations ont été réalisées sur le littoral, à Cannes, dans le contexte des « bibliothèques de plage ». En mars 2022, une première activité à grand déploiement est réalisée dans le cadre du *Tour Science IA*, en collaboration avec les Petits Débrouillards au collège Jean Cocteau à Beaulieu-sur-Mer, puis au collège Les Mimosas dans le cadre de la Semaine de l'IA. Dans ce collège, ce sont tous les élèves de 6e, 4e et 3e qui ont été sensibilisés à l'IA.

Les activités réalisées dans le cadre des quatre chantiers du projet Arc-en-ciel sont innovantes à plusieurs égards. Ainsi est-il apparu pertinent de documenter leur évolution dans une perspective d'amélioration continue, mais aussi pour inspirer et orienter des initiatives futures. Des activités d'observation des élèves, des équipes enseignantes et des équipes de médiation pédagogique ont lieu en collaboration avec des partenaires académiques et communautaires.

D'abord, à l'issue de chaque activité de médiation *in situ* ou hors les murs, un questionnaire de retour d'expérience est proposé à l'équipe enseignante, à remplir avec la classe. Outre le fait que cette action facilite la réalisation d'activités post-visites à l'école, l'objectif est de quantifier et qualifier plusieurs dimensions de l'animation. Les élèves sont invités à se prononcer sur les aspects qu'ils ont le plus appréciés ou qu'ils ont trouvés les moins intéressants, et à suggérer des changements. Ces données sont analysées par l'équipe de la MIA et feront l'objet d'une publication.

Dans le cadre du GTNum Scol_IA, à chaque début et fin de visite de deux heures, les élèves sont invités à répondre à un questionnaire électronique comportant 12 questions sur l'intelligence humaine et l'IA. L'objectif est d'évaluer l'évolution des représentations de l'IA avant et après les activités de médiation. Cette étude scientifique est conduite par le Laboratoire d'innovation pour le numérique en éducation. En outre, lors des stages scolaires ou périscolaires, les stagiaires sont invités à s'autoévaluer par rapport à cinq dimensions : leur connaissance de l'IA, le métier qu'ils aimeraient exercer plus tard, leurs craintes par rapport à l'IA, leur ressenti par rapport à l'IA et leurs attentes par rapport au stage. Cette autoévaluation a été conçue par Alter Egaux.

# Un exemple de projet : la ruche intelligente

À l'occasion de la Fête de la Science 2021, la MIA a proposé la réalisation d'un hackaton, qui a évolué en *IAckathon*[17], sur la thématique de la ruche intelligente, un projet ayant impliqué des élèves de la 6e à la 3e (soit 85 élèves issus de quatre collèges accompagnés d'une dizaine d'enseignants).

Les réalisations issues de cet événement, auquel participaient les FabLabs des collèges en réponse à un appel de la Direction régionale académique du numérique éducatif (DRANE), ont été exposées au *World AI Cannes Festival*.

Dans le Tableau 2, nous décrivons la contribution de chaque collège au projet et dans la Figure 8 partageons un extrait de la présentation du prototype réalisé par des élèves et leur enseignante, qui ont pu bénéficier de l'animation pédagogique des équipes de la DRANE et de la MIA. Grâce à son ancrage local, le projet a pu bénéficier de l'expertise des apiculteurs[18] de l'association Bee Riviera, et ainsi s'appuyer sur un cahier des charges pour concevoir la « ruche intelligente ». Sans oublier l'expertise et le pilotage de la conception par le *SoFAB Telecom Valley* au bénéfice de la réalisation du dispositif.

---

17 Un *hackaton* est un événement où des participants doivent inventer des solutions innovantes, notamment informatiques, à des problèmes d'intérêt public en un court laps de temps, parfois en une seule journée. Ici, le terme *IAckathon* est un jeu de mot original pour souligner l'usage de l'IA.
18 Une entrevue avec un apiculteur, réalisée par la DRANE, a été présentée aux élèves en guise d'introduction au *IAckathon*.

L'objectif du *IAckathon* était de réaliser un projet interdisciplinaire intégrant l'IA et poursuivant une visée de développement durable. La thématique de la ruche intelligente a donc été choisie parce qu'elle est suffisamment porteuse pour s'inscrire dans plusieurs disciplines et pas uniquement dans les matières scientifiques. Le développement durable peut être abordé via l'idée de préservation de la biodiversité, ou encore de l'économie des ressources en utilisant l'*edge computing*[19] (Shi et Dustdar, 2016).

| COLLEGE | CONTRIBUTION |
|---|---|
| Collège Pierre Bertone, Antibes<br>Via leur FabLab (classe de 3ᵉ) | Les élèves ont mis en lumière la problématique interdisciplinaire des ruches intelligentes, en collaboration avec l'apiculteur. |
| Collège du Centre International, Valbonne<br>Via leur FabLab (classe de 6ᵉ) | Les élèves ont expérimenté l'acquisition de données avec des capteurs utiles à la découverte de la vie d'une colonie d'abeilles avec des solutions dites d'*Internet des objets*. |
| Collège Auguste Blanqui, Puget-Théniers<br>Via leur campus connecté (classes de 4ᵉ et 3ᵉ) | Les élèves ont cherché à découvrir et comprendre l'emplacement adéquat de capteurs dans une ruche et ont analysé les besoins en énergie. |
| Collège Les Mimosas, Mandelieu-la-Napoule<br>Via leur FabLab (classes de 6ᵉ à 3ᵉ) | Les élèves ont expérimenté l'apprentissage automatique pour améliorer la connaissance des activités d'une colonie d'abeilles. Une présentation du résultat par les élèves, les enseignants et le personnel de la MIA est disponible en vidéo[20]. |

**Tableau 2. Contributions des élèves de quatre collèges des Alpes-Maritimes au *IAckathon* sur la thématique de la ruche intelligente lors de la Fête de la Science 2021**

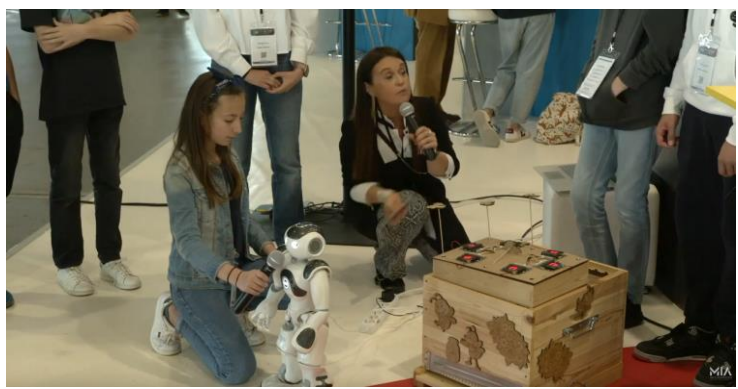

**Figure 8. Présentation du prototype de ruche intelligente par des élèves et leur enseignante**

# Genèse du projet Arc-en-ciel

Le projet Arc-en-ciel[21] n'aurait pas pu se développer et être déployé sans avoir reçu l'appui de nombreux acteurs qui ont su créer les conditions de son succès. Ce projet est une preuve tangible qu'il est possible de coordonner plusieurs administrations et partenaires en vue de réaliser des projets pédagogiques concrets. En moins de deux ans, le projet est passé d'une volonté politique à des activités concrètes sur tout le territoire du département. Cette section présente le plan de transition numérique *SMART Deal* dans lequel dans lequel s'inscrit le projet

---

19 L'*edge computing* est une approche informatique dans laquelle le traitement des données est réalisé au plus près de leur point de collecte pour éviter le transit inutile et coûteux de l'information vers des serveurs centralisés.
20 Pour voir la présentation des élèves du Collège Les Mimosas sur le projet de ruche intelligente, voir https://youtu.be/oaRh8yR4ENU.
21 Ce chapitre a été rédigé en juin 2022.

Arc-en-ciel (*SMART Education*), la structure de gouvernance et les collaborations avec des organismes aux mandats complémentaires.

## LE PLAN DE TRANSITION NUMÉRIQUE *SMART DEAL*

En janvier 2018, le plan de transition numérique *SMART Deal* du département des Alpes-Maritimes a été lancé par le président Charles Ange Ginésy. Ce plan poursuit deux objectifs : (1) moderniser et améliorer les performances de l'administration départementale afin d'offrir un meilleur service aux usagers, et (2) répondre aux enjeux du territoire (p. ex. les risques naturels) tout en améliorant la qualité de vie et le quotidien de la population. Le comité d'experts du *SMART Deal*, piloté par Marco Landi, a retenu deux thématiques de travail en lien avec les compétences départementales : *SMART Territory* et *SMART Education*.

Moins de trois ans après le lancement du *SMART Deal*, la MIA a été inaugurée le 10 mars 2020. Quelques mois plus tard, le projet Arc-en-ciel dont il a été question dans la première partie a été lancé. La vision du projet a été élaborée par Jean-Marc Gambaudo, président de l'Université Côte d'Azur, Marco Landi, président du comité d'experts du *SMART Deal* et Paul Sgro, directeur de la MIA. La coordination du projet a été assurée par Marie Hering, chargée de mission pour le *SMART Deal*, et l'animation par Florence Tressols, experte *SMART Education*. Mentionnons également le soutien indéfectible de la DRANE, pierre angulaire activement représentée depuis les débuts par Jérémy Camponovo. La Figure 9 présente une photo prise lors de la signature de l'entente pour le projet Arc-en-ciel.

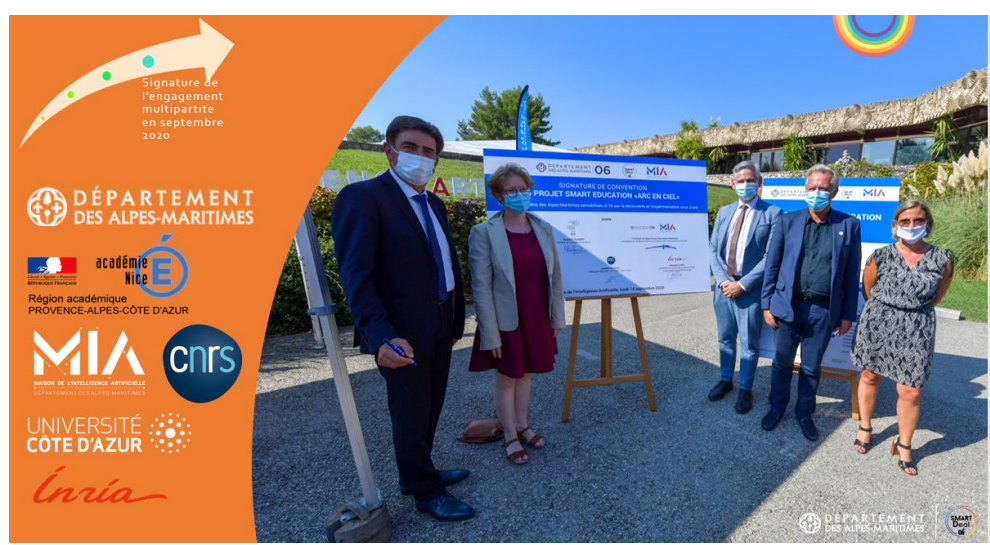

**Figure 9. Signature de l'engagement multipartite pour le projet Arc-en-ciel le 14 septembre 2020 à la MIA**

**De gauche à droite : Charles Ange Ginésy, président du Département des Alpes-Maritimes, Maureen Clerc, directrice de l'Inria Sophia Antipolis Méditerranée, Richard Laganier, recteur de l'Académie de Nice, Jeanick Brisswalter, président de l'Unviersité Côte d'Azur, Aurélie Philippe, déléguée régionale du CNRS Côte d'Azur.**

## STRUCTURE DE GOUVERNANCE

Deux structures parallèles ont donc été mises en place : un comité de pilotage pour les questions stratégiques et politiques, et un groupe projet pour les questions opérationnelles et tactiques. Chaque structure est composée de trois collèges : institutionnel, académique, et industrie & commerce, dont la composition et le rôle sont présentés dans le Tableau 3.

| STRUCTURE | COLLEGE | ROLE | MEMBRES OU PARTENAIRES |
|---|---|---|---|
| Comité de pilotage | Institutionnel | Coordination et financement | Membres issus de la Communauté d'agglomération Sophia Antipolis, du Département des Alpes-Maritimes et de la MIA |
| | Académique | Valider le fil conducteur pédagogique et les contenus | Membres issus du CNRS, de la Délégation régionale académique au numérique éducatif, de la Délégation artistique et culturelle, de la Délégation régionale académique Provence-Alpes-Côte d'Azur à la formation professionnelle, initiale et continue, de la Délégation à la recherche et à l'innovation, de l'Inria et de l'Université Côte d'Azur. |
| | Industrie & commerce | Proposer des ateliers, des retours d'expériences, des conférences, des MOOC | Membres issus de la Chambre de commerce et d'industrie Nice Côte d'Azur, du Cluster IA (une association ayant pour but de réunir l'ensemble des acteurs de l'IA incluant universités, PME et collectivités), de l'*Industrial council of artificial intelligence research*) et *Women Hackers Action Tank 06* (un collectif qui rend visibles les femmes travaillant dans le domaine de la technologie et fait découvrir les métiers de l'informatique aux femmes et aux jeunes filles). |
| Groupe projet | Institutionnel | Suivi et copilotage des quatre chantiers | Membres issus de la MIA et de l'Université Côte d'Azur. |
| | Académique | Suivi et copilotage des quatre chantiers | Membres issus de la Délégation régionale académique Provence-Alpes-Côte d'Azur à la formation professionnelle, initiale et continue, de Terra Numerica et de l'Université Côte d'Azur. |
| | Industrie & commerce | Suivi des quatre chantiers | Membres issus du Club égalité des Alpes-Maritimes[22], du Code Club de Sophia Antipolis. Le Sophia Club Entreprises, représentant les entreprises de la technopole, et le Rectorat des Alpes-Maritimes, représentant les collèges, ont signé une convention de partenariat dont l'objet est d'encadrer les interventions des ingénieurs volontaires au sein des établissements scolaires partenaires. |

**Tableau 3. Structure de gouvernance du projet Arc-en-ciel**

## COLLABORATIONS AVEC DES ORGANISMES AUX MANDATS COMPLÉMENTAIRES

La MIA participe à des activités conjointes avec d'autres acteurs du domaine de la culture scientifique, technique et industrielle. Dès lors, cela permet aux acteurs de partager leurs visions et de sensibiliser de concert les élèves à travers le prisme de diverses initiatives locales, nationales et internationales. Avec pour objectif, et non des moindres, que cette mise en commun enrichisse les méthodes de médiation pédagogique.

Dans cette perspective, la MIA s'est investie dans d'autres événements :

– Lors de la Fête de la Science 2021, en plus du *IAckathon* sur le thème de la ruche intelligente dont il a été question, la MIA a organisé un « village des sciences » et a proposé un programme ambitieux avec 15 partenaires (des académiques, des entreprises innovantes, des grands groupes membres du *Industrial council of artificial intelligence research*).

– Lors de la Semaine du cerveau 2022, sur les villages des sciences de Villeneuve-Loubet et d'Antibes Juan-les-Pins, des ateliers scientifiques ont été offerts aux publics scolaires sur la thématique de l'évolution des méthodes d'apprentissage cognitif à l'ère numérique. Les ateliers étaient animés par le Laboratoire d'innovation pour le numérique en éducation et le programme de MSc SmartEdTech de l'Université Côte d'Azur. Une activité #CreaCube était proposée dans laquelle les participants devaient résoudre des défis créatifs à l'aide de blocs de robotique assemblables ayant des fonctions variées (tels que des capteurs, moteurs, éclairage). Une activité sur l'IA et la biologie était aussi proposée.

---

22 Le Club égalité des Alpes-Maritimes a été créé en 2015 par Alter Egaux et la Délégation aux droits des femmes de la Préfecture des Alpes-Maritimes. Il est sous le patronage d'Élisabeth Moreno, ministre déléguée auprès du Premier ministre, chargée de l'égalité entre les femmes et les hommes, de la Diversité et de l'Égalité des chances. Le Club égalité regroupe plus de 130 membres qui mettent en commun leurs idées et leurs énergies pour travailler ensemble à la réduction des inégalités à toutes les échelles de la société.

– Lors des visites scolaires combinées entre la MIA et Terra Numerica, puisque ces deux lieux sont situés à proximité de Sophia Antipolis et permettent aux établissements d'optimiser les transports en autorisant le déplacement de deux classes en même temps.

# Conclusion

Depuis plus de deux ans, les activités de médiation pédagogique réalisées par la MIA dans le contexte du projet Arc-en-ciel mettent en lumière que la sensibilisation à l'IA est possible pourvu que l'on ait réfléchi au matériel nécessaire, aux algorithmes à démystifier et aux cas d'usage à discuter. Les activités sont soutenues par des acteurs institutionnels, académiques et industriels, ce qui s'inscrit dans la vision holistique de l'IA qu'entretient la MIA. Les thématiques de culture, de territoire et de la vie quotidienne concrétisent les cas d'usage avec un ancrage dans la réalité. Par ses quatre chantiers, le projet Arc-en-ciel met tout en œuvre pour atteindre sa cible : sensibiliser les collégiens du département à l'IA. Les autres projets, dont sa participation à plusieurs événements comme la Fête de la Science ou la Semaine du cerveau, doivent contribuer à élever le niveau de littératie (numérique) de tous les maralpins.

## Six pensées à emporter sur l'IA issues d'une séance de Design Thinking

1. L'IA n'est pas magique : c'est un cas d'usage, du matériel, des données et un type d'algorithme.

2. L'ère de l'IA est une nouvelle révolution et c'est un outil scientifique, technique et économique.

3. L'IA est déjà répandue dans de nombreux domaines d'activité, avec un impact présent et potentiellement futur.

4. Les usages de l'IA peuvent être une chance ou un risque pour l'humanité.

5. Il est nécessaire de légiférer et d'encadrer pour un usage éthique de l'IA.

6. Et moi dans tout ça ? L'IA a un fort potentiel pour mon travail futur et je peux être acteur pour des usages citoyens éclairés.

Les activités et services de la MIA sont référencés dans le catalogue Ac'Educ, une initiative du Département des Alpes-Maritimes. https://aceduc.departement06.fr/ac-educ-06-14518.html

# Références

# 4 /// REFLEXIONS DES ELEVES DU COLLEGE LIFE BLOOM ACADEMY SUR L'IA

Lianne-Blue Hodgkins, professeure d'anglais[1]
Victoire Batifol, professeure de français[1]
Christelle Caucheteux, professeure d'histoire, géographie et d'enseignement moral et civique[1]
Laurent Fouché, professeur de mathématiques, sciences et technologie[1]

[1] Life Bloom Academy, France

Au cours de l'année scolaire 2021-2022, les élèves du collège Life Bloom Academy ont suivi un parcours interdisciplinaire d'acculturation à l'intelligence artificielle (IA). Accompagnés par leurs enseignants dans cette aventure, ils ont débuté par des réflexions philosophiques sur la signification de l'intelligence humaine. Ils ont visité la Maison de l'intelligence artificielle et ont pu profiter de l'animation et de l'expertise du personnel sur place, en plus de découvrir une variété de cas d'usages contemporains et futurs de l'IA. En continuité, d'autres activités ont été réalisées en classe, dans différentes disciplines scolaires. Les données collectées, notamment les traces de leurs activités en classe, suggèrent que les collégiens sont sensibles aux enjeux de l'IA et que la thématique suscite un intérêt qui peut dépasser le contexte scolaire.

## Un projet interdisciplinaire portant sur l'IA

Ce chapitre présente un projet interdisciplinaire auquel ont collaboré différents professeurs du collège Life Bloom Academy situé à Cagnes-sur-Mer, dans les Alpes-Maritimes : mathématiques, sciences et technologie, histoire, géographie et enseignement moral et civique (EMC), anglais et français. Les objectifs pédagogiques étaient de découvrir l'IA et son fonctionnement, puis d'amener les élèves à réfléchir aux questions éthiques qui en découlent et ainsi développer leur esprit critique à ce sujet.

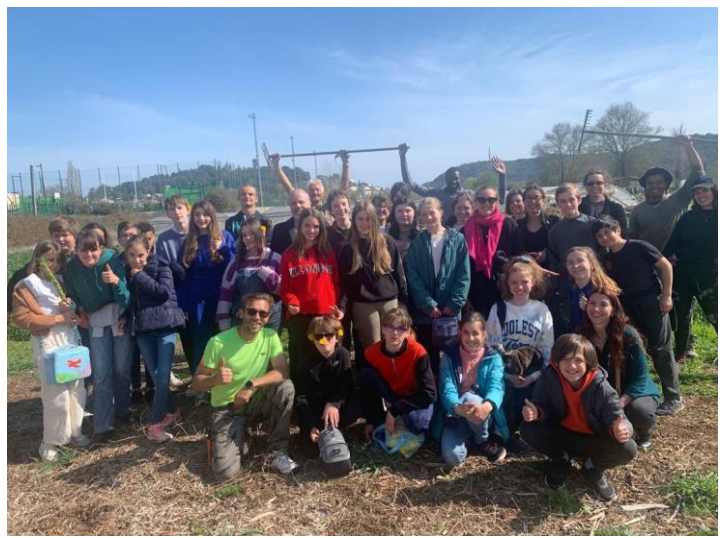

Figure 10. Les élèves du collège Life Bloom Academy et l'équipe enseignante

Le projet a démarré par une séance de préparation en classe d'histoire, géographie et enseignement moral et civique, en amont d'une visite à la Maison de l'intelligence artificielle (MIA). Pendant la séance, les élèves ont participé à un débat philosophique autour de la question « Qu'est-ce que l'intelligence ? ». Cette activité a permis de récolter les représentations initiales des élèves et de les amener à nommer différentes formes d'intelligences. Ensuite, les élèves ont visité la MIA et ont pu bénéficier de l'animation scientifique de l'équipe sur place, notamment des démonstrations de cas d'usage de l'IA et des explications vulgarisées. Les activités scientifiques et culturelles de la MIA ont joué le rôle de prérequis avant de réaliser les activités immersives sur l'IA du livret 5J5IA présenté dans le chapitre 7.

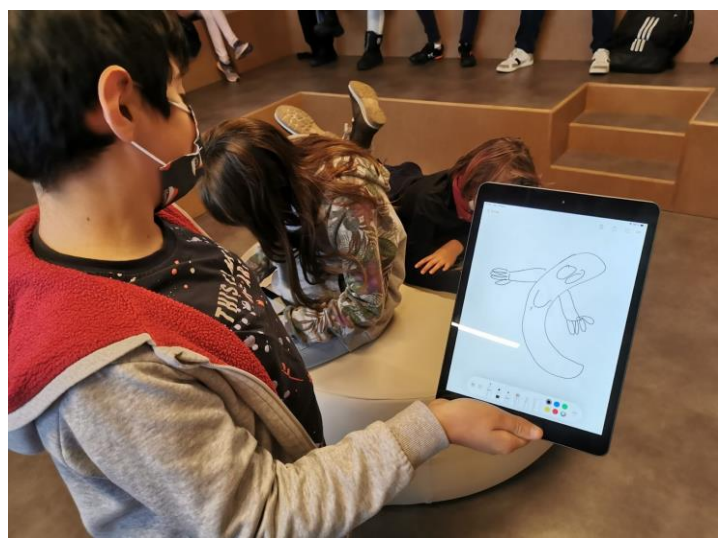

Figure 11. Les élèves du collège Life Bloom Academy lors de leur passage à la Maison de l'intelligence artificielle

Forts de ces nouvelles expériences, et après avoir enrichi leurs connaissances en cours de mathématiques, sciences et technologies, les élèves ont participé à un second débat sur les enjeux de l'IA en classe d'histoire, géographie et EMC. Ce débat portait sur des questions plus précises et visait à leur permettre de mobiliser leurs nouvelles connaissances sur l'IA. Chaque élève, seul ou en groupe de deux, a répondu aux questions suivantes : Quels sont les points positifs et négatifs ? Quels conseils donneriez-vous aux développeurs d'IA ? Le compte-rendu de ces débats a été pris en charge par Louise, élève de troisième. Ce travail d'argumentation devait les conduire à décrire une application pratique de l'IA et ses différentes conséquences dans la vie quotidienne.

# Les récits d'expérience des élèves

La section qui suit présente huit réflexions sur l'IA rédigées par les élèves eux-mêmes, accompagnés de leurs professeurs, et représente un des aboutissements du projet interdisciplinaire. Ils ont d'abord synthétisé les apports de l'IA dans leur vie, puis évoqué les risques et les questions éthiques que l'IA pose. Ils ont bien compris grâce aux modules que l'IA n'est pas si intelligente et qu'elle doit être programmée. Ils évoquent surtout les risques pour leur liberté, leur libre-arbitre et leur capacité de travail.

REFLEXION #1 – QU'EST-CE QUE L'IA ?

« L'IA permet de faciliter et d'automatiser presque tout à l'heure actuelle, de nombreuses choses sont possibles. Par exemple, de plus en plus de voitures Tesla sont équipées d'autopilotes. Par ailleurs, de plus en plus de métiers sont remplacés par des IA comme les caissières dans les supermarchés. »

REFLEXION #2 – L'IA DOIT ETRE ENTRAINEE

« L'IA pourrait aider les gens dans leur travail de tous les jours. Cela représente beaucoup de travail de les programmer, car oui, l'IA n'est rien sans l'humain. Par exemple, Alexa qui répond à nos questions à l'oral et avec qui nous pouvons avoir des conversations est une IA et elle se développe. Je lui ai dit "au revoir" quand je partais et elle a dit qu'elle ne comprenait pas. Donc, je lui ai demandé de dire "au revoir" aux gens qui le lui disaient. Maintenant, elle me sortira des petites expressions gentilles pour me dire "au revoir" ! Elle peut se développer seule, ce qui fait d'elle une IA. »

REFLEXION #3 – L'IA ET LE RESPECT DE LA VIE PRIVEE

« L'IA peut guider des choix grâce à nos données personnelles. Elle nous propose des objets, des services, des biens qui correspondent à nos goûts. Quand on fait des recherches et qu'on navigue sur des sites Web, une IA récolte ces données et les revend. C'est grâce à cela que les publicités que l'on voit sont souvent personnalisées, mais est-ce que cela respecte vraiment notre intimité ? Beaucoup de gens ne sont pas confortables avec le fait que toutes leurs données sont revendues et vont même jusqu'à dire que ce sont des complots du gouvernement ! Et même si c'est peu probable, ces données peuvent être piratées et utilisées pour du chantage. Il y a même eu des hôpitaux piratés pour des rançons, ce qui a mis la vie d'autrui en danger ! Tout cela nous permet de voir que les données utilisées par l'IA mettent en péril notre sécurité et notre liberté. »

REFLEXION #4 – LES DERIVES POSSIBLES DE L'IA

« De nos jours, alors que l'IA s'intègre de plus en plus dans notre vie, de nouveaux risques liés à son utilisation apparaissent. Son utilisation nous fait réfléchir aux notions de vie privée et de sécurité, c'est-à-dire aux risques accrus de piratage de nos données sur la toile. D'autant plus que, si l'IA est utilisée à de mauvaises fins, les risques seraient bien plus importants. Un usage malveillant de l'IA est déjà en place dans certains pays comme en Chine par exemple où le crédit social est mis en place : c'est une sorte de système de surveillance et de contrôle des moindres faits et gestes de chaque habitant avec l'IA. Les nouveaux risques qui sont donc à prendre en compte sont les risques de la sécurité de nos données, notre confidentialité et des usages malveillants de l'IA comme pour la mise en place d'un régime néo-totalitaire. L'IA n'est pas humaine et elle permet de répondre à des besoins. Mais elle peut impacter notre vie et notre liberté. Par exemple, les témoins de navigation (cookies) ou les recommandations nous suivent dans nos recherches internet. Quand on va sur un site, on accepte des témoins et l'ordinateur prend en compte notre goût pour ce site-là et peut nous en proposer des similaires. Cela réduit notre liberté parce qu'on se sent observés et cela nous empêche de faire nos propres choix. On nous propose des choses, donc on est influencés. Bientôt, dans notre vie quotidienne, l'IA nous aidera pour les tâches répétitives et ménagères. Mais nous ne devons pas dépendre complètement d'elle. »

## REFLEXION #5 –L'IA AU SERVICE DU DEVELOPPEMENT DURABLE

« L'IA peut nous sembler une solution pour répondre aux enjeux écologiques et de développement durable. Elle peut nous aider à créer des innovations durables, à gérer l'énergie, à organiser la dépollution et le recyclage ou à résoudre des situations tendues. Grâce à l'IA, il serait possible d'optimiser la gestion des sols et le rendement des terres agricoles en anticipant par exemple l'apparition de maladies, en optimisant l'usage d'eau ou encore en ajustant la production à la demande. Le robot pourrait ressembler juste à une petite puce qui pourrait analyser les terres pour optimiser l'eau, etc. Pour nettoyer les océans, on pourrait faire deux types de robots, un qui reste à la surface de l'eau pour ramasser les déchets qui y sont présents, il aurait la forme d'un petit bateau et il serait formé d'une grande poche pour récupérer les déchets. Pour le deuxième type de robots, celui-là pourrait aller sous l'eau pour ramasser les déchets. Il aurait la forme d'un gros poisson et pour qu'il ne fasse pas peur aux animaux marins, il aurait une très grande poche pour stocker les déchets. Avec les enjeux du changement climatique, les ressources naturelles s'amoindrissent et les risques de famine, de pénuries alimentaires s'intensifient. Un changement dans nos habitudes s'impose alors, et l'IA peut nous aider dans ce changement et dans le contrôle de notre consommation alimentaire afin de préserver nos ressources. Nous pourrions par exemple créer une IA qui permettrait de prévenir le gâchis alimentaire avec un réfrigérateur intelligent qui nous empêcherait de gaspiller de la nourriture. Nous pourrions automatiser et améliorer la production des serres grâce à l'IA qui analyserait la température, l'humidité. Tout ça pour dire que l'IA peut nous aider à régler nos problèmes mondiaux ».

## REFLEXION #6 – L'IA AU SERVICE DE LA SANTE

« Un autre domaine dans lequel l'IA pourrait être utile est celui de la santé et plus précisément des soins et traitements. En effet, l'IA pourrait aider, voire remplacer, les professionnels dans leur travail en effectuant des tâches ou des analyses dans le domaine médical. L'IA pourrait, par exemple, s'occuper d'opérations délicates, sans l'intervention de personnel. On peut imaginer qu'elle serait utile en cas d'accident. On pourrait aussi utiliser l'IA pour optimiser le traitement de maladies. Si l'IA est capable de les détecter et de les reconnaître, il serait possible de faire des diagnostics précis que ce soit pour prévenir dès les premiers symptômes ou pour les traiter à un stade plus avancé, tout en suivant l'évolution et donc en adaptant des traitements. L'avantage de ce système serait de disposer d'une grande base de données qui remplacerait l'expérience limitée des professionnels de santé. »

« Les risques sont multiples et il faut les connaître pour les prévenir et les résoudre, mais déjà quels sont ces risques ? Les risques éventuels sont le traçage de nos données et la divulgation de nos données sur internet. L'autre risque est l'affaiblissement des capacités humaines : si tout est fait par l'IA et qu'elle ne fonctionne plus, alors qu'allons-nous faire, car nous ne saurons plus faire les choses de la vie quotidienne ? Si un robot contenant de l'IA est piraté, cela serait dangereux. Par exemple si un robot médical est piraté, cela pourrait faire des morts. C'est le même problème pour l'usage de l'IA dans l'armée. S'il y a des robots dans une guerre, ils n'auront pas peur de mourir et donc ils pourront faire beaucoup plus de dégâts et de morts. »

« L'IA serait capable d'aider les enseignants pour des tâches administratives. Elle pourra faire toutes les choses administratives vis-à-vis du directeur en l'aidant pour la comptabilité ou l'envoi de courriels. Elle serait aussi très utile au professeur : pour l'appel des présences, la préparation des exercices et des leçons, ou l'évaluation des apprentissages. Cela permettrait d'aider les enfants qui ont plus de mal que d'autres à apprendre une leçon ou faire des exercices. Une IA peut donner des astuces ou des aides aux enfants ou aux adolescents pour avancer dans le programme scolaire. Par exemple, elle pourrait proposer de faire des cartes mentales ou pourrait suivre l'évolution de l'enfant pour se mettre à sa hauteur. »

# Conclusion

Nous observons dans les réflexions des élèves qu'ils ont des attentes élevées par rapport à l'IA et ses différents usages. Mais ils envisagent également des risques comme celui de perdre des opportunités d'apprentissage s'ils font un usage trop important de l'IA pour réaliser leurs activités pédagogiques. Les élèves envisagent également des risques liés à une trop grande dépendance vis-à-vis de l'IA ou au piratage des données qui pourrait détourner les usages initiaux des technologies. Les propos des élèves permettent d'illustrer leur capacité à identifier des enjeux éthiques liés à l'usage de l'IA. Le travail de l'équipe disciplinaire, assis sur cinq entrées pédagogiques, rend les élèves autonomes dans leur capacité à s'exprimer, de manière objective, sur leur citoyenneté à l'ère de l'IA. Les récits d'expérience démontrent leur capacité à appliquer un certain détachement au regard de certaines illusions, tout en percevant qu'il serait difficile de se passer de l'IA pour résoudre des problèmes.

# PARTIE 2 ///
# ENJEUX DE L'IA EN
# EDUCATION

# 5 /// L'IA EN FORMATION PROFESSIONNELLE

Solange Ciavaldini-Cartaut[1][2]
Jean-François Metral[3][4]
Paul Olry[3][4]
Dominique Guidoni-Stoltz[3]
Charles-Antoine Gagneur[4]

[1] Laboratoire d'anthropologie et de psychologie cliniques, cognitives et sociales, Université Côte d'Azur, France
[2] Laboratoire d'innovation pour le numérique en éducation, Université Côte d'Azur, France
[3] Institut Agrosup Dijon, France
[4] Unité de recherche *Formation et apprentissages professionnels*, Conservatoire national des arts et métiers de Paris, France

Créer une IA requiert de disposer de données fiables permettant d'assurer une fonction d'aide à l'enseignement et à l'apprentissage. En formation professionnelle, ces données sont caractérisées par des savoirs pratiques et des processus délibératifs et adaptatifs souvent peu formalisés en milieu de travail où priment les enjeux de productivité.

La seconde difficulté de conception d'une IA et de son usage pour la formation professionnelle est de conserver le potentiel d'apprentissage de l'expérience vécue dans différentes situations, de cheminements pluriels et des erreurs, de pondérer parfois le caractère réaliste des situations de travail au bénéfice d'enjeux didactiques.

La troisième difficulté relative à l'usage de tuteurs intelligents à l'aide de l'IA est d'inclure dans les rétroactions fournies à l'apprenant, les processus d'enquête qui caractérisent notamment le travail « dans et avec le vivant » et qui requièrent une projection d'actions sur du temps long.

## Introduction

L'usage de l'intelligence artificielle (IA) dans la formation professionnelle soulève de nombreuses critiques et questions. Elles sont spécifiques aux contextes de son usage dans l'alternance entre éducation formelle et éducation-formation sur la place du travail. Elles renvoient aussi à la nature même des traces d'apprentissage ou des traces d'activité qui sont susceptibles d'être mobilisées dans une conception continuée dans l'usage, d'environnements d'apprentissage intégrant de l'IA à des fins de renforcement du processus d'enseignement ou d'apprentissage.

Les analyses rapportées dans ce chapitre s'inscrivent dans le paradigme culturel et anthropocentré, mais aussi sociotechnique (Albero, 2019) de l'usage des technologies. Les technologies incluant l'IA y sont, pour les utilisateurs, des instruments (Folcher et Rabardel 2004) subordonnés aux tâches et aux situations dans lesquelles se déploie l'activité professionnelle située. L'IA est considérée comme une aide, en ce qu'elle prend en charge une part des opérations, traite ou donne accès à des informations venant étayer les raisonnements et le processus d'enseignement-formation-apprentissage. Elle peut, à certaines conditions, aider à la coopération entre enseignants-formateurs et apprenants, ou à la régulation des apprentissages par l'enseignant-formateur ou l'apprenant lui-même. Elle peut également soutenir le recueil de traces d'activité dans la réalisation d'une tâche dans un emploi visé et contribuer à leur analyse. Toutefois, l'usage de l'IA dans la formation professionnelle rencontre des freins et des problèmes liés au fait que l'activité productive, délibérative, adaptative, créative, notamment « avec et pour le vivant », ne se réduit pas aux seuls comportements observables et normalisés. Par ailleurs, le recours aux *learning analytics* s'avère peu pertinent pour saisir et analyser les processus cognitifs, adaptatifs et créatifs qui sont requis dans les activités à enjeux productifs, en situation de travail.

Ces processus inhérents à l'activité en lien avec d'autres humains ou le vivant en général (David et Droyer, 2019) conduisent à interroger les traces permettant de constituer des données tangibles pour concevoir des formations professionnelles enrichies des contributions d'une IA. En outre, le public de la voie professionnelle, à l'articulation entre les formations initiales et continues, n'est pas captif des enjeux formels d'éducation. Se pose ainsi aux concepteurs d'instruments et de dispositifs de formation, un second problème qui est de rendre l'IA « acceptable » tant du côté de la formation que du côté du monde du travail.

Dans ce chapitre, nous interrogeons la nature des données initiales et leur qualité (ce qui est à faire apprendre) ainsi que la nature des traces d'apprentissage dans une perspective de forage de données éducatives (*educational data mining*). Ce préalable sera indispensable pour examiner de façon critique les bénéfices consubstantiels de l'IA à la formation professionnelle face aux pressions de la performance organisationnelle. Nous prendrons à cette fin plusieurs illustrations empiriques.

La première est en lien avec la fabrication du fromage de Comté (Chrétien et al., 2020) où le raisonnement adaptatif des professionnels en situation est déconsidéré face aux processus industriels ou semi-industrialisés ce qui *in fine* constitue un frein sociotechnique à la création d'une IA pour la formation professionnelle du secteur.

En ce qui concerne l'usage de l'IA comme outil partenaire de l'apprentissage, la seconde illustration s'appuiera sur les résultats de la recherche e-Fran « Silva numerica » qui portent sur la conception et l'usage expérimental d'un environnement d'apprentissage à la gestion de la forêt (Guidoni-Stoltz, 2019, 2020 ; Chiron, 2018). Même si le prototype de cet environnement virtuel relève actuellement d'une simulation 3D sans recours à l'IA, s'y pose la question d'une modélisation réaliste et authentique du travail du forestier pour l'engagement et le suivi des apprenants dans un parcours de formation. Toutefois, un caractère réaliste peut devenir un frein didactique à l'apprentissage des raisonnements professionnels projectifs sur l'évolution du vivant (la forêt). Dans ce cas, nous nous demanderons comment l'IA en tant que ressource didactique et « outil partenaire » pourrait être susceptible de contribuer à des *feedbacks* de l'activité simulée, tout en accordant une place à l'erreur dans le raisonnement projectif des formés.

La troisième illustration concerne la contribution de l'IA au potentiel d'apprentissage en situation de travail et au développement des compétences des mécaniciens automobiles. À partir des résultats de la recherche menée par Gagneur et Vassout (2019) sur les garages connectés, nous traiterons de l'invisibilisation du travail d'enquête pourtant requis pour l'analyse diagnostic de pannes complexes. L'enjeu pour la formation professionnelle est alors de redonner une visibilité aux processus cognitifs d'inférence opérés par le mécanicien sur l'état observé du véhicule et l'utilisation par son propriétaire. Ce processus dépasse le simple usage d'une valise connectée et requiert d'apprendre en formation professionnelle initiale à dominer l'IA en la replaçant comme un instrument parmi d'autres, à utiliser tant par les tuteurs que par les apprentis.

Nous conclurons ce chapitre autour des enjeux relatifs à la fiabilisation des données construites à des fins d'exploitation par un système d'IA entendu comme un système d'apprentissage automatique. Des suggestions relatives à l'enseignement-formation sont faites dans la perspective du dépassement des problèmes d'usage évoqués précédemment.

# Fiabilité, qualité et tangibilité des données initiales dans la formation professionnelle pour concevoir une IA

Dans les univers sociaux productifs comme dans celui de l'éducation nationale, il existe un discours à propos des compétences professionnelles et des tâches concrètes à réaliser pour les acquérir puis les renforcer. Toutefois, les choses se compliquent quand on prend en compte le travail réel dans l'entreprise et cela pour deux raisons. D'abord, l'entreprise est un lieu de production confidentiel qui ne laisse pas filtrer les données aisément. Puis l'IA suppose de documenter une « matière inflammable » : le travail. En effet, si l'entreprise sait documenter les emplois, la « logique compétence » a individualisé la réalisation des tâches et nombre d'entreprises (ou d'administrations, d'associations, etc.) connaissent mal le travail tel qu'il se fait et ne le souhaitent pas. En effet, documenter le travail dit réel supposerait de collecter les informations à son propos, ce qui correspond à l'exact contraire des politiques RH menées depuis trente ans. D'autres formes de complication s'ajoutent dès lors qu'il s'agit de travailler avec autrui ou bien avec et pour le vivant (Mayen et Lainé, 2014), dans des environnements complexes, dynamiques, à long délai de réponses, comme celui de l'écosystème forestier par exemple. Ce travail qui s'inscrit dans un genre professionnel porteur d'histoire, mais aussi à la croisée de sphères économique, sociale et environnementale est tout à la fois singulier, pris dans des conflits de critères et parcouru d'incertitudes, de dilemmes et de questions socialement vives.

Se pose donc la question des données initiales sur lesquelles faire porter les apprentissages professionnels notamment lorsque le travail est discrétionnaire ou ne se résume pas aux seules prescriptions. Dans l'entreprise, il est assez commun d'évoquer les procédures, mais le savoir auquel elles sont associées est souvent hors du sujet qui le mobilise dans une activité réelle toujours plus complexe que la tâche avec laquelle elle s'éprouve. Ce constat complique le choix des données initiales à implémenter dans les instruments numériques et sur lesquelles vont porter les apprentissages professionnels. Certes, on peut apprendre les bases du management avec un jeu sérieux par exemple, mais l'exercer dans l'entreprise est plus complexe aux niveaux opérationnel et adaptatif. Ce plan adaptatif, voire créatif de l'activité en lien avec l'humain ou le vivant (David et Droyer, 2019), questionne les traces d'activité professionnelle permettant de constituer des données tangibles pour développer des formations enrichies des contributions d'une IA. Or, on ne peut dissocier la question de la fiabilité des données de celle de leur tangibilité qui est forcément corrélée à leur réduction. Dans le monde de l'éducation scolaire, il apparaît plus facile de disposer de données tangibles : les savoirs scientifiques et didactiques sont largement établis (notamment par la recherche en didactique des disciplines et les programmes scolaires) et la traçabilité de l'apprentissage est requise pour l'évaluation de la performance scolaire. Toutefois, il n'en est rien du côté du monde du travail où le curseur de l'efficacité est davantage corrélé au résultat immédiat ou à long terme sur le produit, comme en termes de développement des sujets. Les mesures et les jugements sur les hommes et la qualité de leur travail sont le plus souvent

« maladroits à dire » (Dujarier, 2010). La formation professionnelle se trouve donc à la croisée d'usage des données tangibles d'apprentissage et des données fidèles aux situations de travail (Chiron, 2018). L'usage de la simulation en formation se rapproche des données fidèles aux situations de travail. Toutefois, elle peut paradoxalement ne contribuer que faiblement aux apprentissages si elle est insuffisamment didactisée, ce que documentent fort bien les travaux en didactique professionnelle. En effet, dans toute conception d'activité de simulation se pose la question de la fidélité épistémique : quelles données et fonctionnalités faut-il implémenter pour obtenir quel apprentissage du travail ? Par ailleurs, dans le cas de la transmission professionnelle en situation de travail (mentorat, tutorat, action de formation en situation de travail), les données initiales ne sont pas toujours assez précises pour envisager de créer une IA ou d'alimenter des algorithmes d'apprentissage automatique centrés sur l'activité professionnelle cible afin d'en faciliter l'apprentissage.

## L'EXEMPLE DE LA FABRICATION DES MEULES DE FROMAGE DE COMTÉ

Nous prendrons comme illustration une étude récente sur la fabrication des meules de fromage de Comté (Chrétien et al., 2020). Cette fabrication en France est de loin la plus documentée sur les 165 sortes de fromage. Il existe un centre technique depuis trente ans disposant de productions scientifiques et des tests à la pointe du domaine. Toutefois, ces données fiables et tangibles demeurent incomplètes pour expliquer comment et pourquoi certains fromages sont meilleurs que d'autres alors que les procédures sont standardisées. D'un point de vue anthropocentré (Albero, 2019), l'un des organisateurs de cette recherche (Chrétien et al., 2020) a été d'observer in situ et de questionner l'activité des fromagers pour accéder au réel de leur travail de fabrication.

Cet enjeu d'acquisition de traces tangibles de l'activité productive était de saisir les manières de faire et les liens avec les déterminants du goût du fromage et de sa qualité. Autrement dit, il existait des données nombreuses, fiables, actualisables sur les processus en jeu incluant les types de vaches, la météo, le sol, l'herbe qu'elles mangent jusqu'à la place du fromager et l'affinage une fois que le fromage a été fabriqué. Toutes les conditions semblaient donc réunies pour créer une IA au service de la formation des enseignants de lycées professionnels et des apprentis de la voie professionnelle.

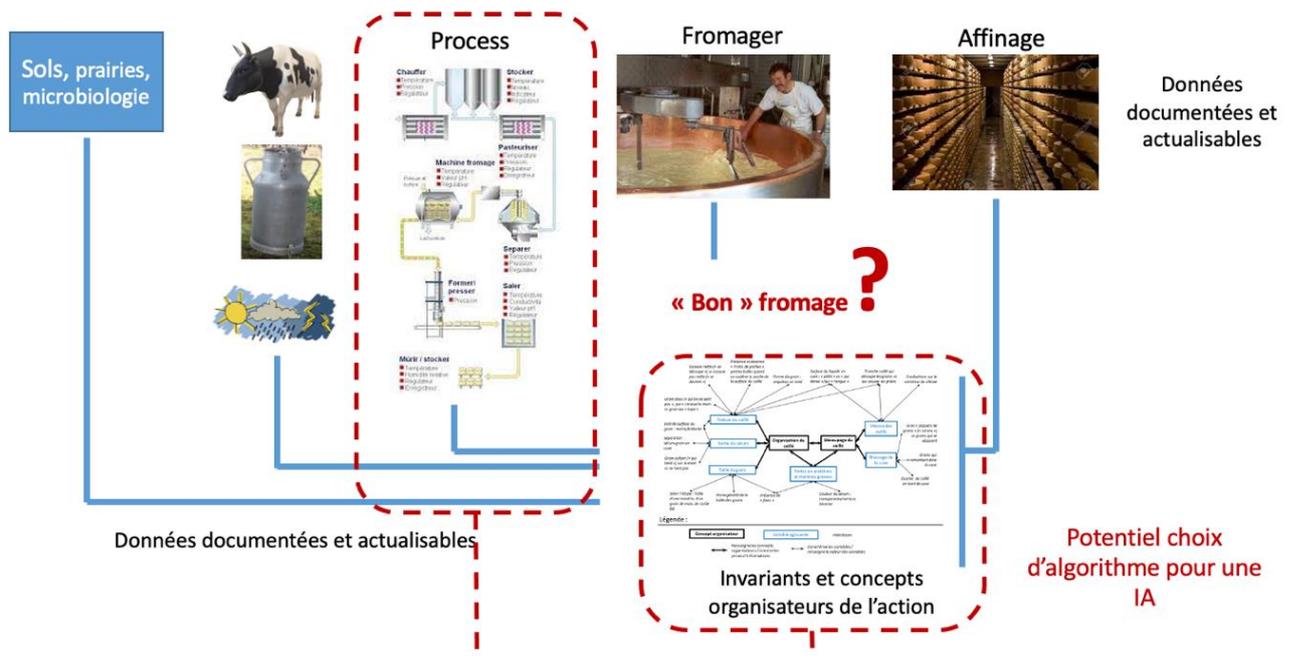

**Figure 12. Fabrication du fromage de Comté et questionnement du choix d'une IA adaptée aux variables d'action des fromagers**

Mais toutes ces données ne permettent pas de comprendre l'essentiel : les différentes variables que le fromager mobilise en action pour fabriquer le fromage selon sa perception de la qualité du lait, la saison, les ferments lactiques dont il a observé depuis quelque temps qu'ils agissaient de telle ou telle manière. Ce sont ces variables qui lui permettent de fabriquer un fromage de Comté qui a du goût et elles peuvent, grâce aux connaissances issues de la recherche menée et à leur formalisation, enrichir les données initiales pour l'apprentissage professionnel. Elles pourraient même permettre de concevoir une IA incluant le raisonnement adaptatif des professionnels en situation. Mais apparaît alors un autre problème : les professionnels, comme les commanditaires de cette recherche, accordent moins de crédits à leurs propres raisonnements singuliers, contextuels, qu'aux processus industriels ou semi-industrialisés. Il s'agit ici précisément du principal frein sociotechnique à la conception d'une IA contributive du perfectionnement des formations existantes. Autrement dit, la place de l'humain dans l'approche anthropocentrée n'est pas défendue à l'intérieur des univers de travail, ce qui est au cœur de la question de la fiabilité des données initiales en contexte professionnel. Cela pose aussi la question du choix des données pour concevoir l'IA autrement que ce qui est fait dans le domaine de l'éducation scolaire et des apprentissages disciplinaires. Le cas de la fabrication du fromage de Comté souligne que l'on apprend aussi au travail et sur les lieux du travail. Dans quelle mesure l'IA pourrait-elle agir sur le potentiel d'apprentissage des situations de travail ? Selon Littlejohn (2017), la première difficulté est inhérente à la modélisation d'un environnement complexe et dynamique avec lequel le professionnel interagit et pour lequel on ne dispose pas de capteurs d'information suffisamment performants. C'est le cas dans la fabrication du fromage de Comté avec les dimensions perceptives et gestuelles mobilisées lorsque le fromager veut savoir où en est le processus de transformation du lait : il plonge la main dans la cuve pour toucher le lait. Or, il n'existe pas de capteurs capables de retourner des informations interprétables par une IA en lien avec des traces d'activité du sujet apprenant en agissant au contact de la matière première en situation de production. De façon abrupte, et hors des enseignements des écoles professionnelles initiales, la question du recueil de traces tangibles est autant une question politique que technique. Quelle entreprise est prête à reconnaître les dimensions objectivant le travail tel qu'il se fait pour documenter un système d'IA ? Quelles références, normes, composantes d'une activité singulière et plurielle, toujours complexe doivent être prises en compte ? Jusqu'à présent, on trouve des réponses pour les industries à risque (chimie, santé, aéronautique, transport ferroviaire, nucléaire, etc.), car y travailler suppose de respecter étroitement la prescription. Mais ce sont précisément ces secteurs qui prêtent la plus forte attention au facteur humain dans les diagnostics de situation et de prise de décision pour assurer la fiabilité des processus. L'enjeu de l'IA en formation professionnelle est de toucher les autres secteurs de travail.

# L'IA en formation professionnelle, un partenaire de l'apprentissage

Un premier usage possible de l'IA dans la formation professionnelle consisterait à l'implémenter et à l'utiliser en tant qu'outil d'aide, partenaire de l'apprentissage. Par exemple, l'un des intérêts serait de documenter des données fiables pour concevoir des simulations à partir de situations professionnelles totalement ou partiellement inaccessibles aux futurs professionnels. Ou encore, ce serait aussi de fiabiliser le traitement de situations critiques, soit pour le produit, soit pour le processus de fabrication, soit pour les personnes qui y sont engagées. Enfin, ces situations peuvent être dangereuses en elles-mêmes, ou au titre des conditions de l'exercice professionnel. Les phases d'atterrissage ou de décollage d'un avion avec des passagers réels peuvent illustrer notre propos : ces moments sont par nature facteurs de risque, ils le sont encore plus lorsque les créneaux de temps sont saturés sur les plateformes aéroportuaires : c'est la complexité de l'apprentissage de ces phases pour un pilote que pourrait documenter une IA. Mais la difficulté pour l'IA est de disposer de ces données, au regard de processus invisibles ou imperceptibles. De même, les actions conduites sur ces processus et ces situations s'étalent parfois dans le temps long, et on ne peut en voir les effets dans le délai par nature contraint de la formation, voire carrément dans le délai d'une vie. C'est précisément le cas de la gestion d'une forêt, organisme vivant d'un écosystème complexe.

LA GESTION DES FORETS ET LE PROJET SILVA NUMERICA

Gérer le vivant, lorsqu'on parle de décisions dont l'impact est à horizon de près d'un siècle (par ex. abattre ou non un arbre, quand ?), opacifie la nature des données susceptibles de documenter une IA. Car simultanément, cette gestion dépend aussi des variables relatives à l'évolution de l'écosystème lui-même, des objectifs de gestion d'un propriétaire, des prix du marché du bois, des risques sanitaires, des évolutions climatiques, etc. Apprendre à gérer une forêt, c'est apprendre à gérer plusieurs processus simultanés qui appellent une expertise à penser un système, une expertise d'ailleurs plurielle à partir de laquelle une IA pourrait nourrir avec pertinence un environnement d'apprentissage à l'aide de la simulation.

Le projet Silva numerica[23] (Guidoni-Stoltz, 2019, 2020) a rassemblé une masse conséquente d'informations pour documenter le travail de gestion forestière et son apprentissage dans le but de concevoir un environnement numérique d'apprentissage intégrant cette complexité. L'environnement virtuel simulateur s'est appuyé sur une modélisation des processus de développement des arbres dans une population d'arbres qui constituent la forêt. Mais cette modélisation, dans un but pédagogique[24], est confrontée à plusieurs limites. La première est que les processus naturels de développement qu'elle vise à reproduire sont tellement complexes qu'il n'existe pas de modèle global prêt à l'emploi pour la formation, ni même de base de données complètes d'autant qu'ils sont très dépendants du contexte (géologique, climatique, etc.) et de la sylviculture pratiquée. Si ces données existaient, une IA pourrait alimenter des scénarios de simulation pour l'apprentissage de la gestion de parcelles forestières très différentes. Il en va de même pour l'étude de Chrétien et collaborateurs (2020) pour la fabrication fromagère, dont le processus de fabrication est très documenté, suffisamment précis pour élaborer une IA à partir de ces données riches, fiables et tangibles.

---

23 Pour de l'information sur le projet Silva numérica, voir https://silvanumerica.net/

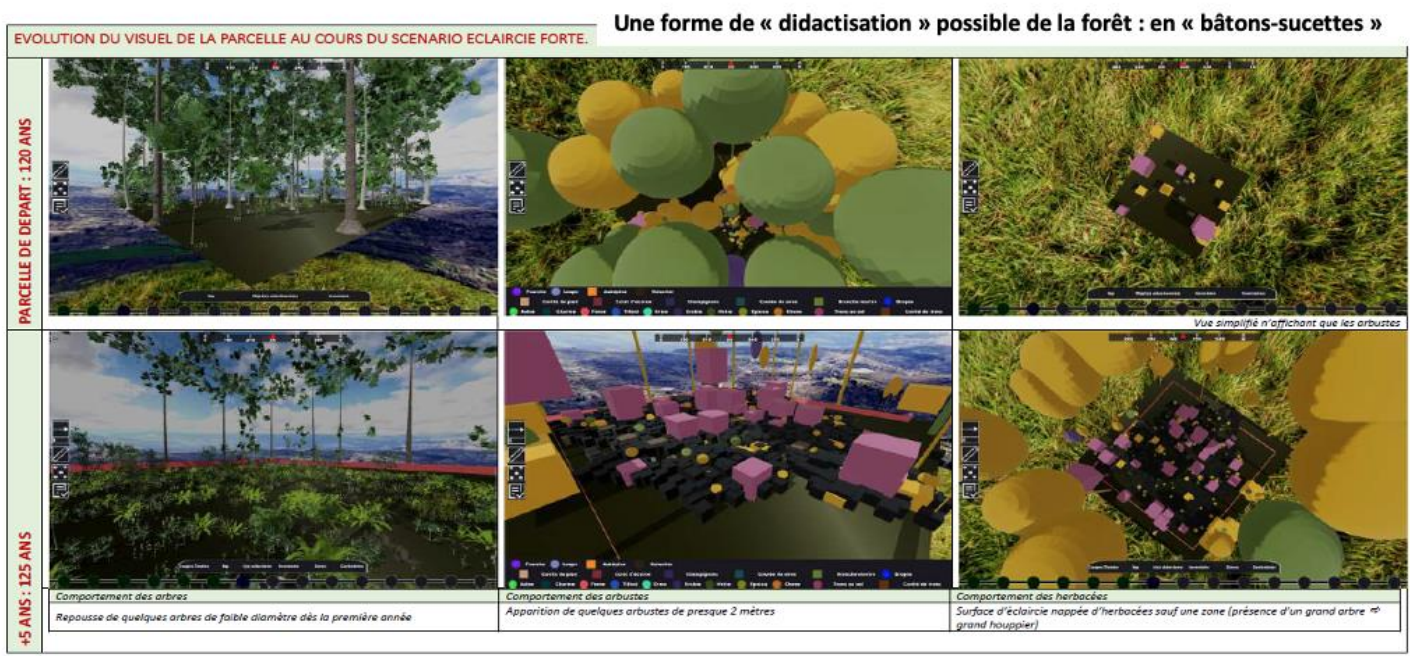

Figure 13. Variabilité du réalisme de la simulation en fonction d'enjeux didactiques (source : projet Silva numerica)

À gauche de la Figure 13, on retrouve une simulation de la forêt qui est relativement réaliste au regard de la sylviculture sur laquelle elle est basée (futaie régulière de chênes sessiles) et à droite une simulation qui symbolise les arbres en « bâtons de sucette ». Cette dernière, bien qu'ayant perdu de la fidélité au réel, constitue une ressource sémiotique pour apprendre par exemple à sélectionner des arbres que l'on veut conserver ou abattre selon leur position sociale dans la forêt.

La limite de la simulation réside dans la variabilité et la diversité des situations de travail. Cela nous conduit à poser les questions suivantes : dans quelle mesure en formation professionnelle l'IA pourrait-elle aider les concepteurs à modéliser un processus de formation à partir d'un ensemble de bases de données ? Comment l'IA pourrait-elle aider à la modélisation et à la simulation du réel si les données existent, alors même qu'un certain nombre de phénomènes et de processus sont très mal documentés ? Certains auteurs, dont les travaux récents de didacticiens (Vadcard 2013, 2019) soulignent que réussir à simuler l'environnement, les situations complexes et dynamiques dans lesquelles les futurs professionnels ont à agir en modélisant au plus près le réalisme ou l'authenticité des situations n'est pas toujours un garant d'efficacité du point de vue de la formation. Ce réalisme est parfois indésirable parce qu'il va justement à l'encontre du fait que pour faire apprendre et pour que les situations puissent devenir des situations de formation, on a besoin de modifier le réel à des fins pédagogiques.

---

24 Il existe des logiciels performants de croissance des arbres mais à usage de recherche ou pour l'enseignement supérieur. Voir par exemple Capsis Presentation : https://capsis.cirad.fr/capsis/presentation

L'intérêt didactique d'une IA est ainsi non seulement de modéliser le réel, mais aussi de conserver un potentiel d'apprentissage d'une activité professionnelle dans une perspective de didactique professionnelle. Les situations proposées dans les scénarios de simulation, les réponses à l'activité simulée devraient être à la fois documentées par l'IA, pour permettre aux futurs professionnels de s'y immerger, mais aussi de s'en échapper pour les analyser, y réfléchir. Car ce sont les possibles évolutions d'une situation problème, et sa gestion qu'il s'agit d'apprendre en développant un raisonnement, une réflexivité à leur propos et cela en sélectionnant les données pertinentes pour y agir. C'est de son usage dans les raisonnements que les données retenues tirent leur tangibilité. Or, ces raisonnements d'une part, et les données sur lesquelles ces derniers s'appuient d'autre part, sont rarement documentés et difficilement modélisables.

Ainsi, documenter une IA, pour concevoir ou optimiser un système formatif suppose de comprendre comment ces raisonnements s'apprennent dans les lieux d'apprentissage et quels en sont les obstacles. L'enjeu de la conception tient alors dans la formalisation de scénarios de formation suffisamment proches des situations de travail et de l'activité professionnelle cible et réduisant les contraintes imposées par les environnements informatiques à la conception didactique.

# Enjeux didactiques de l'IA et de la simulation en formation professionnelle

Dans quelle mesure l'IA peut-elle aider et apporter une plus-value avec une visée didactique ? Quelles sont les contraintes de conception pour un usage créatif de l'IA ? Nous reprenons ici des questions complémentaires à celles déjà posées autour de la plus-value potentielle d'un recours à IA dans la formation professionnelle. Quelles informations relatives à un environnement complexe doivent être données aux formés, sous quelle forme et à quel(s) moment(s), pour étayer le processus ? Contrairement aux experts, l'une des difficultés des novices consiste à repérer les indicateurs et à élaborer des inférences pertinentes pour diagnostiquer la situation. Ainsi, l'environnement numérique à créer doit attirer l'attention des apprenants sur les dimensions essentielles qu'ils doivent prendre en considération pour agir en tant que professionnels : mais à quel point ? Quelles transpositions numériques et didactiques intégrant de l'IA sont requises pour aider, guider, accompagner la bonne décision en faisant en sorte qu'en situation l'erreur soit encore possible ? Comment ne pas réduire la complexité de l'environnement réel et la richesse d'une activité de travail tout en facilitant sa compréhension aux apprenants ?

Peut-on envisager un guidage sémantique ? Faut-il y préférer un guidage iconique ? Dans quelle mesure l'IA couplée à des traces d'apprentissage pourrait-elle non seulement permettre d'adapter le type de guidage proposé (en fonction des connaissances, des apprentissages des utilisateurs), mais aussi faciliter la rétroaction des enseignants formateurs ? Certains chercheurs pensent que l'utilisation de l'IA pour mieux guider les apprentissages est intéressante (voir, *International Journal of Artificial Intelligence in Education*). Dans le projet Silva numerica, s'est posée la question de la nature des informations relatives aux environnements forestiers simulés à proposer aux apprenants, sous quelle forme et à quel moment. Différents types d'informations leur ont été données. Prenons l'exemple des fiches descriptives, par exemple d'animaux, d'insectes, d'oiseaux, mais aussi des données sur les populations d'arbres et les différentes espèces d'arbres en présence, ou encore la construction d'un audit sur la valeur économique et la valeur écologique de la parcelle forestière.

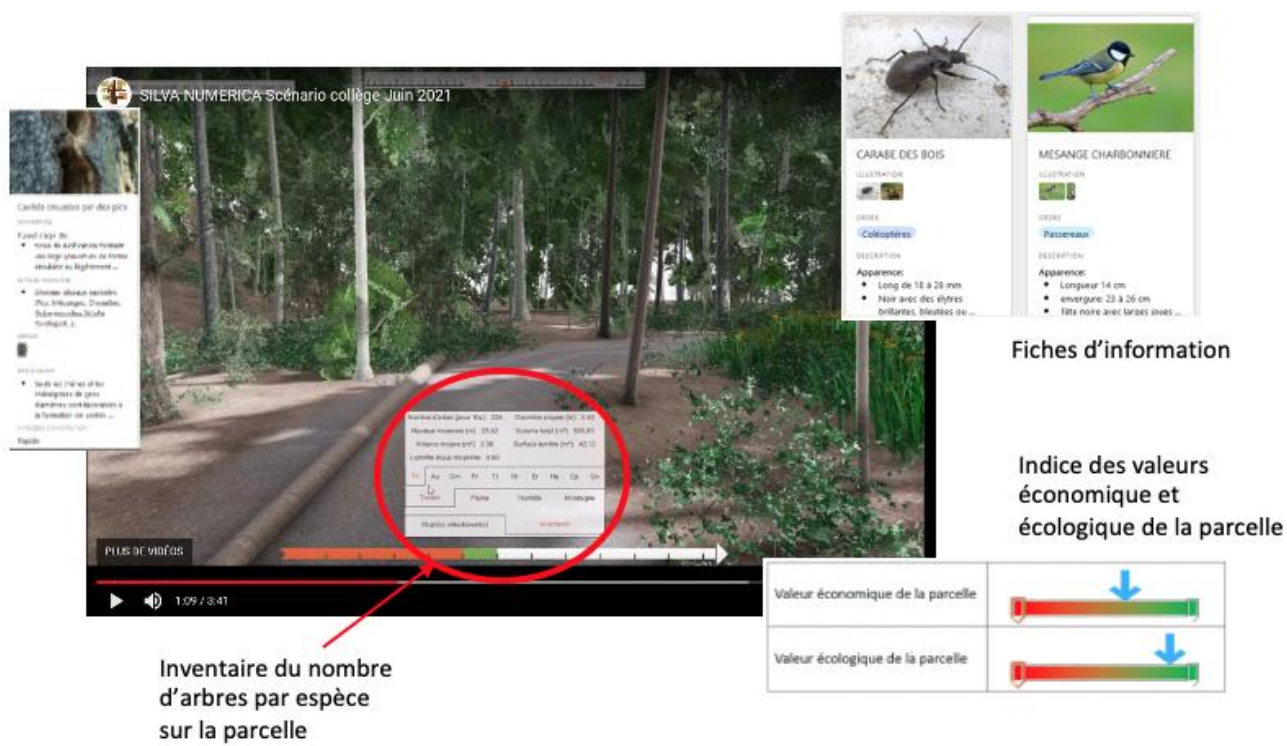

Figure 14. L'IA et les *learning analytics* pour adapter les informations, leur forme et le moment où elles sont accessibles (source : projet Silva numerica)

Faut-il donner accès immédiatement à ces informations, à ces ressources comme le ferait un enseignant dans un cours magistral ? Une IA pourrait-elle aider à adapter les informations, leur forme et le moment où elles sont accessibles pour l'apprenant ?

Les deux points précédents mettent en évidence une maturité insuffisante de l'écosystème de la formation professionnelle vis-à-vis de l'IA et réciproquement, par la méconnaissance de la formation professionnelle par les acteurs de l'IA. Documenter une IA suppose une décision politique lourde pour les organisations productives, comme pour les formations professionnelles initiales et continues : accepter de donner à voir le travail tel qu'il se fait, tel qu'il se pense, tel qu'il peut ou non se dire. Les industries à risque y sont contraintes par la réglementation. Les autres secteurs professionnels demeurent en partie « sourds et aveugles », non par mauvais esprit, mais par méconnaissance. Il y a beaucoup d'enjeux dans le dévoilement du travail et les raisonnements qui le guident.

Documenter une IA suppose forcément une réduction des situations professionnelles à des segments traitables de l'activité professionnelle à faire apprendre. Mais cette réduction doit se maintenir au niveau des variables de l'action mobilisées dans les raisonnements des professionnels. Autrement dit, il convient de penser la tangibilité des données non comme une information ou un fait, mais comme une information qui soit « traitable », soit exprimée dans un format reconnaissable, et qui est susceptible d'être mise à l'épreuve, satisfaisant simultanément les conventions de métier, la pluralité des modèles opératoires et les ajustements à ce qui convient en situation.

# L'IA et les tuteurs intelligents pour les apprentissages professionnels et l'assistance à l'enseignement

Une question cruciale à nos yeux porte sur la capacité de l'IA à renvoyer des rétroactions pertinentes pour les apprentissages. Dans la littérature, comme le montre l'étude de Hwang et collaborateurs (2020), il chemine l'idée d'une IA comme tuteur intelligent. En formation professionnelle, l'IA est-elle en capacité d'analyser l'activité de l'apprenant afin de réaliser des rétroactions pertinentes pour les apprentissages ? Cette question est redoutable, car cela implique une attente envers les technologies basées sur l'IA, qui devraient savoir analyser l'activité de l'apprenant sur la base de ses seules actions sur la machine, opérer une explicitation de la validité ou de l'invalidité de ce qu'il a réalisé et évaluer la qualité des apprentissages liés et cela à partir de l'intégration de différentes informations : celles relatives aux actions réalisées par l'apprenant, puis leur interprétation. En effet, dans l'action professionnelle, il y a souvent plusieurs résultats possibles qui peuvent être pertinents. Dans la gestion forestière, par exemple, chaque forestier en fonction des objectifs (de production, environnementaux, sociaux) peut analyser la situation et prendre des décisions d'action différentes d'un autre forestier. Pour autant, chaque itinéraire sylvicole établi peut être pertinent. Par ailleurs, cette interprétation de l'IA devrait tenir compte du fait qu'une conformité du résultat ne garantit pas du tout la compréhension de l'apprenant, des stratégies, modes opératoires et savoirs professionnels. Cela ne peut pas non plus être une conformité à une bonne pratique ou un raisonnement standard puisque, là encore, dans le domaine professionnel, plusieurs raisonnements différents, pour une même classe de situations, permettent d'arriver à un résultat similaire ou un résultat acceptable. Et enfin, il resterait à préciser quels indicateurs permettent d'inférer que l'apprenant a appris quelque chose ou est en difficulté. Par exemple, un apprenant peut tout à fait obtenir une « belle forêt » virtuelle (de beaux chênes, au diamètre impressionnant et donc avec une grande valeur marchande) par un cheminement intellectuel hasardeux et sans même avoir compris les concepts pragmatiques organisateurs d'une bonne gestion forestière. L'IA en tant qu'assistant de l'enseignant professionnel doit alors lui fournir des informations pertinentes et utilisables afin de pouvoir formaliser, discuter et institutionnaliser des savoirs professionnels.

La question devient alors : dans quelle mesure l'IA, couplée à des traces d'apprentissage permettrait d'adapter le type de guidage proposé en fonction des connaissances, des apprentissages des utilisateurs, comme certains chercheurs le pensent possible ? Pourrait-on s'appuyer sur la capacité annoncée de l'IA à apprendre et à s'autocorriger (apprentissage automatique) et à mimer le comportement humain (apprentissage profond) afin qu'elle détecte des erreurs ou des difficultés récurrentes des apprenants, puis transforme le contenu, la forme ou le moment des rétroactions produites, comme pourrait le faire un enseignant ? Cela nous semble hautement hasardeux, car 1/ la capacité d'autoapprentissage des réseaux de neurones artificiels est limitée par la manière dont ils s'indexent sur le réel, car ils cherchent des régularités non pas dans le réel, mais dans l'image numérique qu'ils en forment et 2/ le moteur du tutorat est une maïeutique des discordances, alors que le principe de construction des IA est de rechercher des concordances entre jeux de données pour extraire ces régularités.

Il est tout à fait possible que l'IA arrive prochainement à un niveau de finesse lui permettant de détecter des discordances pédagogiquement pertinentes au sein du texte (ou de l'activité simulée). Mais les problèmes d'indexation de l'IA sur des éléments contextuels distants demeurent prégnants. En effet, en formation professionnelle, quand bien même la recherche pourrait aider à déterminer une structure conceptuelle pour chaque classe de situation professionnelle et modéliser un mode opératoire expert, l'indexation des apprentissages sur les situations de travail réelles, sur leur dynamique et leurs particularités, constitue une part très importante de l'activité des formateurs et tuteurs. Elle implique bien souvent la connaissance d'implicites professionnels pouvant être très distants des conditions hic et nunc de la réalisation de l'action (culture de l'entreprise, style professionnel de tel tuteur en particulier, inscrite ou non dans le genre professionnel, etc.). Cela pose des problèmes d'acquisition, de sélection et de pertinence que l'avancement de la numérisation du monde ne permet pas encore de résoudre par l'automatisation, fût-elle appuyée sur des capacités de calcul contemporaines. Les exemples supra de la fabrication du Comté, de la gestion forestière ou de l'usage de l'IA dans les garages automobiles nous semblent tout à fait significatifs de cette difficulté d'indexation du virtuel numérique sur le réel. Dans ce type de contexte, il nous semble encore fantasmatique de confier à une IA seule, sans supervision de l'enseignant, la conduite d'une maïeutique apprenante à partir des écarts constatés au sein du discours sur l'activité, entre l'activité racontée et l'activité observée, entre l'activité observée et l'activité réelle, entre l'activité prévue et réalisée ou souhaitable a posteriori. Si cette maïeutique est peut-être accessible à l'IA quand elle s'exerce au sein du texte facilement accessible à la numérisation, il nous semble que les activités téléologiques dont la régulation fait appel à des prises d'informations aux règles complexes et pouvant se faire très à distance de l'activité immédiate vont rester encore hors de portée de ces approches pour un certain temps. On l'aura compris, le problème n'est pas tant du côté de l'IA en soi. Les deux exemples suivants illustrent ces problématiques pour deux configurations de mobilisation de l'IA dans la constitution de la capacité à agir.

## LES SCENARIOS PROJET SILVA NUMERICA

Nous reprendrons l'exemple du projet Silva numérica : l'outil a été conçu pour enregistrer les actions réalisées par les apprenants au fur et à mesure qu'ils avancent dans les scénarios et les réponses qu'ils apportent aux activités demandées. Ces traces d'actions individuelles étaient accessibles aux enseignants à la fin de l'activité, sous une forme extrêmement complexe. Elles se sont révélées difficilement exploitables par elles-mêmes après la formation (texte illisible et chargé d'informations) et carrément inexploitables en cours de formation. Il était donc impossible pour les enseignants formateurs de pratiquer un débriefing sur la base du parcours de chaque apprenant dans les scénarios numériques. Fort de ce constat lors des premières expérimentations, la décision a été prise de plutôt suivre l'activité des apprenants au fur et à mesure de leur travail dans l'environnement virtuel et d'avoir des traces d'activité permettant d'implémenter a posteriori dans les scénarios des questionnaires d'évaluation. Et c'est sur la base des réponses à ces tâches d'évaluation qu'ont été prises en compte des traces de l'activité, des traces des apprentissages. On est loin du tuteur intelligent qui pourrait apporter une assistance pédagogique au professeur. Si l'on veut vraiment savoir si les apprenants ont compris quelque chose, cela implique aussi qu'ils puissent avoir des espaces dans lesquels ils peuvent développer leur raisonnement et donc qu'il y ait des questions ou des activités qui impliquent des réponses ouvertes. Or, les enseignants ont constaté que des éléments issus de questions ouvertes étaient trop compliqués à interpréter et à exploiter collectivement, en formation. Ils en sont donc venus à proposer des quiz et des textes à trous plus faciles à utiliser. Alors est-ce que l'IA serait en mesure d'apporter une plus-value en lien avec des données beaucoup plus larges, beaucoup plus hétérogènes, pour proposer un certain nombre d'indicateurs qui permettraient de situer l'apprenant dans son apprentissage ?

Nous prendrons cette fois-ci le cas des garages connectés interrogeant le potentiel d'apprentissage des situations de travail et la contribution de l'IA au développement des compétences. Gagneur et Vassou (2019) ont mené une recherche sur les garages connectés. Du point de vue des constructeurs automobiles, une révolution serait en marche en lien avec l'omniprésence du numérique. Du point de vue des auteurs, un problème apparaît lié à une focalisation excessive sur les valises diagnostiques utilisées en cas de panne des véhicules. Les capteurs servent à remonter de l'information, qui est ensuite comparée à des bases de données dans lesquelles ont été implémentées les pannes historiques : cela permet d'identifier le type de panne et de fournir aux mécaniciens une procédure pour la résoudre. Le constat est que 90 % des mécaniciens utilisent cette valise diagnostique sur un mode basique, autrement dit sur des pannes simples, parfaitement référencées et pour lesquelles ils suivent ensuite la procédure. Cela marche assez bien. Dans ce cas-là, le processus n'est pas compliqué à apprendre en formation professionnelle. Mais dans 30 à 40 % des cas, ça ne marche pas. À ce moment-là, il faut beaucoup plus de temps pour résoudre ces pannes et pour mener l'enquête. En effet, bien qu'il y ait des capteurs un peu partout dans les voitures, certains endroits ne peuvent pas en être pourvus. Et la finalité de ces capteurs est de faire fonctionner la voiture, mais pas de permettre l'enquête des mécaniciens en cas de panne. De nombreux endroits où il n'y a pas de capteurs échappent donc à l'outil numérique. Et c'est là qu'intervient l'expertise des professionnels dans ce qui n'est pas accessible à l'outil numérique et donc à la mallette diagnostique. D'abord pour faire « capteur humain » et renseigner la machine. Ensuite, pour les pannes plus complexes, les professionnels doivent être capables « de prendre la main sur le diagnostic fait par la machine » pour l'utiliser dans un autre mode. Ce sont des compétences critiques pour la survie des entreprises : il suffit de quelques pannes qui résistent au diagnostic et à l'intervention, et consomment donc du temps de travail sans possibilité de facturation, pour mettre en péril la profitabilité d'un atelier de réparation automobile. Il s'agit de recueillir des données, mais ensuite de les croiser avec d'autres données. Des données d'observation des véhicules tels que les bruits, les traces d'huile, voire les données relatives à l'expérience client, son usage du véhicule, etc. Pour les mécaniciens experts, l'enjeu est de « dominer l'IA et non pas de se laisser dominer par elle ». Et donc, toute la compétence, à ce moment-là sera dans l'enquête et leur intégration des différents points de vue, des différents types de données : cela implique d'avoir des connaissances sur ce qui se passe physiquement et mécaniquement pour limiter les interprétations erronées. Pour la formation professionnelle, le défi est alors de proposer des conditions qui vont permettre d'apprendre à dominer la machine pour en faire un instrument de son activité, dans un système sociotechnique.

En conclusion, avec la mallette diagnostic dans le secteur de la maintenance automobile, en suivant Casilli (2019), on pourrait dire que l'IA « invisibilise une part du travail humain ». N'oublions pas que dans l'organisation du travail, cet usage maîtrisé de la machine est confié souvent aux agents plus experts alors que les mécaniciens de base, dont les apprentis, n'ont pas souvent accès à ces pannes. Apparaît ici le risque d'une déqualification progressive des mécaniciens. Cela signifie aussi que tous les professionnels ne pourront pas être tuteurs en mesure d'assurer des actions de formation en situation de travail (AFEST) par exemple. Ce cas nous invite à rappeler que pour dominer la machine, il faut aussi dominer son objet de travail, c'est-à-dire avoir des connaissances en mécanique, en physique assez pointues, pour pouvoir se prémunir des biais de la médiation numérique. Le risque est grand aussi de conformer les apprenants à un ou quelques modèles professionnels qui ne tiennent pas compte de la richesse du travail réel et de l'intelligence professionnelle (Guidoni-Stoltz, 2019).

En définitive, dans la formation professionnelle, la présence de l'IA ne fait que renforcer la nécessité d'un accompagnement humain des apprentissages par des professionnels tuteurs et par des formateurs qui sont capables de rendre visible la part invisible du travail. Ils doivent mettre en partage les implicites qui fondent l'activité professionnelle.

Il nous semble donc que l'affirmation posée par Hwang et al. (2020) demeure tout au plus une croyance ou une illusion, au moins dans le domaine de la formation professionnelle :

Une application d'IA pourrait jouer le rôle d'un tuteur qui observe les processus d'apprentissage des élèves, analyse leurs performances d'apprentissage et leur fournit une assistance instantanée en fonction de leurs besoins. Sur la base des besoins potentiels des élèves, une équipe interdisciplinaire (composée, par exemple, d'informaticiens et de spécialistes de l'apprentissage) peut développer un système de tutorat intelligent qui permet aux élèves d'apprendre, de s'exercer et d'interagir avec leurs pairs ou leurs enseignants, mais qui fournit également des conseils, des orientations et des aides aux individus en fonction de leur statut ou de leurs besoins. (p. 2, traduction libre)

# Conclusion

Dans ce chapitre, nous avons voulu traiter des intérêts potentiels et des limites actuelles concernant l'usage de l'IA pour enseigner et pour apprendre les métiers dans la formation professionnelle et sur la place du travail. Nous avons pu expliquer comment l'une des clés pour rendre possible une plus-value de tels usages de l'IA réside dans la capacité des évolutions technologiques à fiabiliser et à rendre tangibles et exploitables les données du travail sur lesquelles vont porter les apprentissages professionnels (gestes, situations, circonstances, raisonnements, etc.) ainsi que les données d'apprentissage elles-mêmes. Si la vitesse d'évolution des techniques d'apprentissage automatique rend difficiles les pronostics de développement de l'IA dans les années à venir, l'indexation de ces processus sur le réel nous semble constituer une limite d'ordre logique plus que technique, et donc probablement pérenne. Partant de là, rendre tangibles et exploitables les données du travail implique une démarche partenariale avec les organisations professionnelles sur différents points : la conception des outils d'aide à l'apprentissage fondée sur un accès de telles données auprès des professionnels, notamment par la recherche ; le dépassement du frein sociotechnique du manque de confiance des professionnels quant à de telles données ; l'accès des enseignants, formateurs, apprenants aux outils professionnels embarquant une IA, en mettant en place les conditions pour rendre visible le travail réel, apprendre la part de raisonnements mobilisés et permettre une domination de la machine par une appropriation en tant qu'instrument (Folcher et Rabardel, 2004). Nous avons montré en quoi cela nécessite un travail de conception d'usage des outils d'assistance à l'enseignement avec les enseignants et les formateurs eux-mêmes : c'est une condition de leur intégration de l'IA dans les pratiques et son acceptabilité. L'enjeu est aussi que l'usage de l'IA dans l'accompagnement des apprentissages se donne les moyens d'une reconfiguration ergonomique des outils proposés aux acteurs.

La didactique professionnelle peut contribuer à des rapprochements pour que les instruments numériques éducatifs intégrant de l'IA soient réellement adaptés à la formation professionnelle. Pour en savoir plus, un site internet national existe : https://didactiqueprofessionnelle.ning.com/

# Références

# 6 /// L'IA POUR MIEUX APPRENDRE ET APPREHENDER L'IA

Frédéric Alexandre[1]
Marie-Hélène Comte[1]
Aurélie Lagarrigue[1]
Thierry Viéville[1][2]

[1] Institut national de recherche en sciences et technologies du numérique, Équipe Mnémosyne, France
[2] Laboratoire d'innovation pour le numérique en éducation, Université Côte d'Azur, France

L'IA en éducation peut être abordée depuis trois perspectives parallèles. D'abord, elle peut servir à adapter l'expérience d'apprentissage par la conception d'outils prenant en compte différentes caractéristiques des apprenants ou des traces numériques issues de leur interaction avec des systèmes. Bien utilisés, de tels systèmes pourraient décharger les enseignants de tâches relatives à la transmission des contenus et leur permettre d'intervenir sur des aspects plus complexes de l'apprentissage des élèves. Ensuite, l'IA peut être utilisée comme outil scientifique pour mieux comprendre les phénomènes d'apprentissage humain, par la modélisation de l'apprenant. Finalement, l'IA peut être envisagée depuis la perspective de l'éducation critique à l'IA. Ce chapitre présente succinctement ces trois perspectives qui ne s'excluent pas les unes des autres, mais qui se complètent pour mieux cerner les enjeux de l'IA. Les dernières recherches associant les sciences de l'éducation et les sciences du numérique permettent de comprendre les liens entre l'intelligence artificielle (IA) et l'éducation, y compris leurs limites. Ces recherches nous montrent comment l'IA peut être pensée pour mieux apprendre et développer son esprit critique (Roux et al. 2020 ; Viéville, 2018), pour comprendre l'apprentissage humain lui-même, et enfin comme objet d'enseignement, pour maîtriser de manière éclairée ces outils devenus quotidiens (Viéville et Guitton, 2020).

## L'IA comme outil d'apprentissage adaptatif

Tout d'abord, en utilisant des algorithmes, l'apprentissage peut être *adaptatif*. En analysant les traces d'apprentissage de l'élève, comme des résultats à des questionnaires ou des données d'utilisation d'un logiciel, le système peut modifier son fonctionnement pour s'adapter à la personne, notamment à travers la sélection de contenus et du niveau de difficulté. Il commence à être possible d'analyser son comportement grâce à des capteurs, certains externes comme une caméra, et d'autres plus intrusifs comme une interface cerveau-ordinateur. Ce principe d'adaptation est au cœur de la pédagogie numérique, et se rencontre le plus souvent dans un contexte où sont aussi poursuivis des objectifs de ludification ; l'apprenant s'inscrivant alors dans un jeu pédagogique avec la machine, parfois en collaboration avec d'autres apprenants (Giraudon et al., 2020).

Le projet KidLearn[25], illustré dans la Figure 15, propose une activité d'apprentissage dont les multiples variantes impliquant additions ou soustractions de nombres entiers ou décimaux ont été conçues et mises en place par des didacticiens. Ces variantes sont organisées sous forme d'un graphe de difficultés croissantes, en respectant le concept de zone proximale de développement (Vygotsky, 1978). Ce concept est fondé sur l'idée que, entre un exercice trop difficile qui décourage et un exercice trop facile qui lasse, il existe une zone optimale qui maximise le progrès d'apprentissage, mesuré ici en observant les performances de l'élève au fil du jeu. Ce sont ces éléments qui sont intégrés à l'algorithme, qui va s'adapter automatiquement à la personne apprenante (Oudeyer et al., 2020).

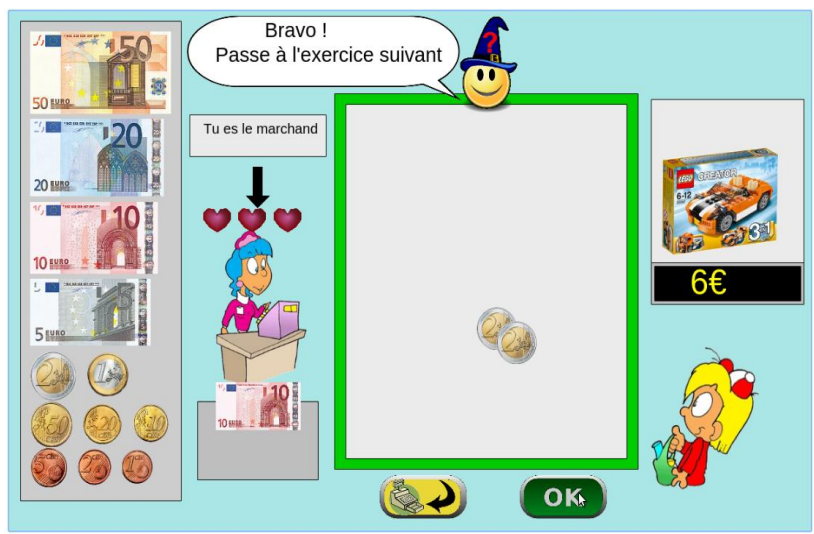

Figure 15. Le projet Kidlearn se base sur des scénarios d'achats de 1 à 2 objets ou de rendu de monnaie (source : équipe Flowers, Inria)

Si le développement de ces usages reste encore limité[26], la recherche scientifique étant toujours en cours, elle offre une étape de réflexion préalable fondamentale, pour comprendre comment fonctionnent l'acquisition et l'appropriation de connaissances. En effet, pour systématiser cette approche adaptative, il faut formaliser les savoirs (connaissances) et savoir-faire (pratiques) à faire apprendre, ce qui oblige à expliciter et structurer les types de tâches et techniques de résolution. Par ailleurs, il faut garder à l'esprit la nécessité de ne pas surcharger l'apprentissage avec des tâches cognitives annexes liées à l'activité elle-même. L'apprentissage adaptatif doit également se faire dans un contexte contraint par les disponibilités du matériel, la formation des personnes, ou les limites à l'usage des écrans.

Les effets positifs de cet apprentissage machine peuvent être nombreux. En tout premier lieu, on observe qu'ils génèrent en général un meilleur engagement de la personne apprenante, car interagir autrement avec les contenus offre une chance supplémentaire de bien les comprendre. En effet, le fait que la difficulté soit adaptée à l'apprenant permet de limiter, voire d'éviter le découragement ou la lassitude. De plus, à la différence de l'humain, la machine ne « juge pas », ce qui peut contribuer à maintenir cet engagement. Pour autant, ce type d'apprentissage peut nécessiter un investissement important pour l'enseignant, si la conception ne prend pas en compte la charge cognitive de l'élève. Il peut aussi induire le risque que l'élève se disperse au lieu de s'investir dans l'apprentissage escompté si l'aspect ludique est prépondérant.

Les outils adaptatifs intégrant des principes considérés comme étant de l'IA doivent permettre aux enseignantes et enseignants de se rendre plus disponibles pour les élèves qui en ont le plus besoin, car la classe est investie dans des activités d'apprentissage autonomes. De même, cela leur permet de se libérer, comme en pédagogie de la classe inversée, d'une partie des transmissions de savoir grâce à ces contenus multimédias autoévalués et à des exercices d'entraînement automatisés, pour se concentrer sur d'autres approches pédagogiques, par exemple fonctionnant par projets. Par rapport à des outils numériques non adaptatifs, c'est-à-dire sans apprentissage automatique, le degré d'apprentissage en autonomie peut être bien plus élevé et s'applique plus largement, avec des parcours complets de développement de compétences. Ces outils répondent par ailleurs à un besoin dans le cadre de situations d'enseignement à distance, et viennent questionner l'organisation du temps de travail scolaire.

Il convient néanmoins de souligner les dérives possibles des usages de ces données : le traçage omniprésent et omnipotent des apprenants, leur catégorisation, la tentation de réduire les effectifs de personnel scolaire, le renforcement des inégalités en lien avec l'illectronisme (Gandon, 2020), d'autant que ces pratiques numériques accroissent le risque d'une intégration aux autres facettes du comportement en ligne, comme les achats, la consultation de vidéos ou les lectures, bref d'un éventuel changement de finalité des traitements.

## L'IA comme modèle pour comprendre l'apprentissage humain

La possibilité de collecter et interpréter des traces d'apprentissage pourrait permettre d'améliorer l'apprentissage, si ces traces sont utilisées par l'apprenant ou l'enseignant pour s'autoréguler ou pour la régulation externe par l'enseignant (Romero, 2019). L'usage des traces pourrait permettre aussi de mieux comprendre sur le long terme les modes d'apprentissage humains. Ces traces peuvent être relevées lors de l'utilisation d'un logiciel par la mesure des déplacements de la souris ou des clics réalisés au doigt, des saisies au clavier, ou encore par des capteurs employés dans des situations pédagogiques sans ordinateur ou avec ordinateur (p. ex. caméra, micro, accéléromètre, GPS). On peut ainsi penser à une activité physique dans une cour d'école, observée avec des capteurs visuels ou corporels. Exploiter ces mesures impose alors non seulement de formaliser la tâche d'apprentissage elle-même, mais en plus, de modéliser la tâche et la personne apprenante, non pas dans sa globalité, mais dans le contexte d'une tâche en particulier.

> L'usage de traces d'apprentissage dans les environnements numériques d'apprentissage permet de de modéliser la tâche d'apprentissage, mais également l'activité de l'apprenant dans la tâche.

Les algorithmes d'apprentissage automatique reposent sur des modèles assez sophistiqués. Ils ne sont pas forcément limités à des mécanismes d'apprentissage supervisés où les réponses s'ajustent à partir d'exemples fournis avec la solution, mais fonctionnent aussi par « renforcement » : le système va inférer les causes permettant d'expliquer les retours positifs (appelés récompenses[27]) ou négatifs au cours de l'apprentissage, en construisant un modèle interne de la tâche à effectuer. Ces modèles sont opérationnels, c'est-à-dire qu'ils permettent de créer des algorithmes effectifs qui ajustent leurs paramètres. On peut alors se demander si de tels modèles permettent de modéliser des aspects de l'apprentissage humain. En neurosciences, ces modèles dits computationnels, représentent déjà les processus fonctionnels de notre cerveau sous forme de mécanismes de calculs ou de traitement de l'information au niveau neuronal pour mieux les expliquer.

Ce domaine d'utilisation des sciences informatiques et de l'IA comme outils de formalisation pour modéliser l'apprentissage humain, qu'on pourrait qualifier de « sciences de l'éducation computationnelles » (Romero et al., 2020), n'en est qu'à ses débuts, mais on perçoit déjà le potentiel qu'il peut représenter pour les sciences de l'éducation : c'est pourquoi des recherches sont menées de manière transdisciplinaire avec les sciences du numérique et les neurosciences cognitives afin d'explorer ces potentialités.

# L'IA, un enseignement citoyen

Afin de « maîtriser le numérique » au sens de Giraudon et al. (2021) et non uniquement le consommer, il est important d'être initié au fonctionnement scientifique et technique des objets informatiques matériels et logiciels, pour éclairer l'usage de ses applications (Atlan et al., 2019 ; Romero, 2018). L'intégration des technologies dites d'IA dans notre quotidien appelle au développement de l'esprit critique des jeunes (Viéville et Guitton, 2020).

Il est par exemple important de comprendre que, en IA, le résultat du traitement des données par les algorithmes n'est pas lié entièrement à leur programmation. Les fonctions souhaitées ne sont pas implémentées seulement à l'aide d'instructions, mais aussi en fournissant des données à partir desquelles les paramètres sont ajustés pour obtenir le calcul souhaité. Selon le degré d'autonomie des programmes, il peut même y avoir des conséquences imprévues comme cela a été le cas dans des systèmes de robots conversationnels qui ont appris, par des corpus de mauvaise qualité, à produire des commentaires non éthiques sur les médias sociaux. Sur le plan juridique, il importe également de se familiariser avec les implications découlant de l'utilisation d'un « cobot », à savoir d'un mécanisme robotique interagissant avec notre vie quotidienne. On désigne ici un système artificiel, comme une machine médicale qui aurait pour fonction de permettre d'éclairer l'aide à la décision thérapeutique dans des situations de différents degrés d'urgence. Ce contexte médical nous montre combien la chaîne de responsabilité entre conception, construction, installation, paramétrage et utilisation est ici infiniment plus complexe que dans une machine dont le comportement n'est pas partiellement autonome.

> La formation à l'IA doit permettre de développer des connaissances et des compétences pour comprendre le fonctionnement de l'IA et pouvoir développer un positionnement citoyen et professionnel sur les potentiels et limites de l'IA.

C'est face à ces enjeux qu'a été créée le MOOC Intelligence artificielle avec intelligence, présenté dans le chapitre 1, afin d'initier de manière citoyenne les éducateurs à l'informatique y compris à propos de comment l'IA peut contribuer à faire développer des compétences (Alexandre et al., 2022). Des outils pédagogiques existent et continuent de se développer pour initier les apprenants progressivement au fonctionnement de l'IA. La Figure 16 montre ainsi une « machine » minimaliste, développée par Pixees et construite par Snzzur.fr, permettant de stocker des boules bleues, pour les succès, ou rouges, pour les échecs, de chacune des parties. Ainsi, la machine « mémorise » la stratégie gagnante à coup sûr. Tous les plans de construction sont accessibles en libre accès et le jeu peut être reproduit à faible coût avec des outils de base[28].

---

27 Dans l'apprentissage par renforcement, les « récompenses » peuvent être tant positives que négatives. Ainsi, elles ne se limitent pas au positif comme au sens classique du terme d'apprentissage par renforcement en psychologie.

On a pu établir qu'apprendre l'informatique de manière « débranchée », c'est-à-dire en s'extrayant de l'interaction avec la machine pour se concentrer de manière active sur les concepts sous-jacents, permet de mieux comprendre un des mécanismes de fonctionnement de l'IA (dans cet exemple, l'apprentissage par renforcement en l'occurrence).

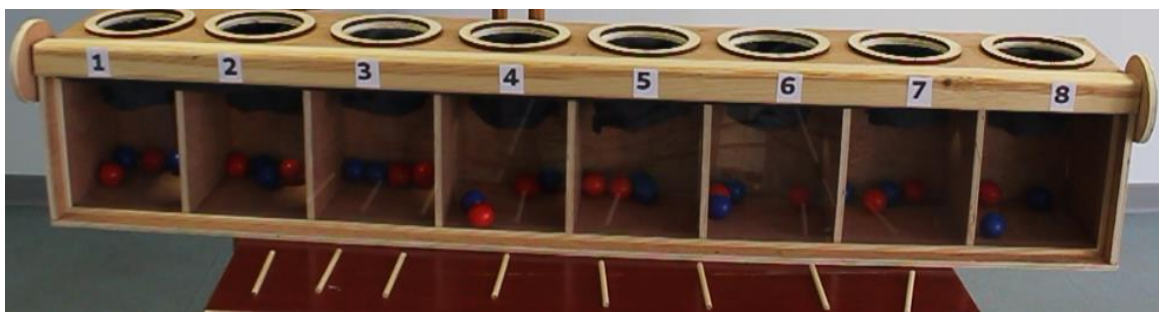

**Figure 16. Un exemple d'activité débranchée permettant d'expérimenter un algorithme d'apprentissage par renforcement**

# Un bouleversement dans notre façon de penser

Dès le début de l'informatique, nous avons vu évoluer notre manière d'apprendre et d'enseigner. Par exemple, faut-il encore apprendre à calculer lorsqu'on peut utiliser une calculette ? Cela est sans doute nécessaire pour développer ses capacités cognitives et comprendre une opération arithmétique, mais de fait, nous avons moins besoin de devenir de « bons calculateurs » qu'à l'époque où le calcul mental était la clé pour s'en sortir au quotidien ou dans ses activités professionnelles. Par contre, il nous faudra toujours être entraînés au calcul des ordres de grandeur, pour vérifier qu'il n'y a pas d'erreur quand on a posé le calcul et obtenu le résultat, ou s'assurer que le calcul lui-même est pertinent.

Ces mutations de l'activité humaine se retrouvent au fur et à mesure que nous automatisons des processus qui relèvent de l'intelligence humaine. Finalement, si nous nous contentons d'utiliser des algorithmes d'IA sans chercher à comprendre leurs grands principes de fonctionnement et leurs implications sur notre vie, nous risquons de perdre de l'intelligence individuelle et collective : nous nous en remettrons à leurs mécanismes en réfléchissant moins par nous-mêmes, et en développant moins l'esprit critique indispensable à la formation de citoyens autonomes et éclairés. C'est là tout le sens de comprendre comment fonctionne l'IA (Roux et al., 2020). Si nous cherchons à comprendre et à maîtriser ces processus, alors la possibilité de déléguer ce qui est mécanisable dans ce que traite de l'intelligence humaine peut nous offrir la chance de nous libérer consciemment de tâches devenues automatiques afin de consacrer notre intelligence humaine à des objectifs de plus haut niveau, et à considérer des questions humainement plus importantes.

---

28 Pour accéder aux plans de construction, voir https://pixees.fr/jouer-au-jeu-des-allumettes-contre-une-machine.

*Une version préliminaire de ce texte a été publiée par les auteurs dans Lecture Jeune en 2021.*

## Ressources complémentaires

Trois projets français utilisant l'IA pour l'apprentissage (Kwyk, LeLivreScolaire et Pix).
www.kwyk.fr
www.lelivrescolaire.fr
https://pix.fr

Le projet KidLearn mené par l'équipe Inria Flowers.
https://flowers.inria.fr/research/kidlearn/

L'outil pédagogique d'informatique débranchée pour apprendre l'apprentissage par renforcement.
https://pixees.fr/jouer-au-jeu-des-allumettes-contre-une-machine.

# Références

# 7 /// LE DISPOSITIF 5J5IA, UN EXEMPLE DE RÉGULATION CRITIQUE DE L'IA EN EDUCATION

Jean-François Céci[1]
Laurent Heiser[2]
Margarida Romero[2]

[1] Techné, Université de Poitiers, France
[2] Laboratoire d'innovation pour le numérique en éducation, Université Côte d'Azur, France

L'IA modifie notre manière d'être au monde et notre rapport à la culture. Une éducation à l'IA, avec pour objectif de contribuer au développement des compétences chez le citoyen numérique, remet en question la forme scolaire traditionnelle de l'enseignement collectif et simultané. Les pédagogies dites actives, de la découverte, du débat et de la mise en action cognitive sur le savoir à acquérir, seront convoquées. Le concept de régulation pédagogique de l'IA renvoie à une réflexion large, autant sur les activités de classe que sur la formation des enseignants, mais également la représentation de leur rôle au sein de l'institution et de la société à l'ère du numérique.

Le livret pédagogique 5J5IA (5 jours, 5 IA) présenté dans le chapitre introduit cinq activités en libre accès utilisables dès le primaire, avec pour objectif principal la démystification de l'IA. Les enseignants sont invités à tester des séquences pédagogiques prêtes à l'emploi autour de la reconnaissance vocale, la classification des données, l'apprentissage supervisé (ou non), l'utilisation d'un agent conversationnel et les enjeux éthiques de l'IA. Enfin, en répondant aux questions à même les activités, les élèves participent à faire avancer la recherche sur la question de l'IA pour l'éducation.

## L'École à l'ère du numérique et de l'IA

Dans ce chapitre, nous cherchons à caractériser l'évolution du numérique[29] à l'École[30] et à analyser sa capacité à préparer les futurs citoyens à vivre dans un monde en pleine mutation sociale, écologique et technologique. Dès lors, face à l'émergence de l'automatisation et de l'intelligence artificielle (IA), nous rencontrons de nouveaux défis éducatifs, tant d'acculturation que de compétences numériques. Cela rend nécessaire le développement de la capacité d'agir pour permettre aux élèves de devenir des citoyens capables d'un positionnement critique, émancipateur et créatif par rapport à ces technologies (Alexandre et al., 2021). De plus, l'École s'adresse à un public qui représente, sans être une catégorie à part (Cerisier, 2011), une génération d'individus qui ont un accès accru au numérique. L'usage pervasif et intensif des écrans de certains jeunes peut changer leur rapport aux savoirs et leur construction identitaire, conduisant à « une nouvelle manière d'être au monde de l'individu scolarisé » (Céci, 2020, p. 407). Dans le registre de la vie courante, apprendre avec et au travers des écrans est à la fois naturel, habituel et intensif puisque ces jeunes passent environ deux fois plus de temps[31] sur écrans qu'à l'École. Ces apprentissages informels (Hamadache, 1993) participent fortement de leur construction identitaire et, pour une part, également aux apprentissages scolaires. Pour autant, il incombe à l'École[32], en plus de la transmission des savoirs savants indispensables, de contribuer à l'émancipation de ces futurs citoyens, en tenant compte de l'écosystème numérique dans lequel nous évoluons.

---

[29] Le substantif « le numérique » est utilisé en référence aux propos de Louise Merzeau (2017) voulant que « c'est dans sa dimension "écologique" qu'il convient aujourd'hui de penser le numérique, c'est-à-dire en tant qu'écosystème ou environnement, [et] c'est dans ses effets d'interactions, de continuum et d'enveloppement qu'on mesurera le mieux comment ce qui n'était d'abord perçu que comme une "nouvelle technologie" a finalement configuré un milieu de vie » (p. 3). « Le numérique » renvoie donc aux concepts de culture numérique et de citoyenneté numérique pour décrire cette vision écosystémique du numérique. Il inclut aussi le numérique en tant qu'outil et support. Pour plus de détails, voir Céci (2020, p.25).
[30] Le mot École avec un É majuscule est utilisé pour qualifier l'école de la République dans son sens le plus large, tous niveaux confondus (et non pas uniquement les niveaux de la maternelle à CM2).

Dans nos recherches, nous qualifions dorénavant ces jeunes de « citoyens numériques »[33] en considérant qu'ils ont besoin de développer de nouvelles habiletés pour vivre en harmonie dans un monde numérisé et percuté par des enjeux sociétaux. Dès lors, la pédagogie doit cibler la capacité des élèves à ne pas subir ce monde et à y participer de manière agentive, c'est-à-dire, en développant à l'École une représentation forte (Pellaud et Eastes, 2020) des enjeux de développement durable. Il s'agit donc d'offrir des occasions de s'acculturer à l'IA et d'en faire un usage créatif en vue de la résolution de problèmes connectés au vécu de l'apprenant, et plus tard à la société.

> La formation doit contribuer à l'émancipation, au pouvoir d'action et à la créativité des élèves, en tant que futurs citoyens capables d'agir dans un monde numérique où les technologies de l'IA sont omniprésentes.

Depuis les États généraux du numérique pour l'éducation[34], nous assistons à un glissement conceptuel du numérique éducatif vers celui du numérique pour l'éducation. Dans le but de développer les compétences numériques des élèves, certains référentiels visent à proposer des repères pour que les enseignants évaluent ces compétences, comme c'est le cas du Cadre de Référence des Compétences Numériques (CRCN) paru en octobre 2019. Des actions concrètes d'intégration de la culture numérique, du cycle 2 au lycée, peuvent être interprétées comme une volonté ministérielle de former le citoyen numérique dans le cadre d'une politique et d'une culture nationale, volontarisme entrant en résonance avec des enseignements existants au collège et au lycée (p. ex. numérique et sciences informatiques). De plus, l'éducation nationale s'intéresse aux usages personnels du numérique à travers des cadrages ou incitations comme le CARMO[35] et le Règlement général sur la protection des données. Enfin, des programmes de recherche, comme les incubateurs numériques (Heiser et al., 2022), rapprochent les enseignants des chercheurs pour former des praticiens à des usages réflexifs du numérique en contexte éducatif. De nombreux autres exemples peuvent être cités.

---

31 La moyenne est de 2 160 heures/an passées sur écrans, tous niveaux confondus, ce qui correspond à 90 jours (le quart de leur vie). Une année scolaire moyenne (36 semaines de 32 heures) représente 1 152 heures/an (Céci, 2020, p. 231).
32 Nous considérons que l'École n'est pas la seule structure capable de travailler à l'émancipation du citoyen numérique de demain, mais la seule en capacité d'adresser – en volume – cette nouvelle jeunesse. Rappelons « qu'un quart du pays est à l'École de la république » (Céci, 2020, p. 68), ou dit autrement, un quart de la population française étudie ou travaille dans le système éducatif français.
33 Nous adoptons la définition de la citoyenneté numérique du Conseil de l'Europe, voir www.coe.int/fr/web/digital-citizenship-education/home.
34 Ces États généraux ont été organisés fin 2020 par le ministère de l'Éducation nationale, de la Jeunesse et des Sports, voir /www.education.gouv.fr/les-etats-generaux-du-numerique-pour-l-education-304117.
35 Il s'agit du Cadre de référence d'accès aux ressources pédagogiques via un équipement mobile, voir https://eduscol.education.fr/1087/cadre-de-reference-carmo-version-30.

Le cadrage de cette éducation au numérique, et donc sa régulation, se développe tant en France que dans les autres pays de l'OCDE (2021), mais il n'en demeure pas moins que la forme scolaire (Vincent, 2021) et universitaire française demeure peu malléable face aux changements potentiels que le numérique peut soutenir en termes d'innovation pédagogique (Céci, 2020, p. 412). Cette stabilité nous conduit à envisager une analyse des contradictions à travers le concept de régulation critique du numérique en éducation et de l'IA, plus particulièrement.

# Fondements théoriques du concept de régulation critique du numérique et de l'IA en éducation

Le Tableau 4 illustre notre vision multiscalaire d'une formation du citoyen numérique, à la genèse d'une pédagogie que nous qualifierons dorénavant de « contemporaine », incluant autant les médiations nécessaires pour une meilleure régulation des sociotechniques (et de l'IA), que les formes pédagogiques supportant ces médiations et menant à une encapacitation raisonnée des apprenants. Notre démarche permettant d'amplifier la régulation de l'IA par la formation est une vision théorique se déclinant autour des trois niveaux de la pensée complexe (Morin, 1994) : macro, méso et micro.

Au premier niveau, macro, nous empruntons la pensée des philosophes de la technique, comme Simondon (1958) ou Ellul (2004). Nous retenons l'idée selon laquelle le progrès technique n'a pas forcément comme conséquence le progrès social ou environnemental. Comment alors réguler en contexte éducatif, pour que cette dichotomie soit moins forte ? Comment concevoir des médiations qui puissent atténuer cette contradiction ? Autant de questions que nous cherchons à résoudre en proposant trois étapes pour résoudre des problèmes de manière créative : la découverte des dispositifs numériques en classe, leur apprentissage et leur appropriation. Cette triade permet de comprendre les enjeux techniques et éthiques de l'IA (Romero et al., 2021), ainsi que ses apports et ses potentielles dérives.

| | ÉTAPES | | |
|---|---|---|---|
| | Découverte des dispositifs numériques en classe | Apprentissage et appropriation | Agentivité des citoyens |
| Niveau micro | Activités de classe spécifiques (reposant sur l'ingénierie technopédagogique et disciplinaire de l'enseignant) | | Encapacitation individuelle et culture numérique partagée |
| Niveau méso | Pédagogie contemporaine (exemple de la techno-créativité pouvant entraîner des tensions avec la forme scolaire actuelle) | | État d'esprit entrepreneur (de sa vie), ou « état d'esprit de développement » (Dweck, 2010, p. 13) |
| Niveau macro | Capacitation citoyenne | Réflexivité sociétale | Formation du citoyen aux compétences numérique (Céci, 2019) |

Tableau 4. Vision multiscalaire de la formation du citoyen aux compétences numériques

Au niveau méso, nous prenons appui sur la pensée de Charlot (2020) selon qui les pédagogies cherchent à résoudre une tension entre désirs et normes. Le XIXe siècle a donné lieu à l'émergence d'une pédagogie traditionnelle essentialiste, reposant sur la discipline. Au XXe siècle, il s'agissait de respecter le statut de l'enfant, orientation donnant lieu à des « pédagogies nouvelles ». Le XXIe siècle, quant à lui, met l'homme face à des technologies pervasives (ou cybertechniques). Cela nous conduit à déduire que ce rapport à la technique doit être enrichi pour accompagner au mieux les jeunes à se construire et à apprendre dans une société connectée. Il est urgent selon nous de participer à l'émergence d'une pédagogie contemporaine à l'ère numérique, alors même que cette émergence se heurte à une inertie importante de la forme scolaire (Vincent, 2021).

Au niveau micro, nous nous attardons sur le travail effectif du professeur dans la classe, entre injonctions et contradictions. Selon Dubet (2002), les enseignants se sont longtemps engagés dans le métier par vocation (appétence pour le travail sur autrui, pour l'accompagnement à grandir intellectuellement). La vision d'un « citoyen numérique de demain » (Céci, 2019, p. 80) appelle à penser le métier d'enseignant également en termes d'ingénierie technopédagogique, pour réfléchir et vivre, à l'Ecole, des situations qui feront sens dans le quotidien du citoyen connecté. Il s'agit donc d'anticiper, concevoir, mettre en œuvre et réguler des médiations pour augmenter les encapacitations (Bernard, 2018) de ce citoyen numérique au sein des dispositifs idoines. Dès lors, il s'agit davantage de scolariser la société selon ses besoins, que de socialiser l'École (Céci, 2020, p. 417) et le professeur-ingénieur technopédagogique devient alors un élément fondamental du développement des compétences de ce citoyen numérique.

> La formation des citoyens aux compétences numériques doit s'envisager tant au niveau micro avec des activités spécifiques en classe, au niveau méso sur une pédagogie techno-créative et au niveau macro sur une encapacitation citoyenne aux compétences numériques.

La Figure 17 illustre ces propos. La régulation de l'IA par le travail en classe donne lieu à un modèle augmenté qui permet, à gauche de la figure, de réfléchir à son impact (agentivité des citoyens) et de mettre en relief les problèmes posés par l'IA.

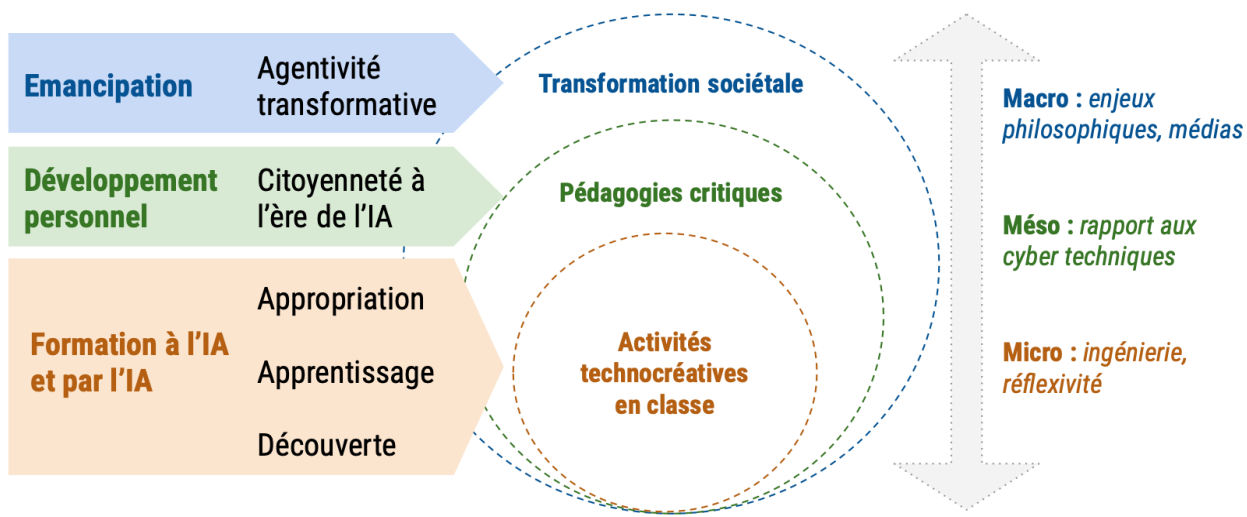

**Figure 17. Les effets de la régulation pédagogique de l'IA**

De ces trois niveaux d'analyse découle une transition du numérique éducatif de ce début de XXIᵉ siècle, essentiellement basée sur une instrumentation de la pédagogie, vers un numérique pour l'éducation avec cette finalité d'émancipation et de formation de ce citoyen numérique, transition que nous allons approfondir, en prenant appui sur l'histoire des pédagogies.

# Une régulation qui ne va pas de soi

Pour comprendre la nécessité de réguler l'IA en contexte éducatif, nous proposons de regarder dans le rétroviseur de l'histoire des pédagogies. Il est alors nécessaire de rappeler l'existence d'une tension permanente entre désirs et normes au sein de la pédagogie (Charlot, 2020). Au XIXᵉ siècle, les désirs de l'enfant doivent être refoulés au sein de l'école pour l'élever à des normes quasi spirituelles. Ces dernières, en outre, s'imposent par la discipline, ce que les normes du citoyen idéal permettent de légitimer. Au XXᵉ siècle, les désirs de l'enfant sont appréhendés comme naturels. Les psychologues de l'éducation demandent que l'on protège l'enfant et que l'on respecte son évolution. La pédagogie dite transmissive se transforme en « pédagogies nouvelles » avec un attrait important pour le constructivisme de Piaget. Les enseignants, dans ce contexte, vont se mettre en quête d'une certaine efficacité dans leur enseignement, à la faveur du meilleur apprentissage possible des élèves.

Précisons qu'il ne s'agit donc pas de questionner la pertinence des normes tant que ces dernières apparaissent comme compatibles avec la conception de la citoyenneté au cours des deux siècles précédents. Ainsi, ces normes devraient être redéfinies, pour tenir compte des évolutions historico-technologiques qui sont en cours au XXIᵉ siècle. Cela nécessiterait de penser une pédagogie qualifiée de contemporaine, à partir de normes qui définissent clairement les compétences numériques que les élèves vont devoir acquérir dans un monde où le numérique, et l'IA, ont fait émerger une nouvelle culture (Doueihi, 2011, p. 9). Parmi ces compétences, il s'agit d'entraîner les élèves à valider ou encadrer une prédiction formulée par une IA, ce qui nécessite de comprendre le fonctionnement de l'apprentissage automatique tout en ayant le choix de se soumettre ou non aux résultats qui en découlent.

Dès lors, une ingénierie pédagogique mettant en lien les référentiels des programmes scolaires avec le développement de la pensée algorithmique et plus particulièrement de cette culture numérique citoyenne intégrant l'IA, est nécessaire. Cependant, l'émergence de ces aptitudes fait apparaître des difficultés ou contradictions. Évoquons, par exemple, la nécessité d'accroître les savoirs et les savoir-faire des élèves, pour les préparer à s'approprier l'IA et en tirer bénéfice de manière intentionnelle. Or, les usages du numérique de ce type, en contexte éducatif, sont peu développés sauf dans des disciplines spécifiques comme les sciences de l'ingénieur en lycée. Les autres enseignants reçoivent peu ou pas de formation à cette ingénierie technopédagogique qui permettrait de développer une pédagogie contemporaine en lien avec les nouveaux enjeux liés à notre rapport à la culture du numérique, depuis leur discipline. Il est également probable que certains enseignants ne se retrouvent pas dans ce rôle d'ingénieur pédagogique, ayant choisi leur métier par vocation (Dubet, 2002, p. 104). Ceci explique, selon Charlot (2020), pourquoi une éducation aux enjeux du numérique, et donc une régulation de l'IA dans et par la pédagogie, nécessiterait de redéfinir de manière concomitante le projet de l'École.

Nous venons de préciser plusieurs éléments. D'abord, le concept de numérique pour l'éducation est désormais appréhendé, mais fait face à des contradictions opérationnelles et structurelles, lesquelles découlent du fait que la pédagogie contemporaine s'appuie difficilement sur de nouveaux référents didactiques et anthropologiques. Ensuite, le travail de l'enseignant est encore trop centré sur la transmission des savoirs et pas assez sur le développement personnel de l'élève dans une visée de développement des compétences pour une citoyenneté développant la capacité d'agir. La conception pédagogique, susceptible de provoquer une bonne régulation des sociotechniques, doit être moins pensée en termes d'efficacité des apprentissages (Charlot, 2020, p. 301), qu'en ceux d'émancipation (versus aliénation). Une telle émancipation est nécessaire au citoyen pour assurer pleinement sa citoyenneté numérique dans un monde connecté.

> L'enseignant doit mettre en œuvre des dispositifs pédagogiques visant autant des apprentissages sociotechniques que l'émancipation citoyenne.

Pour illustrer comment peut se concrétiser ce rôle d'ingénierie pédagogique par l'enseignant, la section suivante présente l'exemple du dispositif pédagogique appelé 5J5IA.

# L'exemple du dispositif pédagogique 5J5IA

Le dispositif 5J5IA (pour 5 jours, 5 IA)[36] est un livret pédagogique prêt à l'emploi, incluant une enquête scientifique. Ce dispositif pédagogique vise une régulation de l'IA par la pédagogie de la découverte, par l'apprentissage et l'appropriation. La philosophie de 5J5IA repose tout d'abord sur ce qui a permis de réunir les chercheurs et auteurs de cet article. Tous questionnent l'usage des technologies en éducation pour les rendre sociales, en se basant sur l'activité réelle, l'expérience authentique vécue et sur une amplification du dispositif pédagogique par usage efficient du numérique en éducation. Nous avons donc élaboré un projet basé sur des activités concrètes et réalistes de l'IA. Par ailleurs, le livret pédagogique explicite l'articulation voulue, entre notre cadre théorique ci-dessus, la recherche et le projet pédagogique support, dont nous allons évoquer, ci-dessous, les grands principes.

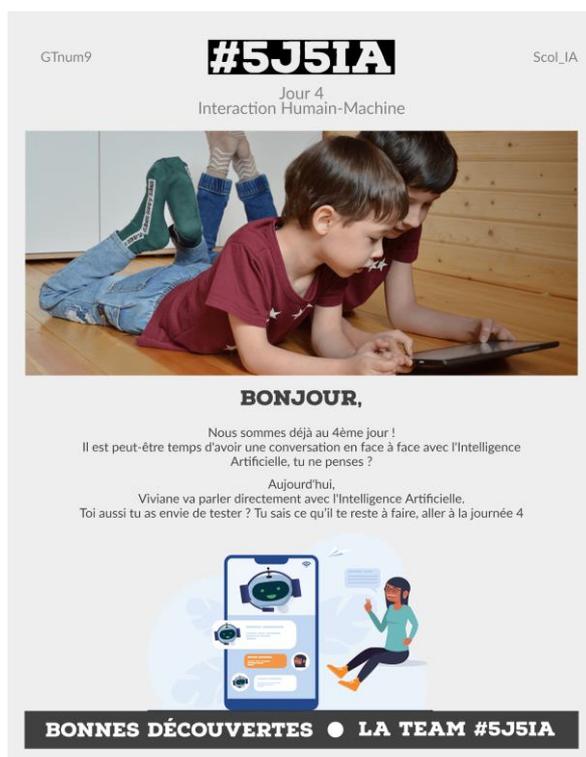

**Figure 18. Le livret pédagogique 5J5IA**

---

## DESCRIPTION DES ACTIVITES DE 5J5IA

Le dispositif a été conçu suivant les sept principes suivants d'ingénierie pédagogique :

1. Proposer une activité facile à intégrer en classe, de moins d'une heure, mais facilement extensible
2. Proposer un livret prêt à l'emploi pour faciliter l'intégration du module par tout enseignant
3. Ludifier l'apprentissage par une trame narrative dans laquelle les apprenants peuvent se reconnaître et se sentir concernés ;
4. Proposer des activités aussi authentiques que possible, porteuses de sens et non stéréotypées
5. Être facilement réalisables en classe, ou à la maison, seul ou en groupe
6. Acculturer au numérique en lien avec le CRCN, considérant l'intégration de l'IA au niveau du référentiel DigComp à l'échelle européenne (Vuorikari, Kluzer et Punie, 2022)
7. Proposer une progression pédagogique permettant de percevoir les représentations des jeunes autour de l'IA dans une visée de recherche (modules 1 à 3 basés sur l'expérience et la découverte, modules 4 et 5 apportant davantage d'informations et un cadre théorique plus solide pour ne pas fausser les représentations avant de les avoir collectées)

Cinq activités qui peuvent se développer sur cinq séances ou plus ont été développées. Chacune est structurée de la même manière : elle commence par une mise en situation (via la narration d'une histoire) suivie d'une activité authentique et d'une phase de questionnements pour alimenter la recherche tout en ancrant les savoirs par l'action réflexive. Des ressources et questions à aborder « pour comprendre » sont offertes pour faciliter l'accompagnement théorique du groupe par l'enseignant. Le Tableau 5 présente les cinq activités de 5J5IA.

| ACTIVITE | DESCRIPTION |
|---|---|
| 1. Reconnaissance vocale | L'activité consiste à dicter une histoire pour comprendre les mécanismes de l'IA à l'œuvre pour la reconnaissance de plusieurs voix, sans apprentissage préalable. Cette activité permet également d'aborder des éléments d'orthographe, de grammaire, de ponctuation ou de prononciation en cours de langues par exemple. |
| 2. Reconnaissance d'objets | Cette activité vise à faire apprendre à une machine à reconnaître plusieurs objets et à les distinguer entre eux. Elle peut être facilement bonifiée pour pouvoir aborder des éléments théoriques – en sciences humaines par exemple – autour des traits caractéristiques des êtres humains. |
| 3. Reconnaissance d'images | Cette activité s'appuie sur l'activité précédente de reconnaissance d'objets, mais l'application apprend au fur et à mesure à reconnaître de nouveaux objets. L'activité ressemble à une chasse au trésor où les élèves sont à la recherche d'objets du quotidien, en un temps limité. Ils doivent montrer les objets trouvés devant la webcam de l'appareil numérique. À la fin de l'activité, l'apport théorique sur l'IA commence via une vidéo ouvrant à un débat. |
| 4. Interaction humain-machine | La situation s'inspire du test de Turing. Il s'agit d'avoir une conversation écrite avec un robot conversationnel (*chatbot*), que l'on nomme Eliza (ami virtuel dans notre scénario), pour voir s'il parviendra à duper les personnes dans une discussion sur de multiples sujets. L'optique est d'aborder l'interaction humain-machine, récupérer des données liées aux représentations des élèves sur l'IA et sur les liens qu'on entretient avec les objets techniques numériques. |
| 5. Régulation de l'IA | Cette dernière activité vise le cadrage théorique de l'IA pour une bonne régulation. Il y a apport de connaissances via plusieurs vidéos et activités d'analyse, ainsi qu'un bilan final. |

**Tableau 5. Les cinq activités du dispositif 5J5IA**

### Un dispositif connecté

5J5IA est un dispositif pédagogique conçu pour être réalisé en ligne, autour des activités IA disponibles sur le Web. Il s'agit d'une contrainte à envisager en premier lieu avant de se lancer dans le dispositif (classe connectée : une tablette connectée pour un à trois élèves). Les activités sont réalisées directement dans un questionnaire en ligne (via *Limesurvey*)[37].

Ainsi, nous pouvons collecter les activités-réponses, donc les connaissances et représentations des élèves autour de l'IA, transformant le dispositif pédagogique en outil de recherche. L'analyse des données vise une meilleure régulation de l'IA par la pédagogie, objectif mis à l'épreuve de notre première collecte de données.

### Résultats préliminaires de la recherche en cours

La diffusion de ce dispositif a été réalisée via des réseaux sociaux et auprès d'INSPE partenaires, ainsi que par l'entremise de la Direction du numérique pour l'éducation (auprès de 36 experts nationaux animant un réseau de 450 enseignants), sur la période allant de juin 2021 à janvier 2022. Le dispositif a aussi été présenté lors de colloques et ateliers pédagogiques, comme le colloque Didapro 2022.

La collecte de données est en cours depuis l'année scolaire 2021-2022, en trois phases successives : (1) déclaration d'intention d'usage par les enseignants, (2) résultats sur *Limesurvey* une fois les activités réalisées, (3) retour d'expériences des enseignants.

En ce qui concerne la déclaration d'intention d'usage par les enseignants (phase 1), nous avons récolté 78 réponses partielles, dont trois intentions d'intégration au second trimestre en classes de CM2, 3e et 2d, ainsi que sept intentions d'intégration au 3e trimestre. Huit enseignants (5 femmes, 3 hommes) ont accepté de nous donner leurs coordonnées pour un témoignage après expérimentation (2xCE2, 1xCM1, 2xCM2, 1x6e, 1x4e, 1x3e). Ces témoignages, portant sur l'expérience d'apprentissage d'élèves de 7 à 15 ans, peuvent constituer une bonne base de départ.

En dehors de l'acculturation à l'IA, les déclarations d'intentions évoquent un usage disciplinaire prévisionnel essentiellement centré sur les langues : vocabulaire, prononciation, ponctuation, intonation, « fluence de lecture » (un enseignant de CM2), et pour « alimenter les débats en enseignement moral et civique sur la place de l'IA dans notre quotidien et plus généralement sur l'impact des technologies » (un enseignant de CM1).

Pour l'heure, aucune intégration complète n'a pu être constatée et produire les résultats selon le scénario prévu (phase 2), via les questionnaires sur *Limesurvey*. Enfin, le témoignage d'un enseignant (phase 3) portant sur la réalisation complète de l'activité n°1 (de dictée vocale) en double classe de CE1-CE2, nous laisse espérer des résultats prometteurs lorsque l'adoption du dispositif sera plus franche. Au rang des difficultés rencontrées, la gestion technique des tablettes (charge, configuration, emprunt) et des difficultés pédagogiques sont relevées, comme la gestion du bruit généré par l'activité et les difficultés d'articulation de certains élèves rendant l'activité de dictée difficile (mais intéressante en termes de plus-value).

L'enseignant mentionne que « l'intérêt des élèves est très fort pour l'activité [de dictée vocale sur tablettes], [captivant les élèves] pendant une demi-heure facilement ». L'enseignant évoque une activité agréable et facile à scénariser en toute ou partie (autonomie) de classe. L'enrichissement en matière de culture numérique est avéré, les « discussions de début d'activité faisaient ressortir le caractère très intelligent et presque sans défaut de l'informatique » alors que « la mise en commun après l'activité était plus nuancée et montrait une réflexion sur les limites de la technologie ». Pour autant, « la plupart des élèves considèrent la tablette et ses fonctions comme un tout, et n'ont pas fait de différence entre l'IA et le matériel qu'ils utilisaient ». Nous constatons un début prometteur d'acculturation aux cybertechniques, certes insuffisant, mais qui serait normalement étoffé par les quatre autres activités, si le dispositif était réalisé dans son entièreté. Finalement, l'enseignant évoque une expérience très positive, engageant une suite, potentiellement dans d'autres classes car « on peut clairement adapter ces activités pour tous niveaux ! ».

---

37 Pour un exemple d'activités, voir https://enquetes.univ-cotedazur.fr/index.php/98944?lang=fr.

Quant aux réponses des élèves (n=17) au questionnaire d'activité, elles montrent que la régulation de l'IA en contexte éducatif (que nous proposons ci-dessus) est nécessaire, tant les représentations sont erronées en matière de TIC et d'IA (à l'âge de 8 ans ici). Les résultats précédents laissent à penser que la réalisation intégrale du dispositif permettrait de corriger (au moins en partie) cela.

# Une responsabilité à hauteur des enjeux

Une régulation pédagogique de l'IA visant l'émancipation du citoyen numérique doit reposer sur un modèle d'intégration raisonnée des usages de l'IA en éducation. Au niveau micro, cette régulation ne peut aller de soi, car il faut envisager trois phases chronophages : la découverte, l'apprentissage et l'appropriation. Quant aux niveaux méso et macro, la régulation pédagogique de l'IA nous invite à penser notre humanisme avec des référents anthropologiques (Charlot, 2020) qui servent à enrichir notre rapport aux objets (Doueihi, 2011). Intrinsèquement l'humain qui œuvre par et avec l'IA doit intégrer des dimensions éthiques avec le souci de développer des liens sociaux, des communautés, des règles et codes sociaux qui assurent son bien-être, l'acquisition de valeurs humaines, sociales, solidaires, fraternelles dans un monde en pleine mutation. En ce sens, découvrir, apprendre et surtout s'approprier l'IA en contexte éducatif, comporte une dimension émancipatrice qui doit favoriser le pouvoir d'agir de chaque personne et non son aliénation numérique.

La régulation pédagogique de l'IA nécessite donc une vision écosystémique des usages de l'IA (et du numérique) en éducation pour tirer le potentiel de la technique sans abandonner le pouvoir de décisions aux machines, sans succomber au « poison » du *pharmakon* numérique. L'IA représente un immense potentiel pour l'humanité, mais entraîne de grandes responsabilités pour la maîtriser. Là semble être l'enjeu du numérique pour l'éducation à l'ère de l'IA.

## Ressources complémentaires

Les activités et le livret pédagogique sont disponibles en ligne. https://scoliablog.wordpress.com/5j5ia/.

# Références

# 8 /// L'ACCULTURATION AU NUMERIQUE

Michel Durampart[1]
Philippe Bonfils[1]
Anne Gagnebien[1]
Audrey Bonjour[1]
Pauline Reboul[1]
Sami Ben Amor[1]
Laurent Heiser[1,2]

[1] Institut méditerranéen des sciences de l'information et de la communication, Université de Toulon, France
[2] Laboratoire d'innovation pour le numérique en éducation, Université Côte d'Azur, France

Les usages des technologies éducatives, dont l'intelligence artificielle (IA) fait partie, demandent une réadaptation et un encadrement pédagogique. Les représentations et la culture des acteurs, enseignants et élèves, peuvent percuter la forme scolaire et sont des points à prendre en compte dans l'intégration de l'IA.

Ce chapitre poursuit deux objectifs : d'une part, définir l'acculturation à l'IA en tant que vaste mouvement d'appropriation des outils numériques et, d'autre part, à la lumière des travaux de l'Institut méditerranéen des sciences de l'information et de la communication (IMSIC), analyser différents usages de l'IA permettant de mettre en perspective l'acculturation à l'IA en éducation.

## La notion d'acculturation au numérique

Les sciences de l'information et de la communication ont contribué à mettre en évidence que la relation entre numérique et éducation est complexe (Moeglin, 2005), itérative et traversée d'enjeux multiples (Durampart, 2018). Dans ce contexte, les chercheurs du laboratoire IMSIC, à travers de nombreux programmes de recherche et publications (Collet et al. 2014 ; Barbagelatta et al. 2014 ; Dechamp et al. 2018 ; Bonfils et Peraya, 2016 ; Bonfils, 2015, Bonfils et Durampart, 2013), ont observé les pratiques des enseignants et des élèves au sein des classes ou encore au sein d'un Incubateur (2016-2018). Ils ont pu constater que les artefacts numériques destinés aux activités d'apprentissage (Bernard et al., 2018) tendent à provoquer des tensions et des apories au sein de la forme scolaire (Vincent, 1994).

Au fil de près de douze années de travaux conduits au sein du laboratoire, la notion d'acculturation nous a semblé pertinente pour qualifier ce vaste mouvement d'appropriation des outils numériques venant percuter la pédagogie et les pratiques éducatives. Cette notion est venue se former comme une heuristique féconde afin de qualifier un mouvement instable, discontinu et hétérogène qui couvre le champ des pratiques où le numérique agit dans le contexte et dans les pratiques éducatives. C'est en ce sens que les chercheurs mobilisés sur ces enjeux et questions proposent de réfléchir en termes « d'acculturation au numérique » et non de « culture numérique » (Durampart, 2016). L'enjeu principal est d'intégrer cette supposée culture numérique dans une perspective de bien-être économique et social, l'inclusion numérique désignant plus largement « la capacité de chacun à savoir mobiliser ces technologies, compétence fondamentale à l'heure du poids croissant de l'économie de la connaissance » (Ministère de l'Économie Numérique, 2013). S'il est impossible de restituer la diversité des terrains et projets dans lesquels nous nous sommes impliqués, il est possible de retracer quelques étapes et jalons clefs qui nous ont permis de mieux cerner cette notion d'acculturation au numérique.

> La notion d'acculturation permet de qualifier le vaste mouvement d'appropriation des outils numériques venant percuter la pédagogie et les pratiques éducatives.

L'élément déclencheur des projets de recherche sur l'acculturation numérique de l'IMSIC réside dans une étude réalisée dans le cadre de l'observatoire des TIC de la région PACA (programme OBTIC, voir Pélissier et al. 2013). Selon cette étude, malgré un bon niveau de connaissances du numérique et un usage intensif des réseaux sociaux dans leurs pratiques de communication, seule une minorité d'élèves exploite sa « culture numérique » dans une perspective d'insertion et d'orientation professionnelle. Même si un apprentissage formel est indispensable pour atténuer la diversité du degré de maîtrise des acteurs sociaux, encore faut-il qu'ils soient capables de mobiliser ces nouvelles pratiques de communication et d'information dans des contextes autres que ceux liés aux loisirs et aux relations interpersonnelles. L'enjeu de l'acculturation au numérique renverrait donc à une forme de capital économique, social et culturel instable, hétérogène et disséminé. Elle est remise sans cesse en jeu et en question, avec de multiples modes différents d'appropriation, dans des contextes divers, renvoyant aussi à des formes successives ou alternées d'apprentissage empilée, qu'elles soient formelles ou informelles.

## Les trois niveaux d'acculturation numérique

Le concept d'acculturation au numérique que nous avons développé se décline en trois niveaux présentés dans les paragraphes qui suivent : l'acquisition de savoirs, l'interaction entre les usages personnels et professionnels, et l'utilisation dans les pratiques pédagogiques.

PREMIER NIVEAU D'ACCULTURATION NUMÉRIQUE : L'ACQUISITION DES SAVOIRS PAR LES ENSEIGNANTS

Le premier niveau d'acculturation au numérique concerne l'acquisition de savoirs par la formation. Parmi les enseignants âgés entre 31 et 50 ans (ce qui représente 77 % de nos répondants[38]), 82 % ont suivi une formation initiale dans les IUFM, avant les INSPE. Ce sont 73 % d'entre eux qui disent utiliser souvent le numérique, tout en déplorant dans l'ensemble une formation insuffisante (68 %). Ils utilisent plus souvent les TIC comme outil personnel plutôt que comme outil d'échange ou de partage. Les apprentissages initiaux (p. ex. essais, bricolage et autoapprentissage), ou partagés (en situation de co-apprentissage et de multi-apprentissage), se déroulant chez soi, dans l'environnement domestique, avec des orientations ludiques ou influençant des besoins et attentes diverses en quête de facilitation de la vie quotidienne, impactent les usages en situation professionnelle, entre difficultés à les

---

retraduire et certaines formes d'aisance. Le manque de formation dans le contexte professionnel rend difficile le développement d'un niveau de maîtrise avancé.

> Au premier niveau d'acculturation numérique, les enseignants utilisent le numérique comme outil personnel plutôt que comme outil d'échange ou de partage.

## SECOND NIVEAU D'ACCULTURATION NUMERIQUE : DES LIENS ENTRE LES USAGES PERSONNELS ET PROFESSIONNELS

Au second niveau d'acculturation numérique, l'intégration du numérique par les enseignants vise à favoriser la relation entre l'école et la maison et met une diversité de supports au service de l'activité pédagogique. Ce niveau se situe dans un lien possible et contrarié entre pratiques et utilisations, apprentissages privés, personnels, domestiques et ceux du monde professionnel ou scolaire. Les projets dits innovants liés au numérique (Incubateur, 2018-2019) s'inscrivent bien dans une volonté globale d'exploiter les supports numériques pour favoriser la continuité entre l'école et la maison.

Pour autant, la transposition des usages du numérique de la sphère privée vers la sphère de l'établissement scolaire demande une réadaptation et un encadrement pédagogique. « Les tablettes rentrent dans l'établissement parce qu'elles sont rentrées dans nos vies », nous dit une enseignante. Elle insiste aussi sur le fait qu'il faut adopter une logique de recherche d'information pour que la tablette soit pédagogique (par le recours à une intense recherche documentaire). C'est ce que nous avons désigné comme un phénomène de porosité, constaté à la place prise par le numérique dans le monde scolaire qui met directement en jeu la « forme scolaire » (Vincent, 1994).

## TROISIEME NIVEAU D'ACCULTURATION NUMERIQUE : LES DISPOSITIFS NUMERIQUES DANS LES PRATIQUES PEDAGOGIQUES

Finalement, le troisième niveau concerne le recours aux dispositifs numériques dans le cadre même des pratiques pédagogiques. Nous avons réalisé un vaste programme d'observations centré sur les apprentissages fondamentaux du numérique entre 2014-2016. Ce programme, nommé Numécole, a permis de réaliser des observations dans des lycées et collèges dits défavorisés de Toulon, des journées d'études avec du personnel enseignant et des projets d'incubateurs numériques. Les questionnements qui émergent de ces observations résident dans le transfert de l'étude des usages vers un regard sur les interfaces. Les enseignants expérimentateurs ayant participé à Numécole déclarent principalement des usages professionnels du numérique liés : au traitement de texte (95 % dont 90 % pour le français), aux documents audiovisuels (86 % dont environ 46 % pour l'histoire, la géographie, les sciences et les arts), aux navigateurs Web (72 % dont 46 % pour la recherche documentaire), à des logiciels ou documents créés par l'enseignant (69 % principalement pour le français et un peu moins pour les mathématiques), à des sites Web éducatifs (62 %, en français, mathématiques, sciences et technologies), à des jeux et applications (autour de 63 %, principalement en mathématiques et français). Les usages professionnels liés à la publication assistée par ordinateur sont moins développés (49 %), de même que ceux liés aux encyclopédies numériques (41 %), aux ressources institutionnelles en ligne (37 %), aux tableurs (18 %), logiciels de PAO (Photoshop, Publisher…) (49 %) et aux espaces numériques de travail (43 %).

Le troisième niveau d'acculturation numérique concerne l'utilisation des dispositifs numériques dans le cadre des pratiques pédagogiques.

Ensuite vient la question de l'orientation et des motivations qui fondent la relation entre dispositifs numériques et pédagogie. De manière générale, dans l'étude Numécole, au-delà de la question du numérique à l'école, les approches pédagogiques des enseignants sont plutôt disciplinaires (75 %) bien qu'ils déclarent aborder des compétences transversales méthodologiques dès que possible (74 %). L'enjeu principal de leur enseignement est l'autonomie des élèves (77 %). Le travail individuel et différencié domine largement (77 % et 56 %) et l'élève préféré est l'élève autonome (82 %). Ils disposent le plus souvent (53 %) de leur classe en alignant les tables face au tableau. Dans ce cadre, les enseignants expérimentateurs déclarent avoir pour principaux objectifs de l'utilisation du numérique en classe, par ordre de fréquence décroissante : la différenciation pédagogique, l'individualisation des apprentissages, la motivation, l'aspect ludique, la variation des supports (couleurs), le la préparation au brevet informatique et internet, l'éducation aux médias et à l'information.

Nous proposons d'aborder un positionnement critique dans l'étude des transformations induites par les nouveaux médias éducatifs et les médiations liées aux technologies numériques. Sous cet angle, l'acculturation numérique induit notre volonté et capacité à analyser les possibles transformations affectives, psychocognitives et pragmatiques (Collet et al., 2021) des enseignants et des élèves face à un phénomène de mécanisation (Mœglin, 1993) et de rationalisation des apprentissages que pourrait relancer l'ère de l'IA. Différentes approches ont été mises de l'avant par les enseignants de Numécole quant à la façon d'intégrer les technologies de l'information et de la communication en enseignement. Parmi celles-ci, nous trouvons : l'approche didactique, la transmission des connaissances, l'approche cognitive (80 % des répondants l'ont classée en premier ou deuxième), l'école comme développement de l'intelligence, l'approche citoyenne (l'école comme lieu de socialisation), l'approche culturelle (l'école comme lieu d'intégration dans une culture) (elles arrivent juste après dans le primaire avec au collège l'approche culturelle qui est placée en deuxième), l'approche professionnelle (l'école comme lieu de préparation à l'insertion professionnelle, largement classée en dernier, au primaire comme au collège).

## Que dit le concept d'acculturation numérique sur l'intégration des technologies à l'école

À la lumière de nos résultats, nous envisageons le concept d'acculturation comme un cadre permettant d'étudier l'intégration des technologies numériques dans l'école. Depuis près de 15 ans, des programmes innovants sont inscrits au sein de démarches éducatives ou dynamiques de projets, accompagnées par l'institution mais aussi présentes dans les pratiques enseignantes. Les élèves sont pris dans des injonctions et volontés contradictoires tissées d'émancipation, d'autonomie, d'individuation, de participation, d'engagement, de soutien et d'aide voire de réinsertion et d'accompagnement perçues comme des dynamiques innovantes, stimulantes offertes par des outils et technologies numériques.

Dans le programme Incubateur, nous constatons que, pour les élèves, les activités recourant au numérique à l'école sont le prolongement du projet hors du cadre scolaire (ils continuent à travailler en groupe en dehors des heures de cours) avec l'idée d'acquérir des compétences transversales. Un degré plus important d'autonomie, essentiel pour l'apprentissage de connaissances disciplinaires, est le fait de savoir agir en équipe. De façon assez constante dans tous les programmes, les enseignants insistent sur l'apport des outils numériques dans la pédagogie pour la remise en cause des routines de comportement, des habitudes de travail, une aide au décloisonnement en dépit d'une chronophagie technologique.

*In fine*, l'acculturation au numérique est aussi une question d'écarts, autant pour les enseignants que les apprenants, entre l'utilisation et la maîtrise du numérique. Elle est aussi une question d'évolution de la forme scolaire, et plus globalement du monde éducatif, et s'inscrit dans des pratiques socioculturelles liées à l'usage de différents systèmes symboliques ou numériques (Kabuto et Harmey, 2019).

Par le biais d'usages créatifs, l'acculturation numérique, c'est aussi la possibilité de trouver de nouvelles modalités pédagogiques permettant d'engager les élèves dans des activités de collaboration. Il s'agit moins d'utiliser le numérique que de trouver, en commun, comment ce dernier peut permettre de développer une représentation forte des objectifs de développement durable (comme l'Agenda 2030). En ce sens, les CurriQvidéos, un dispositif inventé par Faller et Heiser (2022) à l'INSPÉ de Nice, donne à voir plusieurs cours d'action de primo-enseignants qui racontent comment le numérique (des robots pédagogiques, des microcontrôleurs, etc.) et aujourd'hui l'IA (à travers, par exemple, *Google Teachable Machine*, VittaScience ou encore 5J5IA), permettraient d'engager les élèves dans des activités participatives de préservation du patrimoine naturel et culturel (Heiser et al. 2021). En définitive, se situer dans le cours de l'acculturation, c'est bien s'intéresser au processus non stabilisé d'évolution de l'éducation face, avec et par le numérique, sans espérer que le développement de la citoyenneté numérique s'opère sans intervention spécifique, mais en l'accompagnant pédagogiquement.

# Perspectives pour une acculturation face au numérique au regard des perspectives de l'IA dans les apprentissages et l'éducation

Le groupe T2 associant des chercheurs de l'IMSIC et d'autres chercheurs du GTnum #Scol_IA se situe dans cette perspective de l'acculturation numérique. Si nous éprouvons un certain nombre de programmes où l'IA est présente sous forme d'aboutissement d'un projet ou d'une réalisation, nous nous situons dans son antichambre en essayant d'interroger comment des recours et des perspectives liées aux enjeux de l'IA vont venir interroger, réorienter ou redéfinir le niveau d'acculturation numérique observé.

À l'origine, un programme Médipath (Collet et al. 2021) hors contexte éducatif nous a permis de mesurer des enjeux en termes de réorientations d'une expertise médicale liée à des gains de temps et à une facilitation de l'expertise. Un programme au sein de l'enseignement supérieur et de l'apprentissage professionnel dans le domaine naval (E-DEAL) envisage les modalités de transfert et d'acculturation aux pratiques numériques industrielles rassemblées autour des concepts d'industrie 4.0 (Julien et Martin, 2018) et de jumeau numérique (Julien et Martin, 2020). La gestion des données et l'intégration de l'IA deviennent des enjeux de formation à part entière conséquemment aux transformations des métiers par l'intégration de ces technologies. Ce projet de recherche questionne également les représentations et la culture des acteurs de l'industrie concernant les potentialités de l'IA et des traces d'apprentissage (*learning analytics*) (Peraya, 2019) en termes de profilage, d'individualisation des parcours de formation, d'autonomisation et d'accentuation des pratiques d'autoapprentissage pour des communautés d'apprenants en formation professionnelle. D'autres éclairages émanant de l'éducation spécialisée montrent que, même, si les technologies numériques cristallisent des tensions inhérentes au champ du travail social (Bonjour, 2011, p. 52) et conduisent, dans les usages, à produire autant d'évidences que de paradoxes (Bonjour et Daragon, 2017), elles permettent aussi d'introduire davantage d'apprentissages dits scolaires (particulièrement en mathématiques et en lecture) quand de nombreux travailleurs sociaux sont démunis pour mener à bien ces objectifs. L'IA offre alors des possibilités de soutien à la remédiation cognitive et comportementale pour les personnes à besoins spécifiques. En conséquence, l'information, la formation et la mutualisation des pratiques nous paraissent être trois défis majeurs pour le travail sur autrui (Dubet, 2002) à l'ère du numérique (Bonjour et Daragon, 2018).

Du côté des ressources dans les plateformes éducatives, les questionnements orientés vers l'évolution des formes organisationnelles nous amènent à interroger les perspectives éducatives de l'IA. Il s'agit de se focaliser sur le préalable que constitue la question de l'acculturation aux données constituées par les traces d'utilisation que les enseignants et les élèves laissent dans ces environnements numériques. Les enjeux de l'IA en éducation semblent s'orienter vers le repérage des dynamiques et contextes d'apprentissage, le profilage, l'identification et la construction stratégique des données, afin d'engager des formes d'apprentissage centrées sur l'autonomie, l'autoformation, les communautés d'apprenants en lien avec les enseignants. Ces environnements instaurent aussi la vision d'une nouvelle performance d'interfaces et de médiations, supposées expérientielles et efficientes, dans une démarche qui relie l'exploitation des artefacts, des données et de nouveaux processus. Nous partons des enjeux d'une traduction de l'IA dans l'acculturation numérique en cours avec des questionnements sur les limites, contraintes et apories dans le système éducatif ou sur des formes d'apprentissage. Il est possible alors d'envisager que l'IA soit une rupture dans une continuité. Son intégration dans l'école ou la formation reprend encore des discours, débats et invocations passées, qui retrouvent une nouvelle vigueur dans les expériences et démarches liées à l'IA et, en même temps, stimule de nouveau enjeux ou tensions au cœur des mondes et systèmes éducatifs.

# Références

# 9 /// PRESERVER L'AGENTIVITE DES ENSEIGNANTS ET ELEVES : DES PISTES ISSUES D'UNE RECENSION DES ECRITS

Alexandre Lepage[1]
Simon Collin[2]

[1] Université de Montréal, Canada
[2] Université du Québec à Montréal, Canada

Automatiser des actions qui étaient autrefois réalisées par des humains peut poser un certain nombre de risques qu'il est toutefois possible de prévenir dès l'étape de conception. En éducation, le travail des enseignants comme celui des élèves est appelé à se transformer avec l'apparition des systèmes d'IA.

Si aucune précaution n'est prise, certaines technologies éducatives basées sur l'IA risquent de limiter la capacité des enseignants à prendre eux-mêmes des décisions pédagogiques, en plus d'introduire des erreurs de classification ou de jugement. Néanmoins, des solutions existent. Ce chapitre présente les résultats partiels d'une recension systématique des écrits sur les enjeux éthiques de l'IA en éducation, en se concentrant sur les questions relatives à la préservation de l'agentivité des enseignants et des élèves. Il est structuré sous forme de rapport de recherche, avec contexte, méthode, résultats et discussion. Des pistes sont proposées pour se prémunir contre les risques recensés.

## Introduction

L'éducation n'échappe pas au développement rapide des techniques d'intelligence artificielle (IA) et à leur capacité à réaliser des actions de plus en plus complexes. Dans le Consensus de Beijing sur l'IA en éducation, l'UNESCO (2019) entrevoit des bénéfices potentiels de l'IA dans plusieurs activités jusqu'à présent réalisées par des élèves, du personnel enseignant ou du personnel administratif. Les occasions d'utiliser l'IA en éducation sont nombreuses, que ce soit pour le tutorat intelligent, l'évaluation des apprentissages ou bien la prévention de l'abandon scolaire (Zawacki-Richter et al., 2019). Ces avancées nous amènent à nous poser de nouvelles questions : et si le personnel enseignant pouvait enfin être déchargé des fastidieuses heures passées à corriger des copies ? Un élève pourra-t-il obtenir une aide en temps réel, à la maison, lorsqu'il stagne sur un difficile problème de mathématique ? Une IA peut-elle faire aussi bien qu'un enseignant ? Ou même mieux ? Ces questions alimentent des craintes non fondées, mais elles demeurent pertinentes pour faire émerger des enjeux éthiques à considérer dans les déploiements futurs de l'IA en éducation.

Le présent chapitre entend réfléchir à ces questions sous l'angle de la préservation de l'agentivité humaine (Engeström et Sannino, 2013), un des grands enjeux de l'IA en éducation aux côtés de la justice sociale, de la complexité humaine ou de la gouvernance, par exemple. Nous présenterons dans l'ordre quelques éléments contextuels dont certaines définitions de l'IA en éducation, deux repères théoriques, soit le système technicien et le concept d'agentivité, et finalement les résultats d'une recension des écrits sur les enjeux éthiques associés à l'agentivité et l'IA en éducation.

## QU'EST-CE QUE L'IA APPLIQUEE A L'EDUCATION ?

L'IA peut d'abord être envisagée comme un ensemble de techniques aux contours plus ou moins définis. Ses techniques les plus communes relèvent de l'apprentissage automatique (*machine learning*), qui peut être supervisé, semi-supervisé ou non supervisé (Taulli, 2019). L'apprentissage profond par réseaux de neurones artificiels peut être utilisé pour traiter les données dites massives, c'est-à-dire caractérisées par la rapidité à laquelle elles se multiplient, leur volume et leur diversité (Taulli, 2019). Humble et Mozelius (2019) l'abordent en soulignant le caractère interdisciplinaire qui dépasse les sciences informatiques : « l'IA en éducation est un domaine interdisciplinaire intégrant la psychologie, la linguistique, les neurosciences, l'éducation, l'anthropologie et la sociologie dans le but de créer des outils puissants pour l'éducation et de mieux comprendre le phénomène d'apprentissage » (p. 1). L'IA peut aussi être définie autrement que par ses méthodes informatiques, par exemple par les fonctions qu'elle occupe dans un système. Loder et Nicholas (2018) présentent l'IA comme des « ordinateurs réalisant des tâches cognitives, généralement associées au raisonnement humain, particulièrement pour l'apprentissage et la résolution de problèmes » (p. 11). Pour Popenici et Kerr (2017), l'IA en éducation est constituée de « systèmes informatiques capables de s'engager dans des processus humains comme l'apprentissage, l'adaptation, la synthèse, l'autocorrection et l'utilisation de données pour des tâches complexes » (p. 4). C'est cette dernière que nous retiendrons, car elle permet de dépasser l'opposition entre l'intelligence humaine et l'IA et d'envisager des interactions complexes entre les deux.

> L'IA est un terme qui peut vouloir dire beaucoup de choses. Appliquée à l'éducation, elle vise à accomplir des tâches complexes qui étaient jusqu'à maintenant uniquement réalisées par des êtres humains (comme de la rétroaction ou l'adaptation du matériel didactique en fonction du niveau de difficulté optimal pour un groupe d'élèves).

## L'IA VIA LE PRISME DU SYSTEME TECHNICIEN

Nous proposons d'envisager ces techniques sous l'angle de la théorie du système technicien de Ellul (1977). Selon cette théorie, les techniques redéfinissent constamment la réalité de l'expérience humaine. Ellul donne l'exemple de la télévision, rendue possible par accumulation de techniques. En la visionnant, les individus finissent par ne plus voir ces techniques. La télévision a rendu possible une nouvelle forme de communication que nous avons fini par intégrer, puis banaliser au point de ne même plus nous intéresser à son fonctionnement. En somme, les techniques qui l'ont rendue possible, comme l'électricité ou les antennes de diffusion, finissent par s'implanter et redéfinir les actions des individus qui utilisent la télévision, ainsi que les rapports sociaux auxquels ils participent plus largement, etc. En appliquant cette théorie à l'IA, on pourrait en venir à se demander si la complexité des techniques d'IA en éducation nous fera oublier la complexité des techniques et des gestes professionnels propres à l'enseignement. Comment les enseignants occuperaient le temps passé à concevoir et articuler des activités pédagogiques si des systèmes sélectionnaient et adaptaient automatiquement des exercices selon le niveau de leurs élèves ? Allons-nous cesser de nous intéresser à la docimologie, science de l'évaluation, parce qu'une IA le fait à notre place ? Contrairement à d'autres technologies éducatives, l'IA a de particulier qu'elle est développée dans le but d'accomplir des tâches de plus en plus complexes, ce qui permettrait à l'enseignant de se concentrer sur les tâches pour lesquelles l'IA fait moins bien (p. ex. les tâches de haute complexité qui exigent une compréhension fine du contexte, comme les relations avec les élèves).

> L'enseignement est composé d'actions complexes que les enseignants connaissent bien. Introduire des outils basés sur l'IA pour accomplir ces actions ne devrait pas nous amener à oublier ce que nous savons de cette complexité et la façon dont nous la gérons actuellement.

## L'AGENTIVITÉ POUR COMPRENDRE ET SITUER L'ACTIVITÉ HUMAINE

La définition que nous avons retenue de l'IA en éducation plus haut, celle de Popenici et Kerr (2017), introduit l'idée que les systèmes informatiques simulant l'intelligence humaine prennent place dans des systèmes humains. Nous envisageons chacun de ces systèmes comme agents l'un de l'autre. En informatique, le terme agent désigne un système avec une certaine part d'autonomie capable de réaliser des actions qui auront un impact sur l'environnement dans lequel il se trouve : « l'action, qui est un concept fondamental pour les systèmes multi-agents, repose sur le fait que les agents accomplissent des actions qui vont modifier l'environnement des agents et donc leurs prises de décision futures » (Ferber, 1995, p. 13). C'est là une des particularités des systèmes complexes dits intelligents : ceux-ci ne se contentent pas de « raisonner » (Ferber, 1995, p. 13), ils agissent et transforment leur environnement.

En sciences sociales, le concept d'agentivité fait aussi référence à une forme d'autonomie, mais cette fois de la part des personnes. Selon Engeström et Sannino (2013), « l'agentivité est une recherche volontaire de transformation de la part du sujet » et « se manifeste dans une situation problématique polymotivée dans laquelle le sujet évalue et interprète les circonstances, prend des décisions selon les interprétations et exécute ces décisions » (p. 7). Par exemple, un enseignant pourrait souhaiter offrir une rétroaction très précise à ses élèves, mais devoir rendre les notes le plus rapidement possible. Dans une telle situation, deux motifs sont en concurrence et c'est en prenant des actions agentives que l'enseignant parviendra à résoudre ce conflit de motifs. Pour l'IA en éducation, il pourrait s'agir de permettre à l'élève ou à l'enseignant, selon le cas, d'avoir une capacité d'action plus importante par l'utilisation de systèmes d'IA. Mais au-delà de ce scénario idéal, à l'heure actuelle, la plupart des usages de l'IA appliqués à l'éducation tendent plutôt à modéliser des décisions pédagogiques en vue d'une aide à la décision, d'une automatisation de certaines décisions ou d'une analyse des décisions prises.

> Les êtres humains sont caractérisés par leur agentivité, c'est-à-dire qu'ils parviennent à prendre des actions et des décisions dans des situations où seulement de l'information incomplète est disponible.

À première vue, les domaines d'applications actuels de l'IA sont susceptibles d'empiéter sur l'agentivité des élèves et du personnel enseignant, notamment lorsqu'il est question d'évaluation et de choix des ressources ou activités pédagogiques. Pour dépasser ces perceptions initiales, nous poserons la question de recherche suivante : quels types d'usages de l'IA en éducation risquent de limiter l'agentivité des enseignants, et des élèves ?

# Méthode : une recension des écrits sur les enjeux éthiques

Ce chapitre s'appuie sur des données rassemblées lors d'un projet de recension systématique des écrits autour des termes « éthique, IA et éducation », dans les bases de données Google Scholar, Web of Science, Microsoft Academic, EBSCO Education, Dimensions.ai et Scopus, et en procédant par sérendipité (Michel et Le Nagard, 2019). Les documents sont des articles scientifiques ou actes de colloque révisés par les pairs. Ils sont rédigés en français ou en anglais, et publiés entre 2010 et 2021 ($N = 58$). Les articles ont été lus puis des segments codés à l'aide du logiciel nVivo par deux personnes. Alors que la recension fera l'objet d'une publication spécifique (en rédaction), ce chapitre propose une analyse spécifique, approfondie et inédite, des enjeux relatifs à la préservation de l'agentivité humaine. Aux fins du présent chapitre, ce sont 24 documents ($n = 24$) qui ont été retenus pour l'analyse, lesquels comprenaient 62 segments codés.

# Résultats en lien avec l'agentivité des enseignants et des élèves

Cette section présente, dans l'ordre, les résultats relatifs à l'agentivité des enseignants puis ceux relatifs à l'agentivité des élèves. Elle vise à rapporter le plus fidèlement possible, de façon objective et sans interprétation secondaire, les idées véhiculées dans la littérature. La discussion qui suivra permettra d'apporter certaines nuances.

Sans préciser le type d'outils d'IA dont il est question, plusieurs des documents consultés considèrent que ceux-ci risquent de diminuer l'agentivité des enseignants. Le développement de systèmes informatiques complexes fait passer une partie de leur pouvoir décisionnel vers des équipes de conception de logiciels :

> Intégrer des systèmes d'IA en éducation pourrait exacerber un déséquilibre des pouvoirs et créer de nouvelles inégalités. Les systèmes d'IA peuvent faire pivoter le centre d'expertise des enseignants et gestionnaires scolaires vers les programmeurs ou concepteurs de systèmes qui créent les modèles permettant de diagnostiquer les retombées sur l'apprentissage, prédire les accomplissements scolaires et déterminer les recommandations qui seront affichées et à qui. (Berendt et al., 2020, p. 317).

C'est le cas par exemple des tuteurs intelligents qui sélectionnent du matériel didactique à la place de l'enseignant, du diagnostic des élèves à risque ou bien de la prédiction de la réussite. Prenons l'exemple d'une enseignante de mathématiques qui crée elle-même une série d'exercices pour entraîner les élèves à la résolution d'équations du second degré. A priori, on pourrait penser qu'il s'agit là d'une tâche pouvant être automatisée, ou à tout le moins que du matériel pourrait être réutilisé. C'est peut-être le cas, mais on pourrait aussi chercher à comprendre pourquoi l'enseignante le fait, et faire des découvertes intéressantes. Elle pourrait le faire, par exemple, car elle enseigne auprès d'un groupe multiculturel et qu'elle ne trouve pas de problèmes mettant en scène des référents culturels pertinents pour son groupe. Elle pourrait aussi choisir d'utiliser volontairement une série d'exercices trop faciles pour ses élèves pour des considérations purement pédagogiques, par exemple pour renforcer momentanément leur confiance à réussir. Elle pourrait, tout compte fait, exercer son agentivité dans des situations polymotivées là où un système basé sur l'IA ne considérerait qu'un motif didactique.

> Une enseignante pourrait, tout compte fait, exercer son agentivité dans des situations polymotivées là où un système basé sur l'IA ne considérerait qu'un motif didactique.

Selon Berendt et al. (2020), cela pourrait aussi conduire à un recul de compétences des enseignants qui pourraient prendre l'habitude de s'en remettre aux décisions d'un système au détriment de leur propre expertise. Tout porte à croire que le biais d'automatisation (Parasuraman et Manzey, 2010) pourrait aussi s'appliquer, c'est-à-dire qu'il pourrait y avoir une trop grande confiance envers les décisions des systèmes basés sur l'IA. De plus, si ce risque n'est pas reconnu par les établissements scolaires, ceux-ci pourraient exiger l'utilisation de certains outils par les enseignants sous l'illusion de meilleures prédictions ou résultats (Jones et al., 2020).

L'agentivité inscrit toujours l'action dans un contexte élargi. Knox (2017) rappelle que les logiciels, algorithmes et bases de données sont toujours employés dans des contextes plus larges qu'il n'y paraît :

> L'utilisation des logiciels, algorithmes et bases de données, à travers plusieurs acteurs humains, est souvent envisagée comme trop détachée de l'éducation en tant que telle. Les données des étudiants sont soumises, et les enseignants sont encouragés à réagir face à ces données, sans pour autant qu'ils aient fait partie du processus responsable de leur production. (p. 737, traduction libre)

Pour Corrin et al. (2019), l'usage d'outils basés sur l'IA doit toujours impliquer l'intervention humaine, par exemple pour la révision des décisions contestées et des erreurs de classification. Gras (2019), s'appuyant sur le Règlement général sur la protection des données, parle de la « nécessité du maintien du contrôle par l'humain » (p. 4), et Knox (2017) souligne que la possibilité de refuser les recommandations d'un système d'IA doit être préservée et ce, sans conséquences négatives pour les enseignants. Se basant sur une préoccupation similaire, Sjödén (2020) questionne qui devrait avoir préséance en cas de divergence (note à une évaluation, action à poser, diagnostic de risque d'échec par exemple) : l'enseignant ou l'IA ?

L'importance de préserver l'agentivité est soulignée par plusieurs documents consultés. Adams et al. (2021) parle de laisser le choix aux enseignants de recourir ou non à des outils basés sur l'IA. Aiken et Epstein (2000) affirment : « il faut à tout prix préserver la capacité humaine de résoudre des problèmes et de réfléchir de façon rationnelle » (p. 166, traduction libre). Holmes et al. (2021) invite à ne pas tomber dans une glorification des progrès des systèmes informatiques qui diminuerait le rôle de l'humain. Sur la préservation de l'agentivité des enseignants, Smuha (2020) détonne parmi les autres documents consultés en affirmant que, dans la mesure où le choix d'utiliser ou non l'IA et de se fier ou non à ses recommandations demeure possible, l'agentivité peut être accrue :

Parmi tous les documents consultés, il y a convergence sur l'importance de préserver l'agentivité en laissant les enseignants libres d'utiliser ou non les outils basés sur l'IA. Et dès la conception, ces outils devraient être pensés pour augmenter l'agentivité et non la restreindre.

## RESULTATS RELATIFS A L'AGENTIVITE DES ELEVES

Les élèves devraient également pouvoir choisir d'agir conformément ou non aux recommandations d'un système d'IA (Roberts et al., 2017). À l'échelle de la situation d'enseignement-apprentissage, l'usage de l'IA peut retirer de l'agentivité aux élèves. Selon Bulger (2016), c'est le cas lorsqu'un système assigne des tâches scolaires. De façon spécifique à l'enseignement supérieur, Roberts et al. (2017) constatent le risque d'infantilisation des personnes si des systèmes cherchent à ludifier lorsque cela n'est pas nécessaire ou souhaité par les apprenants. West et al. (2020) évoquent aussi ce risque, en insistant sur la pertinence des voix étudiantes dans le processus de régulation des apprentissages. Leurs perceptions, commentaires, expériences sont des éléments à considérer dans les décisions pédagogiques et ne devraient pas être diminués face à un système recourant à l'IA.

> Les élèves devraient également pouvoir choisir d'agir conformément ou non aux recommandations d'un système d'IA.

À l'échelle du parcours scolaire d'un élève, des systèmes prédictifs qui s'appuient sur les données peuvent conduire à recommander des choix de programmes ou de cours (Jones et al., 2020). Ce problème s'inscrit plus généralement dans la problématique des bulles de filtre, où des algorithmes de recommandation parviennent bien à identifier les préférences à partir de données antérieures concernant une personne, mais parviennent mal à éveiller de nouveaux intérêts. C'est le cas, par exemple, de recommandations sur des plateformes musicales ou de visionnement de séries télévisées. Ce faisant, elles limitent l'agentivité (Jones et al., 2020) ou, à tout le moins, participent à redéfinir l'environnement dans lequel elle s'exerce. Regan et Jesse (2019) rappellent que ces usages, même s'ils peuvent sembler banals, ont une incidence sur la capacité des personnes à gérer leur vie librement.

Regan et Jesse (2019), s'appuyant sur Kerr et Earle (2013), présentent trois types de prédictions pouvant affecter l'agentivité : les prédictions qui permettent aux personnes d'anticiper les conséquences négatives, les prédictions qui orientent vers des décisions spécifiques, et les prédictions prescriptives qui réduisent les choix possibles. Selon Regan et Jesse (2019), les prédictions basées sur les conséquences réduisent peu l'agentivité des personnes, alors que les prédictions prescriptives la réduisent beaucoup. Prenons pour exemple la différence entre un système qui recommande automatiquement une série de ressources pour l'apprentissage, sans pour autant masquer d'autres ressources, et un autre qui les sélectionne et les intègre dans un parcours dit personnalisé. Le premier maintient une certaine agentivité alors que le second retire une forme d'agentivité, car il prend des décisions à la place de l'apprenant.

Au niveau des usages didactiques, Sjödén (2020) note que les systèmes basés sur l'IA peuvent intégrer de fausses informations à l'environnement dans lequel agit l'élève. Il présente trois types de procédés que pourraient employer des systèmes d'IA et qui pourraient poser des problèmes éthiques : les cas où les systèmes mentent, c'est-à-dire présentent une information volontairement inexacte, les cas où ils cachent de l'information en sélectionnant celle à présenter, les cas où ils entretiennent des croyances erronées chez les élèves. Sjödén (2020) pose la question : « Jusqu'à quel point est-il justifiable, d'un point de vue éthique, d'entretenir de telles illusions ? » (p. 293). Ici, le lien avec l'agentivité tient à l'authenticité de l'environnement dans lequel l'agentivité est exercée. Une agentivité appuyée par des informations partielles ou fausses est-elle réellement de l'agentivité ?

> Une agentivité appuyée par des informations partielles ou fausses est-elle réellement de l'agentivité ?

Reiss (2021), s'appuyant sur Puddifoot et O'Donnell (2018), souligne que des outils qui poursuivent une intention de facilitation des apprentissages peuvent entraver l'activité intellectuelle nécessaire à la formation des concepts :

> Puddifoot et O'Donnell (2018) défendent l'idée qu'une trop grande dépendance envers les technologies pour stocker l'information à notre place - information que nous devions auparavant mémoriser - peut être contre-productive et donner lieu à des opportunités ratées pour les élèves de former des abstractions et de réaliser des inférences à partir de nouvelles informations. (p. 4, traduction libre)

Même en présence d'outils visant à faciliter la tâche des apprenants, il peut être momentanément pertinent de conserver une activité intellectuelle précise pour le développement de certaines structures logiques de pensée. Par exemple, un outil qui préidentifie les passages importants dans un texte, pour éviter à l'élève de devoir le lire au complet, pourrait ne pas être souhaitable si l'intention pédagogique est le développement de la capacité de synthèse. Un peu comme d'autres l'ont fait avant au sujet de la calculette, Smuha (2020) parle du risque de développement d'une paresse intellectuelle à force d'interagir avec des machines plus performantes pour la réalisation de certaines tâches. Selon des parents d'élèves, certains outils pourraient même être des entraves à l'apprentissage si les élèves ne développent pas de recul critique et leur accordent une trop grande confiance (Qin et al., 2020).

> Par contre, quelques parents sont inquiets que l'usage de systèmes basés sur l'IA en éducation puisse faire en sorte que les élèves en deviennent dépendants et manquent de pensée critique. Cela explique le fait que les parents soient réticents à faire confiance à ces systèmes. (p. 1699, traduction libre)

Certains risques s'ajoutent lorsqu'il est question d'enseignement supérieur. Des systèmes trop guidés qui s'immiscent dans l'organisation du travail scolaire pourraient être perçus comme infantilisants par les principaux intéressés (Roberts et al., 2017). Ce risque est aussi supporté par West et al. (2020) selon qui les apprenants doivent être perçus comme des personnes capables de réguler eux-mêmes leurs apprentissages. Roberts et al. (2017) proposent de s'assurer que les étudiants ne soient jamais obligés d'agir en concordance avec les recommandations d'un système ou sur la base de ses indicateurs de performance. Il faut aussi souligner l'importance que les systèmes utilisés soient valides, fiables et capables d'accomplir les tâches pour lesquelles ils ont été conçus dans un contexte réel (Smuha, 2020).

# Discussion : quelques nuances et pistes pour se prémunir des risques

À la lumière des résultats, cette section revient de manière systématique sur deux des éléments abordés dans le contexte, à savoir le système technicien et le concept d'agentivité. Elle présente également quelques pistes pour soutenir le développement de l'IA en éducation tout en préservant l'agentivité des élèves et des enseignants.

## PAR RAPPORT AU SYSTEME TECHNICIEN

Suivant Ellul (1977), l'enseignement, comme à peu près tout, peut être compris comme un ensemble de techniques (formules pédagogiques, méthodes d'évaluation, etc.). Au fur et à mesure que ces techniques se complexifient, dans ce cas-ci par le recours à l'IA, la complexité sous-jacente aux techniques qui prévalaient jusqu'alors est masquée, opaque ou bien carrément niée. Les outils basés sur l'IA sont rendus possibles par des techniques, au premier plan desquelles se trouve l'apprentissage automatique (*machine learning*). Plus ils se développent, plus ils précisent la réalité dans laquelle les individus évoluent. En ce sens, les systèmes informatiques basés sur l'IA appliquant ces techniques en contexte éducatif ne peuvent pas être considérés uniquement comme des outils. Ils redéfinissent plusieurs paramètres de la situation éducative, par exemple le temps requis pour accomplir une tâche, ou bien la nécessité ou non de mémoriser certaines informations, la nécessité ou non de demander de l'aide pour accomplir une tâche, ou bien les possibilités d'interactions sociales.

> Les outils basés sur l'IA sont rendus possibles par des techniques, au premier plan desquelles se trouve l'apprentissage automatique (machine learning). Plus ils se développent, plus ils précisent la réalité dans laquelle les individus évoluent.

Le recours à l'IA en éducation s'inscrit aussi dans la théorie du système technicien par le déplacement de certains pouvoirs. Certains outils représentent « une forme de privatisation et de commercialisation en transférant le contrôle des programmes scolaires et de la pédagogie des enseignants et des écoles vers les entreprises à but lucratif » (Saltman, 2020, p. 199, traduction libre). Si ce déplacement de pouvoirs peut être souhaitable, par moments, pour des questions d'efficience ou d'innovation, il doit être fait prudemment en évaluant l'ensemble des conséquences qui peuvent en découler. Le développement de l'IA ne peut donc être envisagé que comme une façon d'améliorer l'expérience d'enseignement et d'apprentissage. Cette conception est insuffisante pour décrire la transformation que ces outils pourraient opérer. Insuffisante d'abord car elle ne considère pas l'IA comme un ensemble de techniques qui reconfigurent les actions des élèves et enseignants, et insuffisante aussi car elle omet de considérer que l'enseignement et l'apprentissage sont eux-mêmes alimentés par des techniques.

## PAR RAPPORT A L'AGENTIVITE

L'agentivité, telle que nous l'avons présentée, implique une initiative d'action dans des situations présentant des conflits de motifs. Tout usage de l'IA qui retire de l'espace pour prendre de telles initiatives limite l'agentivité. Il serait toutefois possible d'atténuer ce risque en orientant les systèmes, dès la conception, pour qu'ils augmentent l'agentivité. Selon Kerr et Earle (2013), cela pourrait se faire en concevant des systèmes qui réalisent des prédictions de conséquences et rendent ainsi disponible plus d'information pour que les personnes prennent elles-mêmes des décisions, plutôt que des systèmes basés sur des prédictions de préférence ou prescriptives. Une prédiction prescriptive vise à diminuer les options disponibles, alors qu'une prédiction de préférence permet d'orienter ou classer, sans diminuer les options disponibles. Comme le suggèrent Roberts et al. (2017), il est important que l'étudiant n'ait pas l'obligation d'agir sur certains indicateurs. Dans l'enseignement supérieur, s'ajoute le risque d'un guidage trop important des étudiants qui peut avoir un effet infantilisant sur ces derniers. C'est pourquoi, dans le domaine des traces d'apprentissages (*learning analytics*), le développement des tableaux de bord appelle à prendre en compte les besoins réels des étudiants.

## PISTES POUR LE DEVELOPPEMENT D'OUTILS BASES SUR L'IA PRESERVANT L'AGENTIVITE

Les résultats de notre recension des écrits soulèvent plusieurs risques relatifs au maintien de l'agentivité humaine à travers les usages de l'IA en éducation, autant pour le personnel enseignant que pour les élèves. Nous proposons ici trois pistes pour concevoir des technologies éducatives basées sur l'IA tout en préservant, voire renforçant, l'agentivité humaine.

Premièrement, l'IA devrait d'abord être employée non pas pour prendre des décisions, mais plutôt pour améliorer la qualité de l'information pour permettre aux personnes de prendre des décisions plus éclairées. Cela peut se faire par exemple en présentant les conséquences probables de décisions (Regan et Jesse, 2019) tout en maintenant de la transparence sur ce dont tiennent compte les prédictions et ce dont elles ne tiennent pas compte. L'utilisation de tels systèmes devrait demeurer facultative pour les enseignants (Knox, 2017).

> En éducation, l'IA devrait d'abord être employée non pas pour prendre des décisions, mais plutôt pour améliorer la qualité de l'information pour permettre aux personnes de prendre des décisions plus éclairées.

Deuxièmement, relativement à l'évaluation des apprentissages, il semble prometteur d'utiliser l'IA pour des rétroactions rapides, personnalisées et fréquentes, dans des contextes où la rétroaction de l'enseignant ne peut être réalistement attendue. Bien sûr, ces systèmes doivent être transparents, même auprès de l'élève, sur la façon dont la rétroaction a été établie. En cohérence avec le point précédent, ils doivent contribuer à augmenter l'information disponible auprès de l'élève, sans toutefois prendre de décisions relatives à la notation, par exemple. L'IA ne devrait pas être utilisée pour remplacer le jugement professionnel des enseignants pour des évaluations certificatives. Elle pourrait aussi servir à porter à l'attention de l'enseignant certains biais potentiels dans ses évaluations ou certaines incohérences.

Finalement, en ce qui concerne les aides à la tâche destinées aux élèves, elles pourraient être développées éventuellement, mais ce terrain semble incertain pour le moment. Les intentions pédagogiques des curricula ne permettent pas toujours, à un niveau micro, de distinguer quelle activité intellectuelle est essentielle de celle qui est instrumentale ou déjà acquise. Les usages de l'IA visant à engager des élèves dans des projets technocréatifs (Romero et al., 2017) et développer eux-mêmes des compétences dans le domaine semblent plus prometteurs à court terme et peuvent même être combinés à des usages a priori destinés à l'enseignant.

# Conclusion

En situation éducative, enseignants comme élèves font preuve d'agentivité au quotidien. Malgré tous les efforts pour modéliser des problèmes éducatifs et automatiser des solutions, par exemple pour l'évaluation des apprentissages, il faut garder en tête que ces problèmes peuvent être abordés depuis plusieurs points de vue et qu'ils font intervenir des dimensions humaines parfois intangibles, et des motifs contradictoires. C'est par l'agentivité que sont résolues ces situations au quotidien. Il faut donc éviter le piège réductionniste de concevoir des outils didactiques détachés du contexte dans lequel ils sont utilisés.

Les pistes étant certainement nombreuses, nous en avons proposé trois : (1) utiliser l'IA pour améliorer l'information décisionnelle plutôt que prendre des décisions éducatives, (2) utiliser l'IA pour des rétroactions formatives et non pour l'évaluation certificative, tout en entraînant les élèves et les enseignants à un recul critique, et (3) privilégier les usages qui engagent les élèves et leur permettent de développer un recul critique sur le fonctionnement de l'IA.

# Références

# CONCLUSION

Julie Henry
Université de Namur, Belgique

Ce livre blanc rassemble des textes discutant la place de l'intelligence artificielle (IA) en éducation au sens large. Une distinction a été faite entre l'IA comme un outil pédagogique ou comme objet d'apprentissage. D'abord, comme outil, elle peut servir à améliorer l'apprentissage (le point de vue de l'apprenant) ou bien à améliorer notre compréhension du processus d'apprentissage (le point de vue de l'enseignant et du formateur). Ensuite, comme objet d'apprentissage, l'IA doit être envisagée dans le contexte élargi d'une éducation à la citoyenneté numérique telle qu'elle est présentée dans le cadre de référence pour les compétences numériques DigComp 2.2. Cependant, les deux aspects ne doivent pas être étrangers l'un par rapport à l'autre. Il faut privilégier avant tout les usages qui engagent les apprenants et leur permettent de développer un recul critique sur le fonctionnement de l'IA. C'est l'émancipation de chaque citoyen en termes de pouvoir d'agir et d'engagement qui doit être visée, et non son aliénation numérique.

Il y a donc un enjeu démocratique à amener tous les citoyens à comprendre les principes de base de l'IA pour développer des pratiques éclairées, autonomes et critiques qui leur permettront d'appréhender les questions que cette technologie soulève. En effet, les connaissances que possèdent le grand public de l'IA sont assez limitées : nombreux sont ceux qui ne peuvent pas reconnaître quand ils interagissent avec une technologie faisant appel à l'IA (Siri, Alexa, etc.) et ne savent pas expliquer son fonctionnement (Druga et al., 2018 ; Williams et al., 2018 ; Druga et al., 2017). Cette méconnaissance, couplée aux discours médiatiques qui oscillent entre craintes et fantasmes, alimentent représentations biaisées et mauvaise compréhension des enjeux techniques et éthiques de l'IA. Bien souvent, les technologies faisant appel à l'IA sont réduites aux robots et sont supposées avoir des capacités qui sont loin de leurs capacités réelles.

Ainsi, en ce qui concerne les usages éducatifs de l'IA, il faut prendre conscience des limites tout en travaillant à rendre tangibles et exploitables les données collectées durant l'apprentissage. Pour pallier ces limites, il faut cocréer les technologies à usages éducatifs en impliquant davantage les principaux intéressés, à savoir les enseignants, les formateurs et les apprenants. Cela tient aussi pour la conception de dispositifs d'éducation à l'IA (Henry et al., 2021 et 2020).

Enfin, si jusqu'à il y a quelques années, peu de ressources pédagogiques existaient pour éduquer à l'IA dès le plus jeune âge, ces ressources commencent à exploser comme en témoignent les chapitres de la première partie. La littérature scientifique rapporte aussi plusieurs ressources ou démarches d'introduction à l'IA (p. ex. Sabuncuoglu, 2020 ; Vartiainen et al., 2020 ; Ho et al., 2019 ; Touretzky et al., 2019 ; Williams et al., 2019 ; Eaton et al., 2018 ; Kandlhofer et al., 2016 ; Heinze et al., 2010). Peu d'entre elles sont cependant accessibles sans effort aux enseignants qui ont généralement une expérience limitée de l'éducation à l'IA. Une formation reste donc à envisager, et pas seulement concernant les aspects techniques de l'IA puisque les défis sont également éthiques et sociétaux. L'approche pédagogique envisagée pour cette éducation se doit donc d'être interdisciplinaire et critique (Henry et al., 2018 ; Saariketo, 2014a et 2014b). Malheureusement, si une telle approche fait moins peur aux enseignants parce que moins orientée technique de prime abord, ils n'y sont pas pour autant mieux formés.

Dès lors, si la place de l'IA dans l'éducation se confirme, notamment à travers la lecture de ce livre blanc, les efforts à fournir pour s'assurer de sa bonne intégration restent encore nombreux et doivent faire l'objet d'une attention toute particulière pour garantir le développement personnel et social de chaque citoyen.

# Références